\documentclass[12pt]{JHEP3}

\usepackage{amsmath,epsfig}
\usepackage{amssymb,amsfonts}
\usepackage{latexsym}
\usepackage{bbm}
\usepackage[latin1]{inputenc}
\usepackage{subeqnarray}
\usepackage{xcolor}

\usepackage{graphicx}

\def\thema{\theta}

\usepackage{longtable}

\relax
\renewcommand{\theequation}{\arabic{section}.\arabic{equation}}
\def\be{\begin{equation}}
\def\ee{\end{equation}}
\def\bea{\begin{eqnarray}}
\def\eea{\end{eqnarray}}

\newcommand\fverb{\setbox\pippobox=\hbox\bgroup\verb}
\newcommand\fverbdo{\egroup\medskip\noindent%
                        \fbox{\unhbox\pippobox}\ }
\newcommand\fverbit{\egroup\item[\fbox{\unhbox\pippobox}]}

\newcommand{\bear}{\begin{eqnarray}}

\newcommand{\eear}{\end{eqnarray}}

\newcommand{\bsea}{\begin{subeqnarray}}
\newcommand{\esea}{\end{subeqnarray}}
\newbox\pippobox

\newcommand{\ga}{\gamma}

\newcommand{\ud}{\mathrm{d}}

\def\6{\partial}

\def\a{\alpha}
\def\g{\gamma}

\def\le{\left}
\def\ri{\right}

\def\pa{\partial}

\def\e{\epsilon}
\def\m{\mu}
\def\n{\nu}
\def\r{\rho}
\def\s{\sigma}
\def\t{\theta}
\def\sp{\;\;\;,\;\;\;}

\def\sq
\def\a{\alpha}

\def\l{\lambda}

\def\k{\chi}
\def\hri#1#2{\href{http://arxiv.org/abs/#1}{[ArXiv:#1]#2}}
\def\hre#1#2{\href{http://arxiv.org/abs/#1/#2}{[ArXiv:#1/#2]}}

\def\e{\epsilon}

\def\d{\delta}

\def\g{\gamma}
\def\k{\kappa}
\def\t{\tau}
\def\tb{{\bar \tau}}
\title{A holographic model for the fractional quantum Hall effect}
\author{{\large Matthew Lippert$^{1}$, Ren\'e Meyer$^{2}$, and Anastasios Taliotis$^{3}$}\\
$^1$
\href{http://iop.uva.nl/itfa}{Institute for Theoretical Physics}, University of Amsterdam, 1090GL Amsterdam, Netherlands\\
$^2$
\href{http://www.ipmu.jp/}{Kavli Institute for the Physics and Mathematics of the Universe (WPI)}, The University of Tokyo, Kashiwa, Chiba 277-8568, Japan\\
$^3$
\href{http://www.vub.ac.be/infovoor/onderzoekers/research/team.php?team_code=DNTK}{Theoretische Natuurkunde,} Vrije Universiteit Brussel and \href{http://www.solvayinstitutes.be/}{The International Solvey Institutes,} Pleinlaan 2 B-1050 Brussels, Belgium}

\preprint{\hfill 
IPMU14-0253}

\abstract{Experimental data for fractional quantum Hall systems can to a large extent be explained by assuming the existence of a $\Gamma_0(2)$ modular symmetry group commuting with the renormalization group flow and hence mapping different phases of two-dimensional electron gases into each other. Based on this insight, we construct a phenomenological holographic model which captures many features of the fractional quantum Hall effect.  Using an $SL(2,{\mathbb Z})$-invariant Einstein-Maxwell-axio-dilaton theory capturing the important modular transformation properties of quantum Hall physics, we find dyonic diatonic black hole solutions which are gapped and have a Hall conductivity equal to the filling fraction, as expected for quantum Hall states.  We also provide several technical results on the general behavior of the gauge field fluctuations around these dyonic dilatonic black hole solutions: We specify a sufficient criterion for IR normalizability of the fluctuations, demonstrate the preservation of the gap under the $SL(2,{\mathbb Z})$ action, and prove that the singularity of the fluctuation problem in the presence of a magnetic field is an accessory singularity.  We finish with a preliminary investigation of the possible IR scaling solutions of our model and some speculations on how they could be important for the observed universality of quantum Hall transitions.
}


\begin{document}


\section{Introduction}\label{intro}


Electrons in two spatial dimensions at low temperature and subject to a strong transverse magnetic field exhibit a set of remarkable and robust phenomena known as the quantum Hall (QH) effect.  As the magnetic field is varied, near certain values a mass gap appears, the longitudinal DC conductivity vanishes, and the transverse, or Hall, DC conductivity takes a precisely quantized value.  In units of $e^2/h$, this number is equal to the filling fraction $\nu$, the ratio of the charge density to the magnetic flux.  As a result of these striking observable signatures, the QH effect has been a focus of research for the thirty years since its initial discovery.

When the gapped QH states occur at integer values of the filling fraction, this behavior is well understood in terms of weakly coupled electrons in a magnetic field.  However, the QH states at rational, non-integer values of $\nu$ are less well understood theoretically, as they are inherently strongly coupled.  Some aspects of the fractional QH effect are partially understood in terms of Laughlin wavefunctions \cite{Laughlin:1983fy} or in terms of an effective composite boson description \cite{Zhang:1992eu}.

One of the most remarkable features of the QH effect is the duality transformation relating QH states at different filling fractions.  Such a duality was inferred from the striking observed regularities in the transitions between QH states. For example, varying the magnetic field induces transitions which obey a selection rule:  The filling fractions $p/q$ and $r/s$ of adjacent QH states always satisfy $|ps-qr = 1|$.  While transitioning between such QH states, the conductivity traces out a semi-circle in the $\sigma_{xx}-\sigma_{xy}$ plane.  Furthermore, for all such transitions, the slope of the Hall resistivity between plateaux scales at low temperature as
\be
\frac{d\rho_{xy}}{dh} \propto T^{-\alpha} \ ,
\ee
and the width in magnetic field of the transition region scales as
\be
\Delta h \propto T^{-\beta}
\ee
where measurements find $\alpha \approx \beta \approx 0.42 \pm 0.04$.  A more thorough review of these and other related phenomena is given in \cite{Burgess}.

These experimental results can all be elegantly explained as resulting from a  duality which maps QH states into each other and which commutes with the RG flow \cite{LutkinRoss, KLZ, Dolan:1998vr, Dolan:1998vs, Burgess:1999ug, Dolan:2000ab,Burgess:2001sy,dolan:2006zc}.  The duality group is approximately $SL(2,\mathbbm Z)$, but because the filling fractions of QH states which fit into the $\Gamma_0(2)$ hierarchy always have odd denominators,\footnote{There are also unusual QH states for which the filling fractions have even denominators; the most notable is the $\nu = 5/2$ state.  These enigmatic states are not related by duality to the odd denominator states and may be related to non-Abelian statistics. We will not address them in this work.} it is in fact the subgroup $\Gamma_0(2)$.  Under this duality, the complexified conductivity $\sigma = \sigma_{xy} + i\sigma_{xx}$ transforms as
\be\label{sigmatransform}
\sigma \to \frac{a\sigma + b}{c\sigma + d}
\ee
where $a,b,d \in \mathbbm Z$, $c \in 2 \mathbbm Z$ and  $|ab-cd| = 1$.  When the system is in a QH state, $\sigma = \sigma_{xy}$ which is just equal to the filling fraction $\nu$, implying that $\nu$ transforms in the same way as the conductivity. Predictions for the complex conductivity flow with temperature \cite{Dolan:1998vr, Dolan:1998vs, Dolan:2000ab} arising from the transformation \eqref{sigmatransform} have been experimentally observed in FQH transport measurements \cite{Murzinetal1,Murzinetal2}. There also exist supersymmetric models with the full $SL(2,\mathbbm{Z})$ duality \cite{dolan:2006zc}.

An exact $\Gamma_0(2)$ duality implies a QH state at every rational $\nu$ with an odd denominator.  In experimental systems, of course, there are not an infinite number of QH states.  Instead the duality is only exact at zero temperature and in the absence of impurities.  Disorder leads to localized states, and the QH phase then persists over a nonzero range of magnetic fields, yielding the iconic plateaux in Hall conductivity.

The fractional QHE, a universal phenomenon featuring strong coupling and a striking duality symmetry, presents a tempting target for holographic modeling.  The quest for a string theoretic description of the QHE is long-standing and, in fact, predates the development of gauge/gravity duality \cite{Brodie:2000yz, Bergman:2001qg, Hellerman:2001yv,Freivogel:2001vc}. There also exist several approaches based on effective field theoretical constructions \cite{Bernevig:2002eq,Golkar:2013gqa,Wu:2014dha}. A subset of these works may be motivated \cite{Wu:2014dha} but do not necessarily rely on holographic approaches.  In recent years, two  holographic approaches have attacked the problem in parallel.

One fruitful approach has been to construct bottom-up models in an Einstein-Maxwell-axio-dilaton theory, where QH states correspond to dyonic black hole solutions \cite{Kachru2, Burgess}.\footnote{Other relevant bottom-up QH models include the phenomenological model of \cite{KeskiVakkuri:2008eb} and an approach using a probe Dirac field in a dyonic AdS-Reissner-Nordstom background \cite{Gubankova:2010rc}.}  Charged black holes with running scalars and hyperscaling-violating Lifshitz near-horizon geometries have been extensively studied and employed in a variety of holographic contexts \cite{Charmousis:2010zz, Kachru1, Gubser:2009qt, Brynjolfsson:2010rx, Perlmutter:2010qu, Gouteraux:2011ce, Huijse:2011ef, Dong:2012se}.  Adding an axion allows the bulk theory to be endowed with an $SL(2,{\mathbbm R})$ electromagnetic duality, which is broken quantum mechanically to $SL(2,{\mathbbm Z})$.\footnote{In string theory, the duality group is $SL(2,R)$ only classically.  Because of the quantization of electric and magnetic charges, $SL(2,R)$ is broken to $SL(2,{\mathbbm Z})$.  Even in a bottom-up framework, it is reasonable to assume a similar effect operates here, generating a potential for the axio-dilaton which is only invariant under $SL(2,{\mathbbm Z})$. For more detail, see \cite{Burgess}.}  Although the fractional quantum Hall data can be explained by the action of the subgroup $\Gamma_0(2)\subset SL(2,{\mathbbm Z})$, it is sufficient to focus on the more symmetric case of an $SL(2,{\mathbbm Z})$-invariant theory.\footnote{For the purpose of constructing phenomenologically viable QH plateaux states and analysing their transport properties the difference between $\Gamma_0(2)$ and $SL(2,{\mathbbm Z})$ are minor as in both cases the IR will be governed by the same bulk solution. Differences will arise in the concrete RG flow structure of course and possibly in the detailled structure of the plateaux transitions, though the mechanism of the latter most probably will be the same.}  Using this duality, one can then map the relatively well understood electrically charged black holes into dyonic solutions, where the dyonic charges correspond to the charge density and magnetic field of the QH system.

Largely as a consequence of this $SL(2,{\mathbbm Z})$ duality, the models of \cite{Kachru2, Burgess} capture many observed features of the QH effect.  However, they suffer from the rather serious flaw that the putative QH states lack a hard mass gap. The conductivity, for example, vanishes not as an exponential of the temperature, as with a hard gap, but rather as a power, indicative of merely a soft gap.  In addition, as explained in \cite{Charmousis:2010zz}, even in a bottom-up approach, the parameters of the bulk Lagrangian are restricted by several physicality conditions.  Constraints on the gauge kinetic coupling and scalar potential have yet to be included in any QH model.

A complementary approach to holographic QH model building employs top-down D-brane constructions \cite{Bergman:2010gm, Jokela:2011eb, Kristjansen:2012ny}.\footnote{In addition, other interesting top-down models of the QH effect include \cite{Davis:2008nv, Fujita:2009kw, Hikida:2009tp, Alanen:2009cn, Belhaj:2010iw}.}  In these models, the fermions are represented by open strings at the $(2+1)$-dimensional intersection of a D$p$-brane and D$q$-brane with $\# ND= 6$.  Working in the limit where the D$q$-brane is a probe in a nonextremal D$p$-brane background, the relevant physics is encoded in the worldvolume theory of the probe D$q$-brane.  The two classes of probe brane embeddings correspond to different phases. The solution for a generic charge density and magnetic field is a black hole embedding, where the probe D$q$-brane crosses the black hole horizon, corresponding to an ungapped metallic state.  However, for one \cite{Bergman:2010gm, Jokela:2011eb} or several \cite{Kristjansen:2012ny} special values of the filling fraction, the system assumes a Minkowski embedding and the D$q$-brane smoothly caps off above the horizon, opening a hard gap for both charged and neutral excitations.  The scale of this dynamically induced mass gap is determined by the distance from the horizon to the tip of the D$q$-brane and is, therefore, entirely geometrical in origin.

While these brane systems elegantly model the phase transition between a QH fluid and the ungapped transition states nearby, they only incorporate a single or perhaps a few QH states with particular filling fractions.  Furthermore, they are not imbued with an $SL(2,{\mathbbm Z})$ duality and so do not incorporate all the related observed phenomena.\footnote{In these D-brane models, there is an $SL(2,{\mathbbm Z})$ duality acting on the boundary which yields alternatively quantized theories of anyons \cite{Jokela:2013hta}.  However, this is not a bulk electromagnetic duality, of the type used in the bottom-up QH models of \cite{Kachru2, Burgess}.}

Our goal is to find a holographic QH model which combines the best features of these two different approaches, namely the simplicity of incorporating the modular symmetry in the bottom-up approach with the gapped states of the D-brane approach.  In particular, we want to construct a consistent model with gapped QH states and which exhibits an $SL(2,{\mathbbm Z})$ duality. Our approach is to improve and extend the model of \cite{Kachru2}.  By adding a suitable $SL(2,{\mathbbm Z})$-invariant potential, we obtain charged dilatonic near-horizon solutions.  Imposing the necessary constraints, we find there is, in fact, a sliver of parameter space for which these solutions are both physical and gapped.

Adopting the approach of \cite{Kachru2}, we analyze first the electrically charged system.  We solve numerically for the full RG flow from the UV $AdS_4$ fixed point to the dilatonic scaling solution in the IR.  The $SL(2,{\mathbbm Z})$ duality then allows us to map pure electric solutions to dyonic solutions.  Unlike \cite{Kachru2, Burgess}, these dyonic solutions exhibit a hard mass gap. In particular, by computing their spectrum and conductivity, we argue that they more accurately represent true QH states.

Furthermore, while analyzing the vector fluctuations, we encountered the singularity of Schr\"odinger potential generically found in Einstein-Maxwell systems at a magnetic field, first observed in \cite{Edalati:2009bi}.  In  \cite{Edalati:2009bi}, this singularity was 
shown to not affect the calculation of the low frequency behavior of the current-current correlation functions. 
 We found, after further analysis, that this singularity is actually accessible and poses no obstruction to computing the fluctuations at any frequency at all.

The rest of this paper proceeds as follows.  In Sec.~\ref{axiodilaton}, we present the bulk $SL(2,{\mathbbm Z})$-invariant Einstein-Maxwell-axio-dilaton action and discuss how the $SL(2,{\mathbbm Z})$ acts.  We review in Sec.~\ref{CDBHreview} known results about extremal, charged dilatonic black holes in general Einstein-Maxwell-dilaton theories.  In particular, we focus in Sec.~\ref{constraints} on the constraints that our QH solutions will need to satisfy.  Then, returning the the $SL(2,{\mathbbm Z})$-invariant theory, we find similar black hole scaling solutions in the IR, first in Sec.~\ref{electricsolutions} with purely electric charge, and then, via an $SL(2,{\mathbbm Z})$-transformation, dyonic solutions in Sec.~\ref{dyonicsolutions}.  In Sec.~\ref{RGF} we study the UV fixed points and numerically find RG flows connecting them to the IR scaling solutions.  With these complete solutions in hand, we perform a general fluctuation analysis in Sec.~\ref{EMad}, which we use to compute the conductivity in Sec.~\ref{HallCon} and the mass gap in Sec.~\ref{sec:Gap}.  We conclude in Sec.~\ref{conclusions} by summarizing our findings and with open questions and directions for further research.  Many details of the calculations are reserved for the Appendices.


\section{$SL(2,{\mathbbm Z})$-invariant Einstein-Maxwell-axio-dilaton theory}\label{axiodilaton}

\subsection{Action and equations of motion}

As explained in Sec.~\ref{intro}, we would like to build an $SL(2,{\mathbbm Z})$ duality into our holographic model that maps different saddle points corresponding to distinct quantum Hall states into each other.  String theories invariant under $SL(2,{\mathbbm Z})$ symmetry are well known \cite{ek}.  In our case, the relevant bulk fields are the metric, which is an $SL(2,{\mathbbm Z})$ singlet, a complex scalar field $\tau=\t_1+i\t_2$ transforming by fractional linear transformations,
\be
\hat\tau={a \tau+b\over c\tau+d}\sp \ \ ad-bc=1\sp  a,b,c,d\in \mathbbm{Z} \ ,
\label{e1}
\ee
and a gauge field that is dualized by the $SL(2,{\mathbbm Z})$ action. The axion is given by $\tau_1$, and the imaginary part of $\tau$ can be written as $\tau_2 = e^{\g \phi}$, where $\phi$ is the canonically normalized dilaton.

We have specialized to a (3+1)-dimensional bulk as a holographic dual of a (2+1)-dimensional quantum Hall system.  The most general such action with at most two derivatives is \cite{Gibbons:1995ap}
\be
 S=S_{\hat g}+S_F+ S_V
\label{totalaction}\ee
with
\be
 \label{e3}
 S_{\hat g} = M^2\int \ud^{4}x\sqrt{-g}\le[R-{1\over 2}\left[G_{\t\t}(\partial \tau)^2 +2G_{\t\tb}\pa\t\partial \bar \tau+G_{\tb\tb}(\pa\tb)^2\right]    \ri]
\ee
\be
\label{gaugeaction}
  S_F = -\frac{M^2}{4} \int \ud^{4}x\left[\sqrt{-g}~\tau_2 F^2+\frac{\tau_1}{2}\epsilon^{\m\n\r\s}F_{\m\n}F_{\r\s}\right]
\ee
\be
 \label{potentialaction}
 S_V = M^{2}\int \ud^{4}x\sqrt{-g} \ V(\tau,\bar\tau) \ ,
\ee
taking into account that it is the Einstein frame metric which is $SL(2,{\mathbbm Z})$ invariant and where we have fixed the normalization of the various fields.\footnote{The issue of scalar corrections to gravitational couplings and the associated $SL(2,{\mathbbm Z})$ is subtle and frame dependent, but it is fully understood in string theory \cite{rr}.} Reality of the action implies that $V$ and $G_{\t\tb}$ must be real and that $G_{\t\t}^*=G_{\tb\tb}$.\footnote{Note that in our convention the scalar potential has the opposite sign as usual, such that $AdS_4$ fixed points have positive $V$ and are minima if they have only relevant perturbations.}

The potential $V  (\tau,\bar\tau)$ must be $SL(2,{\mathbbm Z})$ invariant, while the scalar metric $G_{ij}$ must be $SL(2,{\mathbbm Z})$ covariant. In particular this means that $G_{ij}$ must be invariant under $\t\to\t+1$ and must transform as
\be
G_{\t\t}'=\t^4G_{\t\t}\sp G_{\t\tb}'=\t^2\tb^2G_{\t\tb}
\label{e6}\ee
under $\tau\to -{1\over \t} $.

In this work, we will use a simpler version of this action, where the manifold of the complex scalar $\tau$ is the upper half plane.  In this case, the gravitational-scalar action \eqref{e3} specializes to
\be
 \label{gravtauaction}
 S_g = M^2\int \ud^{4}x\sqrt{-g}\le[R-{1\over 2\gamma^2}{\partial \tau\partial \bar \tau\over  \tau_2^2}\ri]
\ee
The remaining free parameters are the real number $\gamma$ and the scalar potential $V$.

Written in terms of the axion $\tau_1$ and dilaton $\phi$, the gravitational-scalar \eqref{gravtauaction} and gauge action  \eqref{gaugeaction} takes the usual Einstein-Maxwell-axio-dilaton form:
 \bea\label{gravphitau1action}
 S_g &=& M^{2}\int \ud^{4}x\sqrt{-g}\le[R-\frac12(\partial \phi)^2-{1\over 2}{e^{-2\gamma\phi}\over \gamma^2}(\partial \tau_1)^2\ri]
\\
 \label{malakies}
S_F&=& -{M^{2}\over 4} \int \ud^{4}x\left[\sqrt{-g}~e^{\g \phi} F^2+{\tau_1\over 2} \tilde{\epsilon}^{\m\n\r\s}F_{\m\n}F_{\r\s}\right]
\eea
Note that, in order to achieve $SL(2,{\mathbbm Z})$ invariance, we have chosen a specific relation between the gauge kinetic function $e^{\g \phi}$ and normalization of the axion kinetic term $\gamma^{-2}e^{-2\gamma\phi}$. Note also that in the absence of a potential, the action has continuous $SL(2,\mathbbm R)$ symmetry, where $a,b,c,$ and $d$ in (\ref{e1}) are real numbers.\footnote{In \cite{Burgess} it was claimed that the action used is the most general compatible with the $SL(2,{\mathbbm Z})$ symmetry. As indicated here, this is not true. Even in the absence of a potential, there remains a single real parameter, namely $\gamma$.}

The equations of motion stemming from (\ref{totalaction}) are the following:\footnote{We use $\epsilon$ to denote the Levi-Civita tensor, such that $\epsilon^{txyr} = \sqrt{-g}$, and $\tilde\epsilon$ to be Levi-Civita symbol, {\it i.e.} ${\tilde\epsilon}^{txyr} = +1$}
\be
	R_{\mu\nu}-{R+V\over 2}g_{\mu\nu}= {1\over 4\g^2\tau_2^2}\left[\partial_\mu \tau\partial_\nu\bar \tau +\partial_\nu \tau\partial_\mu\bar \tau-g_{\mu\nu}\le(\partial \tau\partial \bar \tau\ri)\right]+{\tau_2\over 2}\le[F^{\;\rho}_\mu\ F_{\nu\rho}-\frac{g_{\mu\nu}}4 F^2\ri]
\label{Einstein_eq}
\ee
\be	 \nabla_\mu\le(\sqrt{-g}\tau_2F^{\mu\nu}\ri)+\frac12  \tilde{\epsilon}^{\m\n\r\s}\partial_{\m}\tau_1 F_{\r\s} =0
\label{Maxwell_eq}
\ee
\be
{1\over 2\g^2}\partial_{\m}{\partial^{\m}\bar \tau\over \tau_2^2}-{i\over 2\g^2}{\partial \tau\partial \bar \tau\over \tau_2^3}+\partial_\tau V+{i\over 8}\left[F^2+\frac{i}{2}\tilde\epsilon^{\m\n\r\s}F_{\m\n}F_{\r\s}\right]=0 \ .
\label{tau_eq}
\ee
We will work primarily with a domain-wall Ansatz for the metric,\footnote{See appendix \ref{ABCtodomainwall} for the equations of motion in more general coordinates.}
\be
\label{domainwallmetric}
ds^2 = e^{2A(r)} \left(-f(r) dt^2 + dx^2 + dy^2\right) + \frac{dr^2}{f(r)} \ .
\ee
To study the quantum Hall effect, we must consider dyonic solutions including both nonzero charge density $q$ and magnetic field $h$; the bulk gauge fields are then
\bea
\label{dyonicansatz}
F_{rt} &=& A_t'= \frac{(q-h \tau_1)}{\tau_2} e^{-A} \\
F_{xy} &=& h  \ ,
\eea
where $q$ and $h$ are constants.  The equations of motion \eqref{Einstein_eq}, \eqref{Maxwell_eq}, and \eqref{tau_eq} then become
\be
\tau_2'^2+4 \gamma^2 \tau_2^2A'' + {\tau_1'}^2 = 0
\label{dwEOMA}\ee
\be
f'' + 3A' f' - e^{-4A} \left(  \frac{(q-h\tau_1)^2}{\tau_2} -  h^2\tau_2 \right)= 0
\label{dwEOMf}\ee
\be
\tau_1''+\left[ 3A' + \frac{f'}{f} - 2\frac{\tau_2'}{\tau_2} \right] \tau_1' + \frac{\gamma^2 h \tau_2 }{f}  (q-h\tau_1) e^{-4A} +  \frac{\gamma^2 \tau_2^2}{f}  \frac{\partial V}{\partial \tau_1}  = 0
\label{dwEOMaxion}\ee
\be
(\log \tau_2)''+ \left[ 3A' + \frac{f'}{f} \right](\log \tau_2)'+ \frac{\tau_1'^2}{\tau_2^2}  + \frac{\gamma^2\tau_2}{f}\left[ \frac{\partial V}{\partial \tau_2} + \frac{1}{2}e^{-4A} \left( \frac{(q-h\tau_1)^2}{\tau_2^2} - h^2\right) \right]=0
\label{dwEOMdilaton}\ee
\be
-\frac{1}{2}\left(\frac{\tau_2'}{\gamma\tau_2}\right)^2 +6{A'}^2 + 2A'\frac{f'}{f} - \frac{V}{f} +\frac{1}{2f}e^{-4A}\left( \frac{(q-h\tau_1)^2}{\tau_2}  + h^2 \tau_2 \right) = 0 \ .
\label{dwEOMphi}\ee


\subsection{$SL(2,{\mathbbm Z})$ transformation properties}

\label{axiodilatontransformation}
Putting the potential $V$ aside temporarily, the theory given by \eqref{gravtauaction} and \eqref{gaugeaction}  is invariant under $SL(2,\mathbbm R)$ \cite{Gibbons:1995ap}. However, in string theory we typically expect a breaking of $SL(2,\mathbbm R)$ down to $SL(2,{\mathbbm Z})$ by various instanton effects, which can be captured by including a $SL(2,{\mathbbm Z})$ invariant scalar potential.

The $SL(2,\mathbbm Z)$ symmetry acts covariantly on the gauge field.  Defining ${\tilde F}_{\mu\nu} = \frac12 \epsilon_{\m\n\r\s} F^{\r\s}$ as well as
\be
\label{Gdef}
G^{\mu\nu} = -\frac{2}{M^2\sqrt{-g}} \frac{\partial S}{\partial F_{\mu\nu}} = \tau_2 F^{\mu\nu} + \tau_1 {\tilde F}^{\mu\nu} \ ,
\ee
we can define the complex combinations
\bea
\label{calGFdef}
{\cal F}_{\mu\nu} = F_{\mu\nu} - i {\tilde F}_{\mu\nu} \\
{\cal G} = -{\tilde G}_{\mu\nu} -i G_{\mu\nu}\,.
\eea
With these complex fields, \eqref{Gdef} can be written as
\bea\label{Gdefsimple}
{\cal G} = {\bar \tau} {\cal F}\, ,
\eea
and we can define the transform of the gauge field under $SL(2,{\mathbbm Z})$ to be
\be\label{calGFtrafo}
\left(
\begin{array}{c}
\hat{\cal G}\\
\hat{\cal F}
\end{array}
 \right)
=
\left(
\begin{array}{cc}
a & b\\
c & d
\end{array}
\right)
\left(
\begin{array}{c}
{\cal G}\\
{\cal F}
\end{array}
 \right)\,.
\ee
With this definition, \eqref{Gdefsimple} is invariant.
From \eqref{calGFtrafo}, one can work out the transformation behaviour of the Maxwell tensor,
\be\label{Ftrafo}
\hat F_{\mu\nu} = (c \tau_1 + d)F_{\mu\nu} - c \tau_2 {\tilde F}_{\mu\nu} \ .
\ee
Note that the gauge action \eqref{gaugeaction} is not itself invariant.  The Maxwell equations and the Bianchi identity mix under $SL(2,{\mathbbm Z})$, yielding an $SL(2,{\mathbbm Z})$-invariant theory \cite{Sen:1992fr, Schwarz:1992tn}.

The modular transformation properties of the electric charge $q$ and the magnetic field $h$ in our general dyonic ansatz \eqref{dyonicansatz} can be inferred from the transformations of the gauge field \eqref{Ftrafo}.  For a general dyonic configuration \eqref{dyonicansatz}, the gauge field and dual gauge field are
\bea
F_{rt} &=& \frac{(q- h \tau_1)}{\tau_2} e^{-A}\,,\quad F_{xy} = h \\
{\tilde F}_{rt} &=&  h e^{-A}\,,\quad {\tilde F}_{xy} = - \frac{(q-h \tau_1)}{\tau_2} \ .
\eea
Acting on this solution with an element of $SL(2,{\mathbbm Z})$ yields, via \eqref{calGFtrafo}, a solution of the same form, but with charges $\hat q$ and $\hat h$ given by
 \bea  \label{qhsl2z}
&& \left(
\begin{array}{c}
\hat q \\ \hat h
\end{array}
\right)
=
\left(
\begin{array}{cc}
a & b \\ c & d
\end{array}
\right)
\left(
\begin{array}{c}
q \\ h
\end{array}
\right)
\eea
This transformation of the charges implies that the filling fraction $\nu \equiv \frac{q}{h}$ transforms as
\be\label{fillingfractiontransformation}
\nu \rightarrow \frac{a \nu+b}{c \nu+d}\,,
\ee
{\it i.e.}~$SL(2,{\mathbbm Z})$ acts on the filling fraction by standard modular transformations.


\subsubsection{Conductivity}
\label{conductivitytransformation}

For applications to the quantum Hall effect, we are particularly interested in the behavior of the longitudinal and Hall conductivities under modular transformations.  The conductivity tensor is defined by the equations
\be
\label{conductivitydef}
J^i = \sigma^{ij} E_j \ .
\ee
Our system is rotationally invariant, so $\sigma^{xx} = \sigma^{yy}$ and $\sigma^{xy} = -\sigma^{yx}$.    Then, following \cite{Kachru2, Burgess}, we note that the current $J^i$ is given holographically by the boundary value of $G_{ir}$.  From \eqref{calGFtrafo}, it follows that the current and electric field transform as
\be\label{calJEtrafo}
\left(
\begin{array}{c}
\hat{\cal J}\\
\hat{\cal E}
\end{array}
 \right)
=
\left(
\begin{array}{cc}
a & b\\
c & d
\end{array}
\right)
\left(
\begin{array}{c}
{\cal J}\\
{\cal E}
\end{array}
 \right)\ ,
\ee
where  we have defined ${\cal J} = i(J^x + i J^y)$ and ${\cal E} = E_x + i E_y$.  This implies, from \eqref{conductivitydef}, that the conductivity transforms as
\begin{align}
\label{str}
\hat\sigma=\frac{a \sigma+b}{c\sigma+d}, \hspace{0.1in}\mbox{where}  \hspace{0.1in}
\hat\sigma=\sigma_{xy}+i\sigma_{xx} \ .
\end{align}


\subsection{The $SL(2,{\mathbbm Z})$-invariant potential}
\label{modularinvariantpotential}

We will now return to the scalar potential \eqref{potentialaction} in our gravity system.  In order to preserve the modular invariance of the action, we need to choose a potential invariant under $SL(2,{\mathbbm Z})$ or any subgroup of it. One nice class of $SL(2,{\mathbbm Z})$-invariant functions is the generalized real-analytic Eisenstein series:
\be
E_s(\tau, \bar\tau)={\sum_{m,n\in Z}}' \left({|m+n\tau|^2\over \tau_2}\right)^{-s}
\label{aa14}\ee
where the prime stands for omitting the $m = n =0$ term.  These functions are the natural $SL(2,\mathbbm Z)$-invariant generalizations of the dilaton exponentials $e^{\gamma s\phi}$.
They are eigenfunctions of the Laplacian of the upper half plane with eigenvalues $s(s-1)$,
\be
\tau_2^2\pa_{\t}\pa_{\tb}E_s=s(s-1)E_s\,.
\ee
The functions \eqref{aa14} can be expanded for large $\tau_2$ as
\bea
\label{Eisensteinlargetau2}
E_s&=&2\zeta(2s)\tau_2^s+2\sqrt{\pi}\tau_2^{1-s}{\Gamma(s-1/2)\over \Gamma(s)}\zeta(2s-1) \nonumber\\
&&+{2\pi^s\sqrt{\tau_2}\over\Gamma(s)}\sum_{m,n\not= 0}\left|{m\over n}\right|^{s-1/2}K_{s-1/2}(2\pi \tau_2|mn|)e^{2i\pi mn\tau_1}
\label{aa15}
\eea
Those and further properties of the real analytic Eisenstein series can be found in e.g. \cite{Eisenstein1,Eisenstein2,Eisenstein3,Eisenstein4}.

In principle, we could choose any function of $E_s$ for the potential, but we restrict ourselves to the simplest choice\footnote{Instead of choosing the potential to be a single Eisenstein series, we could also have taken any function $V(E_s)$, possibly even depending on more than one $E_s$ with different values of $s$. There is a good reason to restrict to such a simple form: The Eisenstein functions $E_s$ have UV $AdS_4$ fixed points only at the orbifold points of the fundamental domain of $SL(2,{\mathbbm Z})$. For any $SL(2,{\mathbbm Z})$-invariant potential, the existence of these $AdS_4$ fixed points is ensured  by the transformation properties of the first derivatives of the potential. Sufficiently complicated functions $V(E_s)$, however, can have additional UV fixed points inside the fundamental domain or at other points of the boundary. Hence taking $E_s$ or monomials thereof is the minimal setup in the sense that it has the minimum number of possible holographic renormalization group flows to the IR fixed points and hence a preferable starting point of our investigation.}
\be
\label{EisensteinPotential}
V(\tau, \bar\tau) = E_s
\ee
Note that, for large $\tau_2$, the leading behavior is simply $V \sim \tau_2^s \sim e^{s \gamma  \phi}$ .   Although we will leave the parameters $s$  and $\gamma$ arbitrary for now, their allowed ranges will be significantly restricted by various constraints in Sec.~\ref{constraints}.


\section{Review of charged dilaton black holes}
\label{CDBHreview}

\subsection{Action and scaling solutions}

Since the potential \eqref{Eisensteinlargetau2} at large $\tau_2$ is dominated by a single exponential, this $SL(2,{\mathbbm Z})$ Einstien-Maxwell-axio-dilaton system has a class of solutions 
which flow in the IR to charged dilatonic black hole (CDBH) scaling solutions of the type exhaustively investigated in \cite{Charmousis:2010zz}.  
Therefore, before tackling the problem at hand, we will review the set-up of \cite{Charmousis:2010zz} and some of the relevant results. In addition, we will also review the simplified notation of \cite{Gouteraux:2011ce} which trades the two parameters $(\gamma,s)$ in our action with the dynamical scaling exponent $z$ and the hyperscaling violating exponent $\theta$.

The general Einstein-Maxwell-dilaton action in 3+1-dimensions considered in \cite{Charmousis:2010zz} is
 \be
\label{ActionLiouville}
S=M^{2}\int \ud^{4}x\sqrt{-g}\le[R-\frac12(\partial \phi)^2 - \frac{Z(\phi)}{4} F^2+ V(\phi)\ri] \, . \\
\ee
Focussing on solutions with running dilaton, where $\phi \to \pm \infty$ in the IR, it is assumed in \cite{Charmousis:2010zz} that the leading behavior of the potential for large $\phi$ is $V(\phi) = -2\Lambda e^{-\delta \phi}$ and the gauge kinetic function scales as $Z = e^{\gamma \phi}$.\footnote{Note that in the notation of \cite{Charmousis:2010zz}, $\Lambda$ is the energy scale of the leading IR potential and should not be confused with the cosmological constant.} This action can be obtained from \eqref{totalaction} simply by setting the axion $\tau_1 = \text{const}$ and assuming $\partial_{\tau_1} V = 0$, which is the case as $\tau_2\rightarrow \infty$. In general, the parameters $\gamma$ and $\delta$ are independent.  However, with our choice of potential \eqref{EisensteinPotential}, they are related: the large-$\phi$ behavior of our potential is $V \sim 2\zeta(2s) e^{s \gamma  \phi}$, and so \be\label{deltagammas}
\delta = -s\gamma\quad and \quad \Lambda = - \zeta(2 s)\,.
\ee

For zero-temperature boundary states with a nonzero charge, the goal is to find extremal, charged dilatonic black hole solutions. Taking a domain-wall Ansatz (\ref{domainwallmetric}) for the metric\footnote{For the relation of the domain wall metric to other common coordinates, see App.~\ref{ABCtodomainwall}.} and assuming the only $r$-dependent component of the gauge field is $A_{t}$, the one nontrivial component of Maxwell's equations is
\be
A_t' = \frac{q}{Z e^{A}} \ ,
\ee
 and the equations of motion for $\phi$ and Einstein's equations can be written as
\be
\label{CDBHeomA}
\phi'^2+4A''=0
\ee
\be
\label{CDBHeomf}
f''+3A'f'-\left({q^2 \over Z}\right) e^{-4A}=0
\ee
\be
\label{CDBHeomphi}
-{1\over 2}(\phi')^2+6{A'}^2+2A'{f'\over f}-{V\over f}+{1\over 2fe^{4A}}\left({q^2 \over Z}
\right)=0 \ .
\ee
As usual, the second-order equation for $\phi$ follows from differentiating the constraint \eqref{CDBHeomphi} and replacing $A''$ and $f''$ from \eqref{CDBHeomA} and \eqref{CDBHeomf}, respectively.

In \cite{Charmousis:2010zz}, the following extremal scaling solution to these equations was found (see (8.1a-d) of \cite{Charmousis:2010zz}):
\be
\label{CDBHsolnA}
A(r) = 
\frac{(\gamma - \delta)^2}{4} \log r
\ee
\be \label{putsaf}
\phi(r) = 
(\delta-\gamma) \log r
\ee
\be
\label{CDBHsolnf}
f(r) = \frac{-16 \Lambda}{wu} 
r^{v}
\ee
\be
\label{CDBHsolnAt}
A_t(r) = \frac{8}{w} \sqrt{\frac{v\Lambda}{u}} 
r^{\frac{w}{4}}
\ee
where
{\be\label{wvdef}
w = 3\gamma^2 - \delta^2 - 2\gamma\delta + 4\sp u = \gamma^2 -\gamma\delta + 2\sp
v = -\delta^2 + \gamma\delta + 2 \ .
\ee}
The electric charge is given by
\be\label{qeq}
q = \frac{2 
\sqrt{-v\Lambda}}{\sqrt{u}}\,.
\ee
We have chosen the radial coordinate such that the IR is always at $r=0$, as is obvious by the behavior of the scale factor \eqref{CDBHsolnA}. Note that this corresponds to a particular choice of IR integration constants necessary to match the number of UV integration constants for a well-defined boundary value problem, which have all been suppressed here, but are discussed explicitly in App.~\ref{IRparameters}.

As shown in Sec.~3.2 of \cite{Gouteraux:2011ce}, via the coordinate redefinition
\be
\label{r_in_terms_of_rho}
r = \rho^{-\frac{4}{(\g-\d)(\g+\d)}}
\ee
and an appropriate rescaling of $(t,x^i)$, the above extremal solution is diffeomorphic to a hyperscaling violating Lifshitz geometry:
\be\label{HSV}
ds^2 = \frac{w 
}{\Lambda  (\gamma -\delta )^2 (\gamma +\delta )^2}
 \rho^\theta \left[-\frac{dt^2}{\rho^{2z}} + \frac{d\rho^2 + dx^i dx^i}{\rho^2}\right] \ .
\ee
The dynamical critical exponent $z$ and the hyperscaling violating exponent $\theta$ are related to the parameters $\gamma$ and $\delta$ via\footnote{Here we only quote the results for (2+1)-dimensional boundary spacetimes, {\it i.e.}~$p=3$ in the notation of \cite{Gouteraux:2011ce}. The results for general $p$ can be found in Sec.~3.2 of \cite{Gouteraux:2011ce}.}
\bea
\label{zexponent}
z &=& {(\ga-\d)(\g+3\d)+4\over (\ga-\d)(\ga+\d)}={u+3v-4\over u+v-4}\\
\label{thetaexponent}
\theta&=&{4\d\over \ga+\d}={4(v-2)\over u+v-4}\,.
\eea
In the following subsection we will discuss the constraints on $(\gamma,\delta)$ or, respectively, $(z,\theta)$.


\subsection{Constraints}
\label{constraints}

The action \eqref{ActionLiouville} contains two free parameters, $\gamma$ and $\delta$, which translate into the two exponents $(z,\theta)$ characterizing the scaling properties of the solution. However, not all values of $\gamma$ and $\delta$ lead to acceptable holographic solutions. At zero temperature, many  extremal dilatonic black holes have naked singularities. Demanding that, in spite of these singularities, these solutions yield sensible physical states imposes constraints on the values of $\gamma$ and $\delta$.
\begin{enumerate}
\item Gubser's bound \cite{Gubser} demands that naked singularities become hidden behind a horizon when the temperature is nonzero.  For the charged dilatonic scaling solutions (\ref{CDBHsolnA}, \ref{putsaf}, \ref{CDBHsolnf}, \ref{CDBHsolnAt}), this constraint implies
\be\label{Gubser}
u>0\,,\quad w > 0\,,\quad v > 0\,.
\ee
In terms of $z$ and $\theta$, this translates to\footnote{Note that in the parametrisation \eqref{deltagammas}, the third inequality in \eqref{Gubser} in particular constrains the parameter of the Eisenstein series \eqref{aa14} to be  $s > 1$. This is reassuring, as \eqref{aa14} has a pole in the complex s plane, and hence \eqref{aa14} would not be a well-defined scalar potential. Furthermore, the hyper scaling violating exponent $\theta>0$ for $s>1$.}
\be
{2\thema+2z+4\over 2(z-1)-\thema}>0\sp {z-1\over 2(z-1)-\thema}>0\sp {2z-2\thema+2\over 2(z-1)-\thema}>0\,.
\ee
\end{enumerate}

Beyond requiring that our solutions make physical sense by requiring Gubser's bound above, we need to further narrow the parameter space to obtain the right physical properties for a quantum Hall system.  A key property of quantum Hall states is that they are gapped.  In particular, this implies that the longitudinal conductivity is exponentially suppressed at low temperatures, $\sigma_{xx} \sim e^{-\Delta/T}$, where $\Delta$ is the mass gap for charged excitations (see for example \cite{Panetal}). Existence of a discrete and gapped charged spectrum imposes two more constraints:

\begin{enumerate}
\item[2.]  If the black branes we are using to model the quantum Hall system smoothly settled into an extremal state, {\it i.e.}~exhibited a nondegenerate finite temperature horizon all the way to extremality, the system would be a conductor at any temperature. In order for the solutions to be gapped, it is necessary that the system undergo a first-order, Hawking-Page-like confinement phase transition as the temperature is lowered. In a holograpic RG flow from an $AdS_4$ fixed point in the UV to the IR geometries \eqref{CDBHsolnA}-\eqref{CDBHsolnAt}, such a transition occurs if and only if the temperature of the black brane diverges to $+\infty$ at small horizon radii $r_h\rightarrow 0$. The IR in the electric frame is governed by the nonzero-temperature version of \eqref{CDBHsolnA}-\eqref{CDBHsolnAt} ({\it c.f.}~Sec.~8.1~of~\cite{Charmousis:2010zz}), whose temperature diverges in the $r_h\rightarrow 0$ limit if and only if
\be
\label{TDunstable}
w < 2 (\gamma-\delta)^2\quad \Leftrightarrow \quad  {z\over  2(z-1)-\thema} < 0\,.
\ee
In this case below a certain temperature $T_{min}$, no black hole solution exists, and hence the system undergoes a phase transition to the confining thermal gas solution \eqref{CDBHsolnA}-\eqref{CDBHsolnAt} at a temperature $T_\ast>T_{min}$. The transition typically will generically be first order but can be continuous if the scalar potential and gauge kinetic function are tuned appropriately \cite{Charmousis:2010zz,Gursoy:2010kw}. The necessary relation between the temperature and the horizon radius is summarized in Fig.~\ref{fig:TrhSketch}. Since the metric is $SL(2,\mathbbm{Z})$ invariant, if \eqref{TDunstable} holds for a purely electric solution, then it will also hold for all the dyonic solutions to which it is related by an $SL(2,\mathbbm{Z})$ transformation.

\begin{figure}
	\centering
		\includegraphics[width=0.50\textwidth]{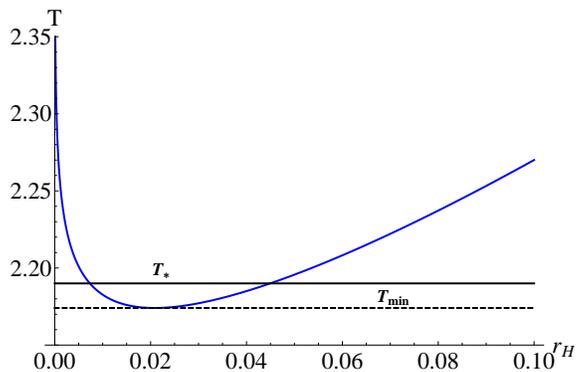}
	\caption{Schematic behaviour of the temperature-horizon radius relation for our choice of parameters {\protect\eqref{gammas}}. At small $r_H$ the temperature follows a power law, $T\sim r_H^{\frac{w-2(\gamma-\delta)^2}{4}}$, while at large $r_H$ it follows AdS-Schwarzschild asymptotics, $T\sim e^{\frac{r_H}{L}}$. Below $T_{min}$ there is only the thermal gas solution, and at $T_\ast$ a (generically first-order) confinement/deconfinement phase transition occurs.}
	\label{fig:TrhSketch}
\end{figure}

\item[3.] In order for the current excitations to have a discrete and gapped spectrum, we need to make sure that the associated Schr\"odinger problem for the vector fluctuations has a potential $V_{schr}$ that diverges both in the IR and the UV and hence supports only bound states.\footnote{See Sec. \ref{EMad} for the recasting of the fluctuation analysis as a Schr\"odinger problem.}

Let us start by imposing the condition that the Schr\"odinger potential diverges to $+\infty$ in the IR. 
For purely electric solutions ({\it c.f.}~Sec.~8.6 of \cite{Charmousis:2010zz}), if we transform to a Schr\"odinger coordinate $z$ defined by
\be
\frac{d z}{dr} = \frac{wu 
}{16\Lambda} r^{\frac{2(\gamma-\delta)^2 - w}{4}-1} \ ,
\ee
then the Schr\"odinger potential takes the form
\begin{align}
V_{s}=\frac{c}{{z}^2}\,,\quad c = \frac{2(\delta^2-\gamma^2-4)((\delta-\gamma)(2\delta+\gamma)-4)}{((\delta-\gamma)(3\delta+\gamma)-4)^2} \,.
\end{align}
This potential diverges to $+\infty$ as $z \rightarrow 0$ if $c > 0$.\footnote{This is the case only for thermodynamically unstable small black holes, which we have selected by imposing \eqref{TDunstable}.}
In terms of the scaling exponents $(z,\theta)$, this requirement reads\footnote{Note that the requirement \eqref{discreteIR} is nontrivial, {\it i.e.}~thermodynamical instability \eqref{TDunstable} does not automatically imply \eqref{discreteIR}. If the condition \eqref{discreteIR} is not obeyed, the Schr\"odinger potential diverges to $-\infty$ in the IR and the spectrum is continuous and ungapped. This case, however, is unacceptable from the point of view of the Sturm-Liouville problem: since the Schr\"odinger potential diverges to $-\infty$ as $V_{schr} \sim -|c|/z^2$, two normalizable modes exist, even for $|c|<1/4$, and hence the physics will depend on the resolution of the singularity. Once this case is excluded, the constraint \eqref{TDunstable} is a necessary and sufficient IR condition for the existence of a discrete and gapped spectrum.}
\be
\label{discreteIR}
c=(\theta -2) (4 z +\theta -8) > 0 \ .
\ee

For dyonic solutions, we will find in Sec.~\ref{sec:Gap} that the behavior of the Schr\"odinger potential in the IR \eqref{VsIR} implies a different constraint
\begin{equation}
\label{cdypos}
c=\frac{3 \theta ^2-68 \theta +44 \theta  z-8 z (4 z+7)+124}{4 (\theta -2 z+2)^2} > 0 \,.
\end{equation}
\textit{A priori} we need to impose both conditions \eqref{discreteIR} and \eqref{cdypos} separately. However, it turns out that \eqref{discreteIR} together with conditions 1, 2, and 4, imply \eqref{cdypos}, rendering the additional constraint redundant. 
In fact, we will show later in Sec.~\ref{sec:Gap} that a gapped spectrum for a purely electric solution implies a gapped spectrum for any dyonic $SL(2,\mathbbm{Z})$ solution. 

For any dyonic solution, the Schr\"odinger potential approaches a constant in the UV ({\it c.f.}~Sec.~\ref{Univ}). Hence, once \eqref{cdypos} is fulfilled, the spectrum of the current-current correlator is discrete and gapped for any dyonic solution. For purely electric solutions, on the other hand, conditions similar to the one derived in Sec.~4 and App.~F of \cite{Charmousis:2010zz} could be necessary to ensure a gapped spectrum.\footnote{The analysis of \cite{Charmousis:2010zz} cannot be valid here, as the so-derived condition on the operator dimension has no overlap with the region allowed by our constraints, while we have shown by explicit numerical integration that the spectrum of the current correlator is discrete and gapped in our case as well. We believe that the analysis there is not applicable in our case either because our ground state is a thermal gas solution and not a black hole settling into extremality smoothly as assumed in \cite{Charmousis:2010zz}, or because our form of the gauge kinetic coupling does not fulfill the there-assumed condition $Z'(\phi)=0$.} In principle, a general analysis as in the IR would be useful here as well. In view of the non-generic nature of the UV behavior of $SL(2,\mathbbm{Z})$-invariant scalar potentials and gauge kinetic functions such an analysis, however, is plagued by issues of model dependence, and we therefore defer it to a future work.  For the purpose of this work, we instead make do with the observation that we numerically checked the electric frame  Schr\"odinger potential for our case \eqref{gammas}, and found that it diverges in the UV as well. We hence have a discrete and gapped current-current correlator spectrum in all possible $SL(2,\mathbbm{Z})$ frames.

\end{enumerate}

Note that the constraint on the well-defined nature of the spin 1 fluctuation problem derived in Appendix D of \cite{Charmousis:2010zz} does not yield any additional restrictions in the case at hand.\footnote{Due to the combination of Gubser's constraint \eqref{Gubser} with the thermodynamic instability constraint \eqref{TDunstable}, only the third possibility mentioned in App.~D of \cite{Charmousis:2010zz} ($b<1$ and $2>\tilde a>-1$ in the notation used there) applies. This restriction is, however, trivially fulfilled once one includes a possible IR volume factor in front of the $dx^2+dy^2$ term in the IR geometry given by \eqref{domainwallmetric} and \eqref{CDBHsolnA}-\eqref{CDBHsolnAt}: Such a factor will replace the LHS of \eqref{qeq} by $q^2/Vol$, and hence the LHS of (D.18) of \cite{Charmousis:2010zz} in the same way. On the other hand, once we complete the IR geometry \eqref{CDBHsolnA}-\eqref{CDBHsolnAt} to an asymptotically AdS RG flow, this volume factor will become a function of UV data such as the sources of the scalar operators, the chemical potential and the magnetic field. We then can always choose the UV data of our RG flow to be large enough such that the modified version of (D.18) of \cite{Charmousis:2010zz} is fulfilled. Hence, there is no additional restriction from the well-definedness of the spin 1 fluctuation problem.}

Finally, we need to ensure that the purely electric solutions consist of a complete holographic RG  flow from an $AdS_4$ fixed point in the UV to the CDBH at $\tau_2 \to \infty$ in the IR.  These solutions will then correctly map under $SL(2,\mathbbm{Z})$ to dyonic solutions with the necessary properties to model quantum Hall states.  We will therefore need to impose two additional constraints:
\begin{enumerate}

\item[4.] From the behavior of the dilaton in \eqref{putsaf}, it is clear that, depending on the sign of $\gamma(\delta-\gamma)$,  in the IR both $\tau_2 = e^{\gamma \phi} \rightarrow 0$ or $\infty$ are possible.  For the electric solutions to flow to $\tau_2 = \infty$, the IR behavior of our potential \eqref{EisensteinPotential} requires us to impose the condition
\begin{equation}\label{t2infcond}
\gamma(\delta-\gamma) = -\gamma^2 (s+1) < 0\quad \Leftrightarrow s > -1\,.
\end{equation}
%
%

\item[5.] Although we are interested in the universal IR physics, we require our potential admit a valid UV completion.  In particular, we need $AdS_4$ solutions which can act as UV fixed points from which our solutions can begin to flow to the IR.  That is, there must be extrema of the scalar potential $\tau^{UV}$ that have relevant perturbations which can generate an RG flow to $\tau_2 \to \infty$.  Requiring that at least one direction be relevant implies that, for at least one direction $\tau_i$, we have $\partial^2_{\tau_i} V |_{\tau^{UV}}< 0$.  We will analyze fixed points and perturbations of our potential \eqref{EisensteinPotential} in Sec.~\ref{UVfixedpoints}.


\end{enumerate}

\begin{figure}[htp]
\centering
\includegraphics[height=0.5\textheight]{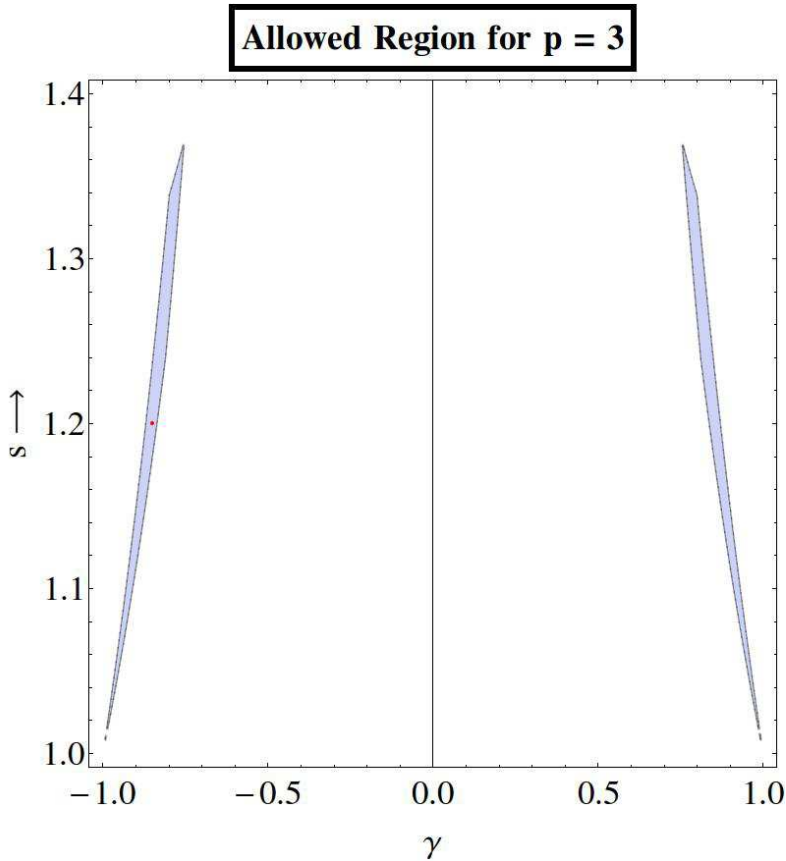}
\caption{The region in the $(\gamma, s)$-plane satisfying constraints 1-5 of Sec.~{\protect\ref{constraints}}. This plot is for the number of boundary space-time dimensions $p=3$ considered here. The red dot marks the values \protect\eqref{gammas} used in this work. Note that the allowed region actually extends to $\gamma=\pm 1$, $s=1$, where the Eisenstein series {\protect\eqref{aa14}} has a single pole.}
\label{fig:constraints}
\end{figure}

Overlaying these constraints severely restricts the allowed values of $\gamma$ and $s$; see Fig.~\ref{fig:constraints}. There is, fortunately, a tiny  allowed region, and in the rest of the paper we choose the following values:\footnote{If we also impose an acceptable Sturm-Liouville problem for the graviton sector ({\it c.f.}~Sec.~4 of \cite{Charmousis:2010zz}), then only the value $\gamma=-1$, $s=1$ is allowed. However, since we do not know how to regularize the singular behavior of the $E_1$ Eisenstein series, and since we are only interested in the charged sector, we drop this requirement for the time being. 
}
\be
\label{gammas}
\gamma = - 0.85\,,\quad s=1.2\quad\Rightarrow\quad\delta=-\gamma s=1.02\,.
\ee

\section{$SL(2,\mathbbm{Z})$-invariant charged dilaton black holes}
\label{IRscalingsolutions}

\subsection{Electric infrared solutions with constant axion}
\label{electricsolutions}

We now restore the axion and return to the $SL(2,\mathbbm{Z})$-invariant action \eqref{totalaction}.  We would like to generalize the charged, dilatonic scaling solutions from Sec.~\ref{CDBHreview} to the axionic case.  We will first consider only solutions with nonzero charge density, that is, radial bulk electric field, and defer the addition of magnetic fields to Sec.~\ref{dyonicsolutions}.

Turning off the magnetic field, $h=0$, the equations of motion (\ref{dwEOMA}, \ref{dwEOMf},  \ref{dwEOMaxion}, \ref{dwEOMdilaton}, \ref{dwEOMphi}, \ref{dyonicansatz}) reduce to
\be
\tau_2'^2+4 \gamma^2 \tau_2^2A'' + {\tau_1'}^2 = 0
\label{electricEOMA}\ee
\be
f'' + 3A' f' - \frac{q^2}{\tau_2} e^{-4A} = 0
\label{electricEOMf}\ee
\be
-\frac{1}{2}\left(\frac{\tau_2'}{\gamma\tau_2}\right)^2 +6{A'}^2 + 2A'\frac{f'}{f} - \frac{V}{f} + \frac{q^2}{2f\tau_2} e^{-4A} = 0
\label{electricEOMphi}\ee
\be
\tau_1''+\left[ 3A' + \frac{f'}{f} - 2 \frac{\tau_2'}{\tau_2} \right] \tau_1'  +  \frac{\gamma^2 \tau_2^2}{f}  \frac{\partial V}{\partial \tau_1}  = 0
\label{electricEOMaxion}\ee
\be
F_{rt} = A_t'= \frac{q}{\tau_2} e^{-A} \ .
\label{electricEOMgauge}\ee
As usual, the second-order equation for $\tau_2$ follows from differentiating the constraint \eqref{electricEOMphi} and replacing the second derivatives of the other fields from the other second-order equations.

To find IR scaling solutions, the simplest approach is to take the axion-free solution (\ref{CDBHsolnA}, \ref{putsaf}, \ref{CDBHsolnf}, \ref{CDBHsolnAt}) and see if we can also satisfy the $SL(2,\mathbbm{Z})$-invariant equations of motion (\ref{electricEOMA}, \ref{electricEOMf}, \ref{electricEOMphi}, \ref{electricEOMaxion}, \ref{electricEOMgauge}) with this Ansatz and a constant axion.

Focussing on the axion equation of motion \eqref{electricEOMaxion}, if we can find values of the axion which extremize the potential, $\frac{\partial V}{\partial \tau_1}=0$, then the axion can be set to a constant at one of these values.  In terms of the axio-dilaton, the exponentially running dilaton \eqref{putsaf} implies to $\tau_2 \to \infty$.  From the large-$\tau_2$ form of the potential \eqref{Eisensteinlargetau2} and the asymptotic expansion for the Bessel functions, we can compute
\be
\frac{\partial V}{\partial \tau_1} = i\frac{2\pi^{s+1}}{\Gamma(s)} \sum_{m,n\not= 0} \frac{|m|^{s-3/2}}{|n|^{s+1/2}}~mn~e^{-2\pi \tau_2|mn|} e^{2i\pi mn\tau_1} \ ,
\label{aa16}\ee
which vanishes exponentially as $\tau_2 \to \infty$ for any value of $\tau_1$.  Therefore, we find that $\tau \to \tau_1^{IR} + i\infty$ is a electrically-charged solution for any constant value of the IR axion $\tau_1^{IR}$.


\subsection{Dyonic infrared solutions with constant axion}
\label{dyonicsolutions}

For the the quantum Hall effect we need, in addition to a charge density, a background magnetic field.  Having found the pure electric solutions above, we will now restore the magnetic field $h$ to the equations of motion (\ref{dwEOMA}, \ref{dwEOMf}, \ref{dwEOMphi}, \ref{dwEOMaxion}, \ref{dyonicansatz}).

Our goal now is to find the dyonic analog of the electrically charged IR scaling solutions found in Sec.~\ref{electricsolutions}.  In principle, one could now attempt to find solutions to the equations of motion directly, but instead, we can use the $SL(2,\mathbb{Z})$ duality to generate dyonic solutions from the pure electric ones.

Let us start with the axio-dilaton of the pure electric solution, $\tau \to  \tau_1^{IR} + i\infty$.  Acting on $\tau$ with a modular transformation (\ref{e1}) maps it to
\be
\hat\tau = \frac{ac \tau_2^2 + (a\tau_1+ b)(c\tau_1 + d) + i \tau_2}{c^2 \tau_2^2 +(c \tau_1 + d)^2}
\ee
Taking the limit $\tau_2 \to \infty$, we obtain $\hat\tau = \frac{a}{c}$.  The pure electric axio-dilaton is therefore mapped under $SL(2,\mathbb{Z})$ to any real, rational value.  Note that the resulting $\hat\tau$ is independent of $\tau_1^{IR}$.

We turn now to the transformation of the gauge fields to see how the electric and magnetic charges behave.  Using \eqref{Ftrafo}, we obtain a dyonic gauge field as a modular transformation of an electric solution \eqref{electricEOMgauge}:
\be\label{Fxytransformation}
\hat F_{xy} = -c\tau_2 {\tilde F}_{xy} = -c q
\ee
\be\label{Frttransformation}
\hat F_{rt} = (c \tau_1 + d) F_{rt} =  (c \tau_1 + d)  \frac{q}{\tau_2} e^{-A}
\ee
We see directly from \eqref{Fxytransformation} that a dyonic solution has a magnetic field $\hat h = -c q$. Comparing \eqref{dyonicansatz} with \eqref{Frttransformation}, we find that
\be
\label{comparingcharges}
\hat q - \hat h \hat \tau_1 = (c \tau_1 + d) q \frac{\hat \tau_2}{\tau_2} \ .
\ee
Because $\tau_2 \to \infty$ and $\hat \tau_2 \to 0$, the right hand side of \eqref{comparingcharges} vanishes, implying that, for a dyonic solution, the axion is given by
\be\label{qoh}
\hat\tau_1 = \frac{\hat q}{\hat h}  \ ,
\ee
which we recognize as the filling fraction $\nu$.  We can see that the charge of a dyonic solution is then related to the charge of the pure electric solution by $\hat q = -a q$.

A dyonic solution, although it has both electric and magnetic charges, is a dilatonic scaling solution with only a nonzero magnetic field;  the electric field has been completely screened by the axion.  These solutions are in a sense S-dual to the pure electric solutions;  that is, the electric scaling solutions (\ref{CDBHsolnA}, \ref{CDBHsolnf}, \ref{putsaf}, \ref{CDBHsolnAt}) map under $\tau_2 \to 1/\tau_2$ and $q \to h$ into the pure magnetic solutions.

To satisfy the axion equation of motion \eqref{dwEOMaxion}, a dyonic solution must, of course, extremize the potential.  Since the potential is modular invariant, the $SL(2,\mathbb{Z})$ image of any extremum will also be an extremum.  We can also show this explicitly as follows:
\be
\label{chainrule}
\frac{\partial V}{\partial \hat\tau_1} = \frac{\partial V}{\partial \tau_1} {\partial \tau_1\over \partial \hat\tau_1} \ .
\ee
From \eqref{aa16}, we know the first factor vanishes exponentially at large $\tau_2$.  Computing the derivative from (\ref{e1}) at large $\tau_2$,
\be
{\partial \hat\tau_1\over \partial \tau_1} 
\to \frac{1}{c^2 \tau_2^2} \ ,
\ee
 shows that the second factor of \eqref{chainrule} diverges only quadratically.  Therefore, $\frac{\partial V}{\partial \tau_1}$ vanishes at all points where $\tau$ is real and rational.

We have now shown that the image of a pure electric solution under $SL(2,\mathbb{Z})$ is a dyonic solution with $\tau=\nu = q/h$ taking rational values.  We should note that $\tau$ is rational precisely because $SL(2,\mathbb{R})$ is broken to $SL(2,\mathbb{Z})$.  Recall that in Sec.~\ref{axiodilatontransformation}, we argued that the $SL(2,\mathbb{R})$ symmetry of the action is in fact limited to $SL(2,\mathbb{Z})$ due to the quantization of electric and magnetic charges.  The same reasoning implies that the filling fraction, and therefore the value of the axion, take rational values.


\section{Zero-temperature RG flows to quantum Hall states}
\label{RGF}

In this section, we will find zero-temperature renormalization group flows from AdS fixed points in the UV to dyonic, dilaton scaling solutions in the IR.  These solutions will serve as our model quantum Hall states.

We first construct flows inside the fundamental domain of the $SL(2,\mathbb{Z})$ action on the complex $\tau$ plane. These flows start from the conformal UV fixed points on the boundary of the fundamental domain and end in the IR at the charged dilatonic black hole geometry at $\tau \to \tau_1^{IR}+i\infty$.  These electric solutions holographically correspond to states with nonzero charge density but zero magnetic field.

Solving the equations of motion (\ref{dwEOMA}, \ref{dwEOMf}, \ref{dwEOMphi}, \ref{dwEOMaxion}, \ref{dyonicansatz}) for full flows from the UV to the IR is necessarily done numerically.  We will  reserve some of the more technical details of this calculation for App.~\ref{RGdetails}.

In the second step, these flows can then be mapped by $SL(2,\mathbb{Z})$ transformations to solutions starting from images of the $AdS_4$ fixed points and flowing to dilatonic black holes carrying both electric and magnetic charges at $\tau = \frac{a}{c}$.  We will show that these dyonic solutions are holographically dual to quantum Hall states with filling fraction $\nu = \frac{a}{c}$.


\subsection{UV fixed points}\label{UVfixedpoints}

The holographic renormalization group flows start at UV fixed points of relativistic conformal symmetry, corresponding in the bulk to asymptotically uncharged $AdS_{4}$ with no magnetic field near the boundary.   $AdS_4$ solutions with radius $L$ exist at extrema of the scalar potential,
\be
\label{AdSasymptotics}
A = \frac{r}{L}\,, \quad f = 1 \,,\quad L=\sqrt{\frac{6}{V(\tau_\ast,\bar\tau_\ast)}}\,, \left.\partial_\tau V\right|_{\tau_\ast} = \left.\partial_{\bar\tau} V\right|_{\tau_\ast} = 0\,,
\ee
where the AdS radius $L$ is determined by the value of the potential $V$ at a particular extremum $\tau_\ast$.\footnote{Note that in our convention the scalar potential in \eqref{potentialaction} is positive. The effective UV cosmological constant is then given by $2\Lambda_{eff} = - V(\tau_*,\bar\tau_*)=-6/L^2$.}
These solutions satisfy (\ref{dwEOMA}, \ref{dwEOMf}, \ref{dwEOMphi}, \ref{dwEOMaxion}, \ref{dyonicansatz}) provided the gauge fields are trivial and the scalars $\tau$ are constant.

\begin{figure}[htp]
\centering
\includegraphics[width=0.75\textwidth]{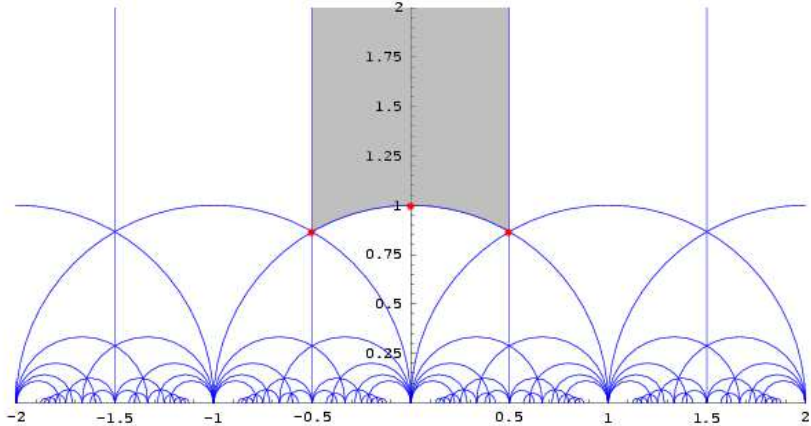}
\caption{The fundamental domain of $SL(2,\mathbb{Z})$ in the complex $\tau$ plane and the location of the $AdS_4$ critical points.}
\label{fig:AdS4CP}
\end{figure}

The problem now reduces to finding the extrema of $V(\tau,\bar\tau)$.  Invariance under $SL(2,\mathbb{Z})$ allows us to focus our attention on the fundamental domain (shown in Fig.~\ref{fig:AdS4CP}) and, in particular, on points on the boundary of the fundamental domain which are fixed points under some $SL(2,\mathbb{Z})$ transformation.  At such fixed points, continuity of the potential demands that $\partial_\tau V = 0$.

There are three such critical points:
\begin{equation}
\tau^{(0)} = i\,,\quad \tau^{(1)} = e^{\frac{\pi i}{3}}\, \quad \tau^{(2)} = e^{\frac{2\pi i}{3}}.
\label{AdS4FundDomain}
\end{equation}
The point $\tau^{(0)}$ is a fixed point of the inversion $\tau \mapsto -1/\tau$.  The points $\tau^{(1)}$ and $\tau^{(2)}$ are mapped into each other under both $\tau \mapsto -1/\tau$ and $\tau \mapsto \tau \pm 1$, and so they are fixed points of the combined mapping of an inversion followed by a shift, $\tau \mapsto \frac{-1}{\tau} \pm 1 =  \frac{\pm \tau - 1}{\tau}$. These three points, corresponding to $AdS_4$ solutions, are shown in Fig.~\ref{fig:AdS4CP}.

Having found the $AdS_4$ fixed points, we now need to investigate the spectrum of perturbations.  We will relegate some technical details to App.~\ref{UVperturbations}.  UV fixed points correspond to maxima of the potential and have relevant operators generating nontrivial flows toward the IR.  After gauge fixing, the general expansion around an asymptotically $AdS_4$ solution is
\begin{eqnarray}
\label{UVA}
A &=& \frac{r}{L} + \dots\\
\label{UVf}
f &=& 1 + f_1^{UV} e^{-3 r/L} + \dots \\
\label{UVtau1}
\tau_1 &=& \tau_1^{UV} + J_1 e^{-(3-\Delta_1) r/L} + O_1 e^{-\Delta_1 r/L} +\dots  \\
\label{UVtau2}
\tau_2 &=& \tau_2^{UV} + J_2 e^{-(3-\Delta_2) r/L} + O_2 e^{-\Delta_2 r/L} +\dots \\
\label{UVgauge}
A_t &=& \mu - \frac{L(q - \tau_1^{UV} h)}{\tau_2} e^{- r /L} + \dots\,.
\end{eqnarray}
There are three independent perturbations, $J_1$ and $J_2$ for the axion and dilaton and $q$ which adds charge,\footnote{Note that we have chosen to work in the canonical ensemble.} of which only dimensionless ratios are physically meaningful.

The operator dimensions $\Delta_1$ and $\Delta_2$ are related in the usual way to the masses of $\tau_1$ and $\tau_2$ at the UV fixed points; for $i=1,2$,
\be
\label{Deltapm}
\Delta_i = \frac{3}{2} + \sqrt{\frac{9}{4} - \gamma^2 \tau_2^2 L^2 \partial_{\tau_i}^2 V|_{\tau^{UV}}}
\ee
Note that the factor of $\gamma^2\tau_2^2$ arises because $\tau$ is not canonically normalized.\footnote{In a system with more than one scalar, the mass matrix {\it a priori} has to be diagonalized. The eigenvalues then yield the corresponding operator dimensions, and the eigenvectors yield the linearly independent scalar fields. We checked numerically that for the Eisenstein Series \protect\eqref{aa14} the mass matrix is diagonal at each of the fixed points \protect\eqref{AdS4FundDomain}, and hence \protect\eqref{Deltapm} applies.}
Due to the noncanonically normalized scalars, the Breitenlohner-Friedmann bound and the window in which alternative quantization is possible now becomes  $\gamma$-dependent,
\bea\label{BFbound}
&& \frac{9}{4} \geq \gamma^2 \tau_2^2 L^2 \partial_{\tau_i}^2 V|_{\tau^{UV}} \quad\text{Breitenlohner-Friedmann}\,,\\\label{AlternativeQuantization}
&& \frac{5}{4} \leq \gamma^2 \tau_2^2 L^2 \partial_{\tau_i}^2 V|_{\tau^{UV}} \leq \frac{9}{4}\quad\text{Alternative Quantization}\,.
\eea
Note that although we switch the identification of leading and subleading pieces in the asymptotic expansions of the scalars with the source and VEV of the dual operator for alternative quantization, the operator dimension is still given by \eqref{Deltapm}.

Computing $\Delta$ numerically, we find that at $\tau^{UV} = \tau^{(0)} = i$:
\bea
\Delta_1 &=& 3.100 \\ 
\Delta_2 &=& 2.449 \, .
\eea
Since $\Delta_2 < 3$, this implies $\tau_2$ is a relevant direction, while $\Delta_1 > 3$, so $\tau_1$ is irrelevant; $\tau^{(0)}$ is therefore a saddle point.

For $\tau^{UV} = \tau^{(1)}$ or $\tau^{(2)}$, we find
\be
\Delta_1 = \Delta_2 = 2.815
\ee
The dimensions are equal because, at these fixed points, the curvature is rotationally invariant.  Both directions are relevant, and so $\tau^{(1)}$ and $\tau^{(2)}$ are stable UV fixed points.  


\subsection{Electric solutions}
\label{electricflows}

Our next task is to construct RG flows from the UV fixed points to the IR scaling solutions, the charged dilatonic black holes.  We will first turn off the magnetic field and find flows to the purely electric solutions discussed in Sec.~\ref{electricsolutions}.  Although physically the RG flow proceeds from the UV to the IR, operationally, it is much more convenient to begin in the IR and solve out toward the UV.

We will only sketch the numerical method here;  the details of the perturbation calculation and the numerical shooting procedure are given in  App.~\ref{RGdetails}.  In order to generate initial data in the IR, we calculate the first-order correction to these scaling solutions, and find there are two independent perturbations.  We numerically integrate the equations of motion (\ref{electricEOMA}, \ref{electricEOMf}, \ref{electricEOMphi}, \ref{electricEOMaxion}) starting from an IR cutoff, having chosen an initial value of the axio-dilaton and the amplitudes for the two perturbations.  Arbitrary choices of the IR boundary data generate flows to the $AdS_4$ fixed points but, in general, will not yield the asymptotic behavior \eqref{AdSasymptotics}.  As explained in detail in App.~\ref{RGdetails}, we must adjust the IR boundary conditions until we find the correct UV behavior.

\begin{figure}[htb]
\begin{center}
	\includegraphics[width=0.65\textwidth]{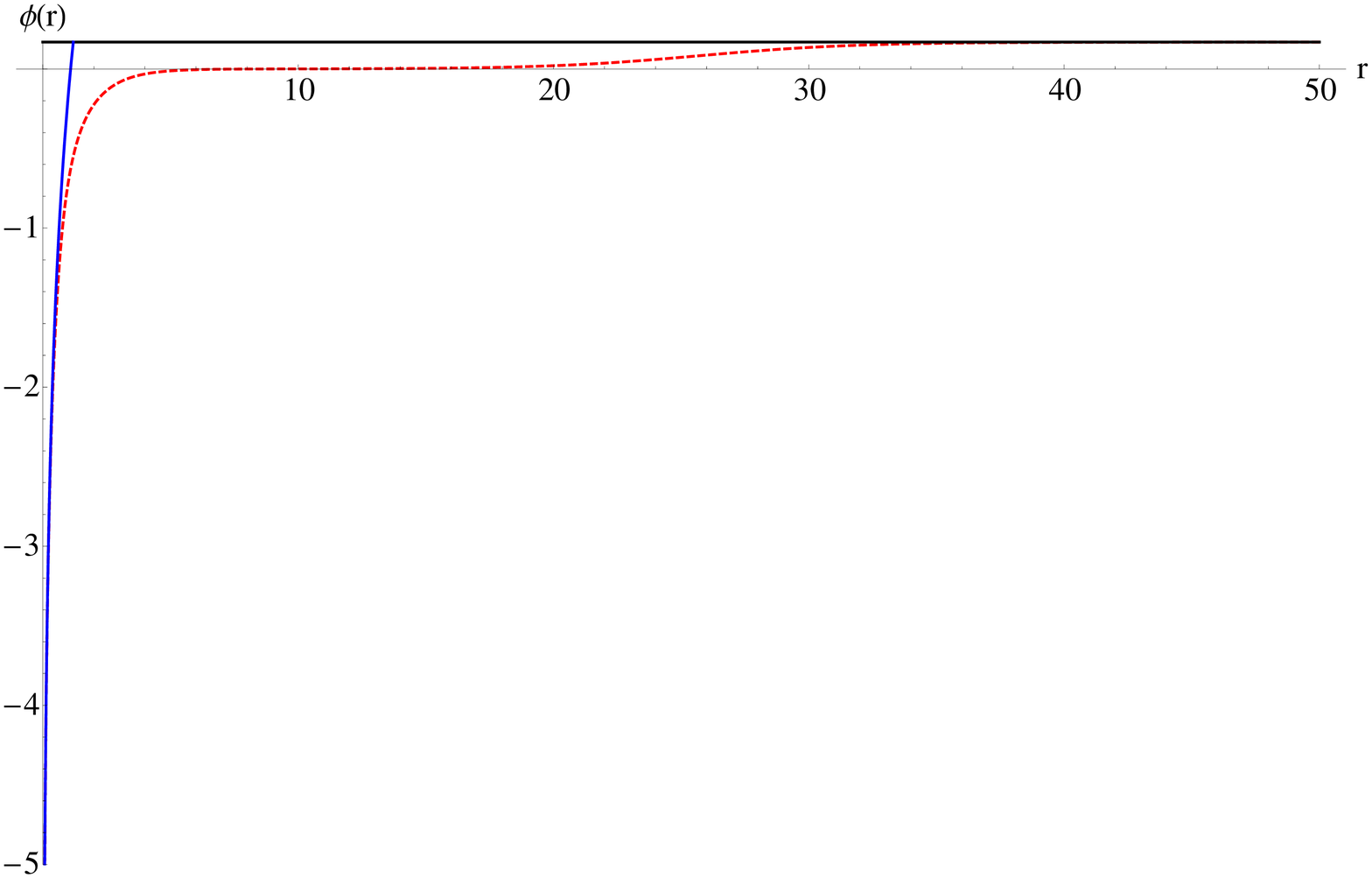} \\
	\includegraphics[width=0.65\textwidth]{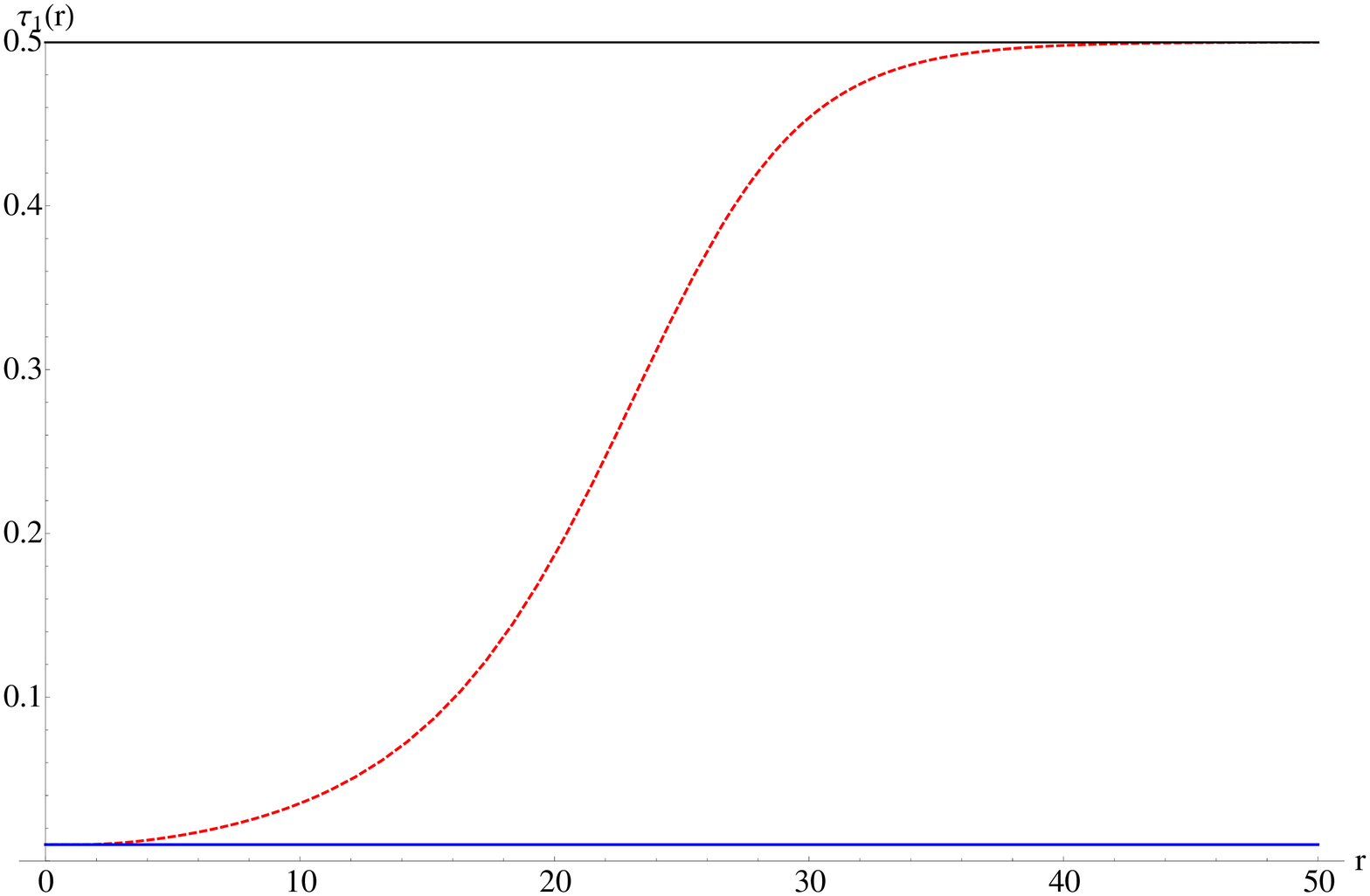}
	\caption{The evolution of (top) the dilation $\phi$ and (bottom) the axion $\tau_1$, shown in dashed red, for a representative flow from the UV $AdS_4$ fixed point $\tau=\tau^{(2)}=e^{\frac{2\pi i}{3}}$ (at large $r$) to the electrically charged dilatonic black brane at $\tau=i \infty + 0.01$ (at small $r$).  The solid blue curves represent the IR scaling solution, and the solid black curves are the UV fixed point.  For $5 \lesssim r \lesssim 20$, the solution is in a walking regime near the saddle point $\tau^{(0)} = i$, where $\phi \approx 0$.}
	\label{fig:flows2}
	\end{center}
\end{figure}

Fig.~\ref{fig:flows2} shows a typical flow starting from the stable UV $AdS_4$ fixed point $\tau^{(2)}$. We display only the more interesting behaviour of the axion and the dilaton field.  The blackening factor $f$ and the scale factor $A$ interpolate smoothly between the CDBHs \eqref{CDBHsolnA}-\eqref{CDBHsolnAt} in the IR and the $AdS_4$ asymptotics \eqref{UVA}-\eqref{UVgauge} in the UV. For Fig.~\ref{fig:flows2} we choose initial conditions such that the scalars, starting from the UV fixed point at the cusp, first flow close to the $\tau^{(0)}$ fixed point. Because the axionic direction at $\tau^{(0)}$ is irrelevant, the flow does not reach it and bends over towards the IR at $\tau_2=\infty$. In the vicinity of $\tau^{(0)}$ however the dilaton undergoes a walking regime, while the axion varies continuously from the UV value $\tau_1^{UV}=1/2$ to the IR value $\tau_1^{IR}=0.01$. 

By comparing the blue curves, which are the IR scaling solutions (\ref{CDBHsolnA}, \ref{CDBHsolnf}, \ref{putsaf}) with the red dashed curves of Fig.~\ref{fig:flows2}, we see that the IR scaling solutions are an excellent approximation to the full geometry for $\phi \lesssim -1$, corresponding to $\tau_2 \gg 1$.

\begin{figure}[htb]
	\begin{center}
	\includegraphics[width=0.75\textwidth]{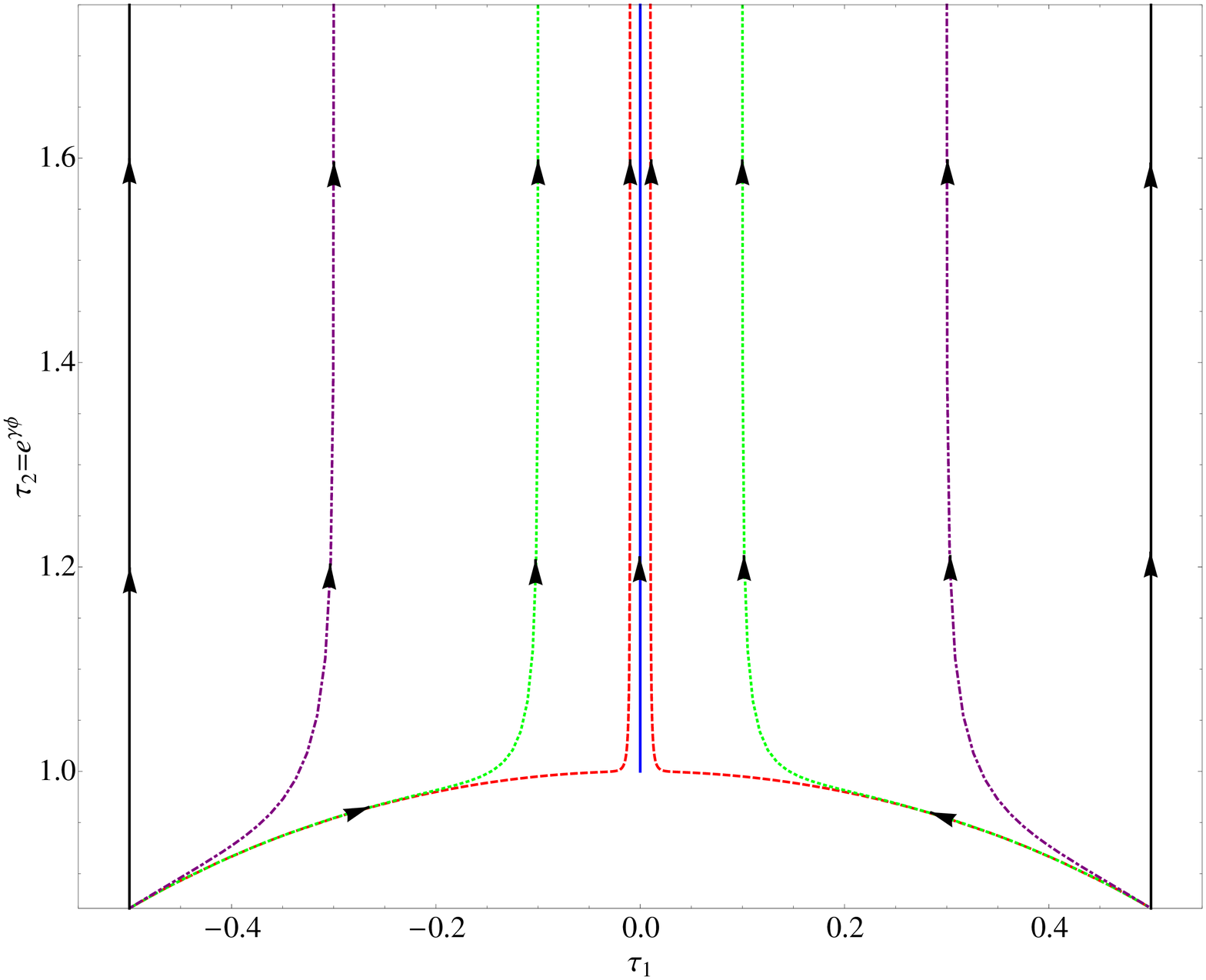}
	\end{center}
	\caption{The RG flow lines in the fundamental domain of the complex $\tau$ plane starting from $\tau^{(1)}$ and $\tau^{(2)}$ and flowing to the IR scaling solution at $\tau \to \tau_1^{IR} + i \infty$. We can label the solutions by their axion value $\tau_{1}$ in the IR: $\tau_{1}^{IR}=0$ (blue, solid), $\tau_{1}^{IR} = \pm 0.01$ (red, dashed), $\tau_{1}^{IR}=0.1$ (green, dotted), $\tau_{1}^{IR}=\pm 0.3$ (purple, dot-dashed), and $\tau_{1}^{IR} = \pm 0.5$ (black, solid). Note that due to the symmetry $\tau_{1}\leftrightarrow -\tau_{1}$ for purely electrically charged flows we could have restricted ourselves to e.g. the left hand side of the fundamental domain.}
	\label{fig:flows3}
\end{figure}

Fig.~\ref{fig:flows3} shows the family of flow lines generated by varying the IR value of the axion and foliating the fundamental domain of the complex $\tau$ plane.  Except for $\tau_1^{IR} = 0$, which flows from $\tau^{(0)}$, the different flows correspond in the UV to different directions in which the $AdS_4$ solutions $\tau^{(1)}$ and $\tau^{(2)}$ are perturbed.  As $\tau_1^{IR} \to 0$, the flows get closer to $\tau^{(0)}$ and the walking regime grows longer.

\subsection{Dyonic solutions}\label{dyonicflows}

Having obtained the RG flows to the pure electric dilaton black holes in the IR, we can now exploit the $SL(2,\mathbb{Z})$ duality of the equations of motion to map these flows into flows inside any image of the fundamental domain using \eqref{e1}.

As explained in Sec~.\ref{dyonicsolutions}, a pure electric scaling solution with charges $(q, 0)$ maps under $SL(2,\mathbb{Z})$ to a dyonic solution with
\be
(\hat q, \hat h) = (- a q, -c q)
\ee
and axio-dilaton $\hat \tau = a/c = \nu$.  These dyonic IR geometries are again the CDBH solutions (\ref{CDBHsolnA}, \ref{CDBHsolnf}, \ref{putsaf}) but with $\tau_2 \to 1/\tau_2$ and $q \to h$, with the axion completely screening the electric charge.  The UV $AdS_4$ fixed points $\tau^{(0)}$, $\tau^{(1)}$, and $\tau^{(2)}$ are simply mapped by $SL(2,\mathbb{Z})$ to other UV fixed points on the boundaries of the image of the fundamental domain.  The transformed RG flows connect the UV and IR fixed points, foliating the image of the fundamental domain.

\begin{figure}[htb]
	\begin{center}
	\includegraphics[width=0.75\textwidth]{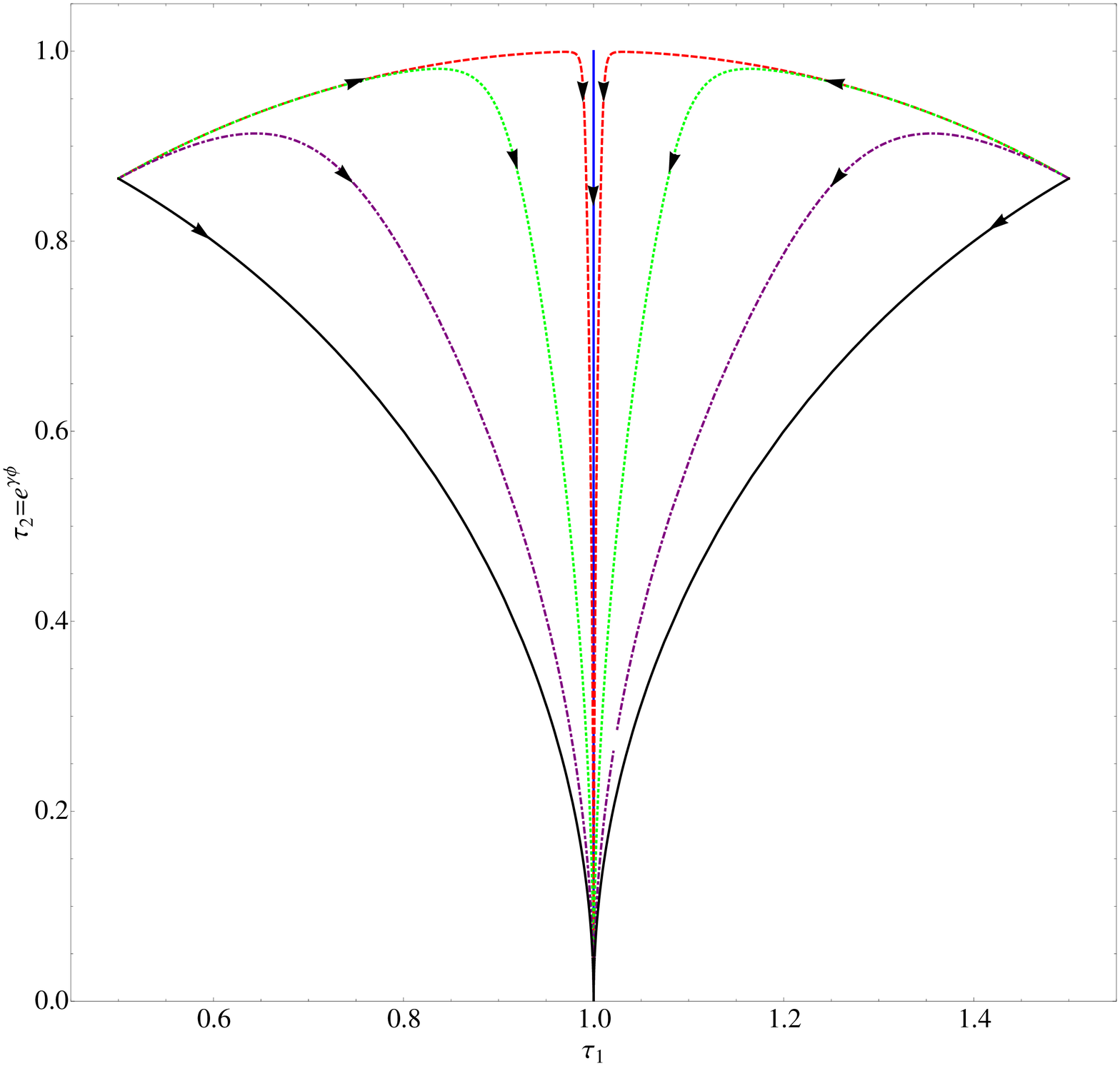}
	\end{center}
	\caption{Dyonic flows which end at $\tau^{IR} =1$ corresponding to filling fraction $\nu = 1$. The flows are obtained by transformation via \protect\eqref{e1} with $a=c=-b=1$, $d=0$.  The color coding is the same as in Fig.~\protect\ref{fig:flows3}.}
	\label{fig:flows4}
\end{figure}

As an example, we depict in Fig.~\ref{fig:flows4} the image of the flows of Fig.~\ref{fig:flows3} under the transformation \eqref{e1} with $a=c=-b=1$, $d=0$. The solutions begin in the UV at the fixed points $\tau = 1/2 + i \sqrt{3}/2$, $3/2 + i \sqrt{3}/2$, and $1+i$ and flow to the IR scaling solution at $\tau = 1$.

\section{Generic analysis for fluctuations in dyonic Einstein-Maxwell-axio-dilaton backgrounds}
\label{EMad}

In this section we investigate some rather generic features of time-dependent fluctuations on Einstein-Maxwell-axio-dilaton backgrounds that carry electric and magnetic fields.
In the first subsection, we show how the fluctuation equations for the spin-1 and spin-2 fields decouple. The resulting equations appear to have a singularity. The same singularity has already been encountered in the literature \cite{Edalati:2009bi}, where the authors 
employed a limit $\omega \ll h$ to calculate the low-frequency behavior of the conductivity, where $\omega$ and $h$ are the frequency of the fluctuation and the external magnetic field, respectively. 
We show in Sec.~\ref{access} that this singularity is an accessible singularity.  In fact, no $\omega \ll h$ restriction is needed, and the electric response can be calculated at any frequency directly. The results of this section will be useful in Sec.~\ref{HallCon} for computing the AC conductivity in the presence of a magnetic field. We will also rely on these results in our analysis of the mass gap in Sec.~\ref{sec:Gap}.


\subsection{Decoupling gauge and metric fluctuations}\label{dec}

In order to keep the analysis more general, we work with the choice of metric (\ref{generalmetric}), in which $g_{rr}=B(r)$ encodes the freedom of gauge choice for the radial coordinate $r$. 
We perturb the solutions of $A_{\mu}$ by time-dependent perturbations $e^{i\omega t}  a_x(r)$ and  $e^{i \omega t} a_y(r)$. Taking into account the backreaction on the geometry $e^{i \omega t}  h_{tx}$ and $e^{i \omega t} h_{ty}$ and raising the spatial indices of $h_{tx}$ and $h_{ty}$ using the background metric,\footnote{This simplifies the algebra considerably.} we obtain the following coupled set of equations to first order in the perturbations
\begin{subequations}\label{ahfla}
\begin{align}
\frac{\omega B}{D} \left(\omega a_x-ihh_t^y \right) +\frac{\sqrt{B/D}}{\tau_2}
\left( (q-h \tau_1)h'{_t}^{x}+(i\omega a_y-hh_{t}^{x})\tau_1' \right)
+   \frac{\left(\tau_2 \sqrt{D/B}a_x'\right)'  }{\tau_2 \sqrt{D/B}}=0, \label{axh}
 \\
\sqrt{DB}(\omega a_y+ih h_t^x) \left( q -h \tau_1  \right) +\left(\omega C^2 h'{_t}^{y} +i hD\tau_2a_x' \right)=0 \label{ahry}.
\end{align}
\end{subequations}
The first equation comes from the $x$-component of the Maxwell equations, while the second comes from the $ry$-component of the Einstein equations. A second set of equations arises from the $y$-component and the $rx$-component of  Maxwell and Einstein equations, respectively. In fact, these equations can be obtained from the set of equations (\ref{ahfla}) under the simultaneous interchanging of $i\leftrightarrow -i$ and $x \leftrightarrow y$.

The decoupling of the four equations is then obtained as follows. First, we define the variables
\begin{align}\label{ExEy}
E_x=\omega a_x-ihh_t^y  \hspace{0.1in}\mbox{and}  \hspace{0.1in} E_y= \omega a_y+ihh_t^x .
\end{align}
The second step is to substitute $h_t^x$ and $h_t^y$ from (\ref{ExEy}) in (\ref{ahry}) and solve for $a_x'$, obtaining
 \begin{align}\label{axp}
a_x'=\frac{\omega C^2 E_x'-i h\sqrt{DB} E_y (q-h\tau_1) }{\omega^2 C^2-h^2D\tau_2}
\end{align}
with a similar equation for $a_y'$ obtained from (\ref{axp}) under $i\leftrightarrow -i$ and $x \leftrightarrow y$. After some tedious but straightforward algebra, we obtain
\footnote{The calculation involves the following steps: One differentiates (\ref{ahry}) with respect to $r$ and substitutes to the resulting equation the $a_x''$ from (\ref{axh}) in order to get a simple second order equation for the metric fluctuation. The next step is to take a complex linear combination of the $h''$$_{t}^{x}$ equation together with (\ref{axh}) such that the variable $E_z''$ appears. The resulting expression contains the variable $E_z$ and its derivatives and also $a_x'$ and $a_y'$. Then, the equation of previous step can be expressed in terms of $E_z''$, $E_z'$, $E_z$ by exchanging $a_x'$ (and $a_y'$) from (\ref{axp}), resulting in (\ref{Ezz}).}
\begin{subequations}\label{Ezzg}
\begin{align}
g_0 E_z+g_ 1E_z'+  E_z''=0,\label{Ezz}\\
g_0=\frac{1}{D C^2 \tau_2 (D h^2 \tau_2 -
   C^2 \omega^2)}
  \Big [-B (D h^2 \tau_2 - C^2 \omega^2) \left(D \left((q - h \tau_1)^2 + h^2 \tau_2^2\right) -
      C^2 \tau_2 \omega^2\right) \notag \\
+  \sqrt{D B} C \omega \left ( h(q - h \tau_1) \left(C \tau_2 D' -2D \tau_2 C'+DC\tau_2'\right) +
     C (D h^2 \tau_2 - C^2 \omega^2) \tau_1' \right)\Big],\label{go}\\
g_1=\frac{
 2D \left( 2 D  h^2 \tau_2 C' -   C^3 \omega^2 \tau_2'\right)
    -  C \tau_2 (D h^2 \tau_2 + C^2 \omega^2) D' }{2 D   C \tau_2 (D h^2 \tau_2 - C^2 \omega^2)}-\frac{B'}{2B}\label{g1}
\end{align}
\end{subequations}
where 
\begin{align}\label{Ez}
E_z\equiv E_x+i E_y.
\end{align}
Note that the decoupled variable $E_z$ is a linear combination of both gauge field and metric fluctuations (see (\ref{ExEy})). 
One can now solve these equations numerically using appropriate boundary conditions in the IR in order to compute, for example, transport coefficients as a function of the frequency, the temperature, the magnetic field, and the charge density. 

For these calculations, it is convenient to recast the fluctuation equation (\ref{Ezz}) into Schr\"odinger form by the redefinition
\be
\label{measure}
E_z(r) = d(r)\psi(r) = e^{-\frac{1}{2}\int^r g_1 dr} \psi(r)\ .
\ee
Equation \eqref{Ezz} then becomes
\be
\label{psi}
\psi''-V_{schr}(\omega,H;r)\psi=0, \hspace{0.1in}V_{schr}=\frac{1}{4}\left( g_1^2+2g_1'  \right)-g_0.
\ee

These equations (\ref{ExEy}, \ref{axp}, \ref{Ezzg}, \ref{psi})  are generic in the sense that they apply for any Einstein-Maxwell-axio-dilaton configuration with a constant magnetic field, a metric ansatz that has the form of (\ref{generalmetric}), and time-dependent (but not spatially-dependent) transverse vector gauge and metric fluctuations.

Note that equations (\ref{axp}) and (\ref{Ezz}) seem to have a singularities when $D h^2 \tau_2 - C^2 \omega^2 = 0$, which at first glance might indicate some difficulty.  We elaborate on this issue extensively in the next subsection.


\subsection{Accessible singularity in dyonic backrounds}
\label{access}

Evidently, equations (\ref{axp}) and (\ref{Ezz}) for the gauge and metric fluctuations have a seemingly singular point $r=r_1$ which is defined by the root $r=r_1$ of equation
\be \label{bo}
b_0(r=r_1)\equiv D h^2 \tau_2 - C^2 \omega^2=0.
\ee
This is a generic feature of the fluctuation problem for Einstein-Maxwell-axio-dilaton theories in a constant magnetic field.  The same type of singularity has been encountered previously in \cite{Edalati:2009bi}. 
It was suggested in \cite{Edalati:2009bi} that the limit $\omega \ll h$ should be 
taken in order to push this singularity out towards the UV boundary and hence be able to solve the fluctuations from the IR to the UV.

However, we will show that the singularity in both equations, the $E_z$ and the $a_z \equiv a_x+ia_y$ equations, is accessible. This means that both of the two independent solutions of either of the fields is regular at the singularity.  This implies that neither of the two independent solutions of both fields is sensitive to the potential singularity and therefore, no extra limit such as $\omega \ll h$ is in fact needed.  In particular, the Schr\"odinger problem is well defined for any magnetic field and any frequency.  We sketch the proof here and relegate further details for App.~\ref{pf}.

We begin by re-writing $g_0$ and $g_1$ from (\ref{go}) and (\ref{g1}) as
\begin{subequations}\label{gog1p}
\begin{align}
g_0&=B \Big[- \frac{h^2 D \tau_2-C^2 \omega^2}{DC^2}-\frac{( h\tau_1-q)^2}{C^2 \tau_2} \notag \\
  &+ \frac{\omega \tau_1'}{\sqrt{D B} \tau_2} + \frac{
  (\tau_1-\frac{q}{h}) \omega}{\sqrt{D B} \tau_2} \left(\frac{ -h^2 }{C}\frac{ \tau_2 D'C +
      D C \tau_2' - 2 D C'^2 \tau_2 }{
    D h^2 \tau_2 - C^2 \omega^2}\right)\Big] \label{gop} \\
    g_1&=\log \left( \sqrt{\frac{D}{B}}\frac{C^2 \tau_2}{C^2\omega^2-D\tau_2h^2}   \right)' \label{g1p} ~ .
\end{align}
\end{subequations}
Expanding the previous expressions in the neighborhood of $r=r_1$ and using (\ref{bo}) yields
\begin{subequations}\label{gs}
\begin{align}
\hspace{-1in}g_0& \sim \frac{k_0(r_1)}{r-r_1}+k_1(r_1)\,\,\, \mbox{where}\notag\\
k_0&=-\frac{\omega B (\tau_1-q/h)}{\sqrt{DB}\tau_2}, \,\,\,
k_1=k_0'+k_0\left(\frac{b_0''}{2b_0'}-2\frac{C'}{C}\right)+\frac{B}{\tau_2}\left( \frac{\omega \tau_1'}{\sqrt{DB}\tau2}-  \frac{B (h\tau_1-q)^2}{C^2\tau_2}\right)
 \label{goex}\\
    g_1& \sim -\frac{1}{r-r_1}+m_1(r_1), \,\,\, \mbox{where} \,\,\,\,m_1=-\frac{b_0''}{2b_0'}+\log \left( \sqrt{\frac{D}{B}} C^2 \tau_2 \right)' .
    \label{g1ex}
\end{align}
\end{subequations}

As we show in App.~\ref{pf}, the statement that the singularity of (\ref{Ezzg}) is accessible is equivalent to $k_0^2+m_1 k_0+k_1\big|_{r=r_1}=0$,\footnote{The proof fails when the root $r_1$ of $b_0$, equation (\ref{bo}), is a multiple root. 
} {\it i.e.}
\be
k_0^2+m_1k_0+k_1\big|_{r=r_1}=0 \,\,<=>\,\, \mbox{accesible singularity}.\label{kmk}
\ee
One can then verify that
\be
k_0^2+m_1k_0+k_1\big|_{r=r_1}=\frac{B}{D C^2 \tau_2^2} (\tau_1-q/h)^2 b_0\big|_{r_1}=0\,\,\,\mbox{because}\,\,\, b_0(r_1)=0.
\ee

We thus have shown that the two independent solutions $E_z$ of (\ref{Ezz}) are completely regular at $r=r_1$. We still have to show that $a_z$ and $h_t^z$ are separately regular at $r=r_1$. In order to show this, one begins by writing the general solution of $E_z$ in the neighborhood of $r_1$. Using (\ref{kmk}) to eliminate $k_1$, the general solution becomes
\be\label{Ezr1}
E_z=C_1 e^{k_0 (r-r_1)}+C_2 e^{-\frac{(k_0+m_1)(2 k_0+m_1)}{2 k_0+r_1} (r-r_1)} \left(1+(2 k_0+m_1(r-r_1))\right).
\ee
where $C_1$ and $C_2$ are arbitrary constants. As expected the general solution is completely regular at $r=r_1$.
The next step is to substitute (\ref{Ezr1}) inside $a_z'=(a_z+i h_t^z)'$, which can be obtained by using (\ref{axp}) and its $y$-counterpart. We then note that the numerator of $a_z'$
is proportional to $b_0$ and hence proportional to the denominator of $a_z'$. This implies that in the neighborhood of $r=r_1$, the potential singularity of $a_z'$ cancels out. Since $E_z$ and $a_z$ are regular and at $r=r_1$, equations (\ref{ExEy}) and (\ref{Ez}) imply that $h_t^z$ is also regular at $r=r_1$. 

In addition, we can show that the equivalent Schr\"odinger potential $V_{schr}$ (see (\ref{psi})) in the neighborhood of $r=r_1$ exhibits a universal behavior.  By substituting the leading behavior of $g_0$ and $g_1$ from (\ref{gs}) into (\ref{psi}), we find 
\be \label{Vsr1}
V_{schr} \sim \frac{3}{4(r-r_1)^2}, \,\,\,\,  r \sim r_1.
\ee


\subsection {$SL(2,\mathbb{Z})$ covariance of the fluctuations}

Here we discuss how the fluctuations and their equations of motion transform under $SL(2,\mathbb{Z})$. Given that the metric is an $SL(2,\mathbb{Z})$ invariant and using (\ref{Ftrafo}), we  obtain
\begin{subequations} \label{newaxy}
\begin{align}
\hat{a}_x=-& \frac{i}{\omega} \left(i\omega a_x(d+c\tau_1) -c\left( h_t^y(q-h \tau_1) +\sqrt{\frac{D}{B}}\tau_2 a_y' \right)\right)\label{newax}\\
\hat{a}_y= -&\frac{i}{\omega} \left(i\omega a_y(d+c\tau_1) +c\left( h_t^x(q-h \tau_1) +\sqrt{\frac{D}{B}}\tau_2 a_x' \right)\right)\label{neway}\\
\hat{h}_t^x=&h_t^x, \,\,\,\,\,\,\,\, \tilde{h}_t^y=h_t^y
\end{align}
\end{subequations}
where $c$ and $d$ are $SL(2,\mathbb{Z})$ parameters and where the hats denote the fields in the new $SL(2,\mathbb{Z})$ frame.



The fluctuation equations transform as follows. The two scalars' equations of motion transform as linear combinations of the old equations of motion in the initial $SL(2,\mathbb{Z})$ frame. The same happens with the gauge field. In particular, the Maxwell equations transform as linear combination of the Maxwell equations and the Bianchi identities of the initial $SL(2,\mathbb{Z})$ frame. Hence the equations of motion for the scalars and for the gauge field transform covariantly. On the other hand, the Einstein's equations are invariant under $SL(2,\mathbb{Z})$ and hence these equations, with some abuse of terminology, transform as scalars.

As a cross check of the consistency of (\ref{newaxy}) with the equations of motion (\ref{ahry}) and the $SL(2,\mathbb{Z})$ transformation rules, we perform the following exercise. We consider the hatted version of (\ref{ahry}) (in a new $SL(2,\mathbb{Z})$ frame). This implies that not only the gauge field transforms but one should also transform the scalars and the $q$ and $h$ as well. Then, we substitute the new fields and parameters in terms of the old ones using (\ref{e1}), (\ref{qhsl2z}) and (\ref{newaxy}). The resulting expression contains
the second derivative of $a_x$ because (\ref{neway}) contains $a_x'$ and (\ref{ahry}) is already a first order in $\tilde{a}_y$. The last step is to exchange $a_x''$ from the Maxwell equation (\ref{axh}) and use the fact that $a d-b c=1$. The resulting expressions amazingly simplify and yield equation (\ref{ahry}) with all the tilde symbols dropping out, as would be expected.


\section{Conductivity}
\label{HallCon}

We now turn our attention to the task of computing the conductivities of the solutions found in Sec.~\ref{RGF}.  Our objective is to show that the DC conductivities of the dyonic solutions match the expected results for quantum Hall states.  However, following the general strategy we have employed so far, we will first tackle the relatively easier problem of calculating the conductivity of the pure electric solutions.  We can then use $SL(2,\mathbb{Z})$ duality to find the results for the dyonic solutions.  Our procedure is quite similar and yields results comparable to those found in \cite{Kachru2}.

In order to keep the analysis more general, we will continue to work here and in Sec.~\ref{sec:Gap} in the general coordinates presented in App.~\ref{ABCtodomainwall}.

\subsection{The Conductivity of the electric solutions}\label{cef}

The conductivity tensor $\sigma_{ij}$ is defined via Ohm's law:
\be
J^i = \sigma^{ij} E_j \ .
\ee
In the regime of linear response, the conductivity is equivalently given by the retarded current-current correlator:
\be
\label{Kubo}
\sigma_{ij} = -\frac{i}{\omega} <J_i J_j>_R \ .
\ee
We will compute linear conductivity following the standard holographic approach \cite{Son:2002sd, Hartnoll:2007ai, Horowitz:2009ij, Kachru1}.  Although we are primarily interested in the DC conductivity, we can not compute it directly due to translation invariance and the lack of dissipation.  We will therefore compute the AC conductivity $\sigma(\omega)$ in the low frequency limit and deduce from this, using Kramers-Kronig relations, the $\omega = 0$ behavior.

As before, we perturb the spatial components of the bulk gauge field by a harmonic time-dependent fluctuation. The perturbation away from the probe limit is consistent if the back-reacted perturbations on the metric in the vector channel are also considered. The vector gauge and metric field perturbations decouple (at zero momentum) from the rest of the fluctuations, so we can consistently ignore all other perturbations,\footnote{Note that $a_i(r,\omega)$ and $g_{ti}(r,\omega)$ is the generally complex Fourier transform of the real fields $a_i(r,t)$ and $g_{ti}(r,t)$.}
\be
\delta A_i = a_i e^{i\omega t},\,\,\,\, \delta g_{ti} =   h_{ti} e^{i\omega t} \
\ee
where $i = (x,y)$. Near the boundary where $r\gg L$, the gauge field has the form
\be
\label{gaugefluctuationUVexpansion}
a_i=a_i^{(0)}+a_i^{(1)}e^{-\frac{r}{L}} + \mathcal O\left(e^{-\frac{2r}{L}}\right) \ .
\ee
The applied boundary electric field is $E_i = - \partial_t \delta A_i|_{r\to \infty} = -i \omega  a_i^{(0)} e^{i\omega t}$.
The linearized Maxwell and Einstein equations in the general coordinates (\ref{generalmetric}) are \eqref{axh}, \eqref{ahry} with vanishing magnetic field $h=0$,
\begin{subequations}
\label{ah}
\begin{align}
\frac{\omega^2 B}{D} a_i+\frac{\sqrt{B/D}}{\tau_2} \left( q(h{_t}^{i})'+i\tilde\epsilon^{ij} \tau_1' \omega a_j \right) +  \frac{\left(\tau_2 \sqrt{D/B}a_x'\right)'  }{\tau_2 \sqrt{D/B}}=0 \ , \label{Maxwellpert}\\
-4DBq\omega a_i  - 4\sqrt{DB}\omega C^2 (h{_t}^{i})' =0 \ . \label{Einsteinpert}
\end{align}
\end{subequations}
Using the second equation \eqref{Einsteinpert} to solve for $(h{_t}^{i})'$ in terms of $a_i$, \eqref{Maxwellpert} becomes two coupled equations for the gauge perturbations $a_x$ and $a_y$
\begin{subequations}
\label{gaugeperturbations}
\begin{align}
\tau_2 \sqrt{B/D} \left( \omega ^2-\frac{\tau_2 (A_t')^2}{B}  \right)
 a_x &+\left( \sqrt{D/B}\tau_2 a_x'  \right)'+i \omega a_y \tau_1'=0, \label{gaugeperturbationsx}\\
\tau_2 \sqrt{B/D} \left( \omega ^2-\frac{\tau_2 (A_t')^2}{B}  \right)a_y
&+\left( \sqrt{D/B}\tau_2 a_y'  \right)'-i \omega a_x \tau_1'=0 \ .
 \label{gaugeperturbationsy}
\end{align}
\end{subequations}
Using the transformation (\ref{zzb}), last equations decouple and take the form
\begin{subequations}
\label{azzb}
\begin{align}
\tau_2 \sqrt{B/D} \left( \omega ^2-\frac{\tau_2 (A_t')^2}{B}  \right)
 a_z &+\left( \sqrt{D/B}\tau_2 a_z'  \right)'+ \omega \tau_1'a_z=0,\\
\tau_2 \sqrt{B/D} \left( \omega ^2-\frac{\tau_2 (A_t')^2}{B}  \right)a_{\bar z}
&+\left( \sqrt{D/B}\tau_2 a_{\bar z}'  \right)'-\omega  \tau_1'a_{\bar z}=0.
\end{align}
\end{subequations}

In this section we will be interested in the behavior of the conductivity at low frequencies, {\it i.e.}~the $\omega$ dependence in the $\omega \to 0$ limit where the $a_x$ and $a_y$ components of (\ref{ah}) decouple. In particular, if the retarded correlator $\langle J^x(-\omega) J^x(\omega)\rangle$ does not vanish linearly in $\omega$ as $\omega\rightarrow 0$, the DC electric conductivity will be infinite due to translation invariance in the presence of a finite charge density. This divergence of $\sigma_{xx}$ is important in order to recover the correct result for the Hall conductivity in the dyonic frame, {\it i.e.}~in our fractional QH states. 
We want to perform the necessary transport calculations at $\omega=0$ directly, {\it i.e.} to lowest order in a small $\omega$ expansion. For this we need to ensure that the limits $\omega \rightarrow 0$ and $r\rightarrow 0$ commute. In other words, the IR boundary conditions used to numerically solve \eqref{gaugeperturbations} must be frequency-independent, which in fact holds in our case ({\it c.f.}~App.~\ref{IRsolelectric}). As already noted in \cite{Charmousis:2010zz}, the $A_t'{}^2$ contribution in the first term from the left of \eqref{gaugeperturbations} dominates over the $\omega^2$ contribution in the $r \rightarrow 0$ limit, if the constraint \eqref{TDunstable} is satisfied.\footnote{In the opposite case in which the black holes are thermodynamically stable and \eqref{TDunstable} is not fulfilled this is not the case, and the IR boundary conditions are $\omega$-dependent. This was studied for an extremal $AdS_4$ Reissner-Nordstr\"om black hole in detail in  \cite{Faulkner:2009wj} (see also \cite{Edalati:2009bi}), where it was pointed out that the low frequency limit has to be taken carefully as a scaling limit of both $\omega$ and $r$.}
This imples that the IR boundary conditions 
will be frequency independent. We can therefore solve \eqref{gaugeperturbations} directly in the zero-frequency limit, in which in particular the equations for $a_x$ and $a_y$ decouple from each other, even with a running axion.\footnote{They also decouple at any frequency for flows with constant axion, $\tau'_1=0$.}

\begin{figure}
\centering
\includegraphics[width=8.cm]{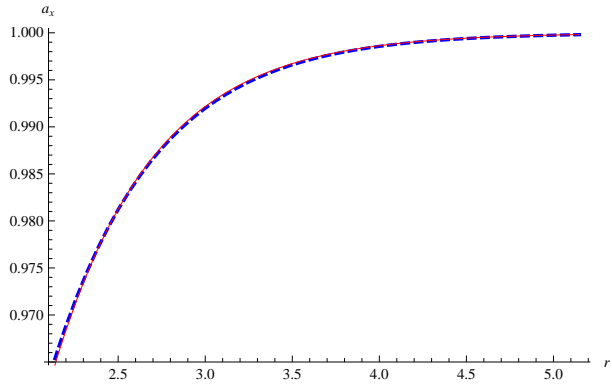}
\caption{A sample of the gauge field fluctuation $a_z$ as a function of $r$, normalized by the boundary value $a_z^{(0)}=1$,  for $\omega = 0$ and a running axion.  The numerically constructed background corresponds to input parameters $q=1$ and $\tau_2^{(IR)}=0.2$ and hence to a running axion (see App.~{\protect\ref{RGdetails}}). The numerical solution found by solving ({\protect\ref{gaugeperturbationsx}}) is shown by a dashed blue curve.  The solid red curve is a fit to the asymptotic UV solution {\protect\eqref{gaugefluctuationUVexpansion}}. The fitting is achieved via the curve $1+a_z^{(1)}e^{-r/L}$ for some negative $a_z^{(1)}$ where $L$ is the $AdS_4$ radius. Although not visible from this figure, in the IR, the numerical curve maps precisely onto the analytical IR-regular solution (see App.~D in {\protect \cite{Charmousis:2010zz}}). The two plots show that the numerical curve has the right IR and UV behavior and provides the numerical specification of the right hand side of equation ({\protect\ref{vos}}) and thus the specification of Im $ \sigma_{xx}$ (see ({\protect\ref{Imsxx}})).}
\label{axflalarger}
\end{figure}

Using the background solutions found in Sec.~\ref{RGF}, one can now solve (\ref{gaugeperturbationsx}) and (\ref{gaugeperturbationsy}) to determine how $a_i^{(1)}$ depends on $a_j^{(0)}$.  Taking as boundary conditions the IR normalizable scaling solutions from App.~D in \cite{Charmousis:2010zz}, equations (\ref{gaugeperturbationsx}) and (\ref{gaugeperturbationsy}) can be integrated numerically using the numerically obtained background fields.
For illustration, we show in Fig.~\ref{axflalarger} the numerical solution for $a_x(r)$ in the limit where $\omega \to 0$.  


\subsubsection{The holographic AC conductivity at zero magnetic field}

In the electric frame, formula (\ref{szzb}) simplifies in the two following ways: (i) The functional differentiations can be replaced by simple ratios because, according to (\ref{azzb}), the VEV and generally the whole gauge field is completely independent of the source of the metric.\footnote{Such a simplification, as will be seen in Sec.~\ref{dyons}, does not apply when there is a magnetic field present.} (ii) According to the same equation, there is a $a_{\bar z}(-\omega)=a_{ z}(\omega)$ symmetry. Therefore, using (\ref{szzb}) and (\ref{sxy}), the conductivity in the electric frame becomes 
\begin{subequations} \label{selw}
\begin{align}
&\sigma_{xx}(\omega)=\sigma_{yy}=-\frac{i}{2\omega L}\tau_2^{UV} \left( \frac{a_z^{(1)}}{a_z^{(0)}} + \frac{a_{\bar z}^{(1)}}{a_{\bar z}^{(0)}}  \right), \\
&\sigma_{xy}(\omega)=-\sigma_{yx}(\omega)=\frac{1}{2\omega L}\tau_2^{UV} \left( \frac{a_z^{(1)}}{a_z^{(0)}} - \frac{a_{\bar z}^{(1)}}{a_{\bar z}^{(0)}}\right)+ \tau_1^{UV}  ,
\end{align}
\end{subequations}
where the ratios are assumed for positive $\omega$. 


\subsubsection{The DC conductivity at zero magnetic field}

In particular, to lowest order in $\omega$, according to (\ref{azzb}), we have $a_z(r,\omega \rightarrow0)=a_{\bar z}(r,\omega \rightarrow 0)$ and hence 
(\ref{selw}) becomes
\begin{subequations} \label{sel0}
\begin{align}
&\sigma_{xx}(\omega)=\sigma_{yy}=-\frac{i}{\omega L}\tau_2^{UV} \frac{a_z^{(1)}}{a_z^{(0)}} , \\
&\sigma_{xy}(\omega)=-\sigma_{yx}(\omega)= \tau_1^{UV}.
\end{align}
\end{subequations}
In particular, the second equation in (\ref{sel0}) shows that there is an anomalous Hall conductivity in the electric frame when the axion runs and does not asymptote to zero at the UV. 

To leading order in $\omega$, the imaginary part of the longitudinal conductivity is
\be \label{Imsxx}
{\rm Im} \ \sigma^{xx} = -\frac{\tau_2^{UV}}{\omega L} \frac{a_z^{(1)}}{a_z^{(0)}}+O(\omega^0) \ .
\ee
Note that the ratio $\frac{a_z^{(1)}}{a_z^{(0)}}$ is real since $a_z(r)$ and $a_{\bar z}(r)$ fulfil real second order differential equations \eqref{azzb} with real IR boundary conditions \eqref{azIR} and UV boundary conditions (source terms) $a_z^{(0)}$ and $a_{\bar z}^{(0)}$. Hence the full solutions $a_z(r)$ and $a_{\bar z}(r)$ can be chosen as really, in which case the subleasing coefficients $a_z^{(1)}$ and $a_{\bar z}^{(1)}$ in the expansion \eqref{gaugefluctuationUVexpansion} must be real. 

We have performed this numerical calculation in a number of cases with different charges and input parameters and have found generically for $O(1)$ input parameters that
\be \label{vos}
\left. \frac{a_z^{(1)}}{a_z^{(0)}}\right|_{\omega=0} = O(1)\,.
\ee
In specifying the ratios $a_{ z}^{(1)}/a_{ z}^{(0)}$ required in (\ref{selw}) one needs to solve (numerically) the fluctuation equations (\ref{azzb}) in the electric frame. The necessary boundary conditions are taken from the IR by demanding regularity of the solution. Fortunately, as we will show in App.~\ref{IRsolelectric}, the IR solutions can be found analytically and read
\be \label{azIR}
a_z^{el;IR}= C^1_z r^{\frac{w}{4}}+ C^2_z r^{-v} = C^1_z r^{1.715}+ C^2_z r^{-0.926}\,,
\ee
where $C^1_z$ and $C^2_z$ are arbitrary constants, and where we inserted our choice of parameters \eqref{gammas} in the second inequality. The ${\bar z}$ components have the same IR solutions. 
The positive power provides the regular solution and hence the boundary condition for the subsequent numerical evaluation. 

From the Kramers-Kronig relations
\be
\label{KramersKronig}
{\rm Im} \ \sigma(\omega) = - \frac{2 \omega}{\pi} {\mathcal P} \int\limits_{0}^{\infty} d\tilde\omega \frac{{\rm Re} \ \sigma(\tilde\omega)}{{\tilde\omega}^2-\omega^2}\,,
\ee
this pole at $\omega = 0$ in the imaginary part of $\sigma_{xx}$ implies the existence of a $\delta$-function in the real part.  The complex longitudinal conductivity is therefore
\be\label{sxel}
\sigma_{xx} =-\frac{ \tau_2^{UV}} {L}  \frac{a_z^{(1)}}{a_z^{(0)}} \left( \pi \delta(\omega) + \frac{i}{\omega } \right)+ \mathcal O(\omega)^0 \ .
\ee
Note that $Re \sigma_{xx} > 0$ is required by unitarity \cite{Jensen:2011xb}, translating into $\frac{a_z^{(1)}}{a_z^{(0)}}<0$, which we have found to be fulfilled in our numerically obtained solutions. 
Moreover, as is shown in App.~\ref{familyq}, under a certain scaling transformation of the equations of motion \eqref{restr3}, the conductivity scales with the charge density (see equation (\ref{cd}))  as
\be\label{sxelk}
\sigma_{xx}(\omega;q) =\sigma_{xx} (\omega;q_0) \sqrt{k}, \,\,\, k=\frac{q}{q_0}
\ee
where $\sigma_{xx}(\omega;q)$ and $\sigma_{xx}(\omega;q_0)$ are the conductivities corresponding to $q$ and $q_0$ respectively for the same $\omega$.

The singularity in $\omega$ of equation (\ref{sxel}) at zero frequency is entirely expected and was the reason we could not just compute the DC conductivity directly:  any translationally invariant, charged system is a perfect conductor simply because momentum conservation prevents any dissipation of the current. Note that this also explains the different scaling of the conductivity with charge density, which is \eqref{sxelk} in our model, while it would be linear with charge density at least at low frequencies in a model where Drude theory is applicable at low frequencies. 
 In order to obtain a finite DC conductivity and Drude-like behavior, we would have to break translation invariance, for example, by adding impurities or a lattice.  We would then expect to recover the standard Drude behavior, as has been seen in \cite{Horowitz:2012ky, Vegh:2013sk}.


\subsection{The conductivity of the dyonic solutions} \label{dyons}

Having determined the DC conductivity of the pure electric solutions, we can employ $SL(2,\mathbbm{Z})$ duality to find the result for the dyonic case.   We calculated the transformation of $\sigma$ under $SL(2,\mathbbm{Z})$ in section \ref{conductivitytransformation}, and the general result is give by \eqref{str}.
%

For the purely electric background in the DC limit $\sigma_{xx}$ diverges while $\sigma_{xy}$ is finite. As a result, equating real and imaginary parts in equation (\ref{str}) to leading order in small $\omega$, we  deduce
\begin{subequations} \label{sh}
\begin{align}
\hat\sigma_{xy} &= \frac{a}{c}+O(\omega^0)=\frac{q}{h}+O(\omega^0) \label{sxyh}, \\
\hat\sigma_{xx} &= 0+O(\omega) \ 
\end{align}
\end{subequations}
where the second equality in (\ref{sxyh}) follows from (\ref{qhsl2z}) when the magnetic field in the initial $SL(2,\mathbbm{Z})$) frame vanishes. 
Note that the precise values of $a_z^{(0)}$ and $a_z^{(1)}$ in (\ref{sxel}), as long as they are different from zero, are irrelevant for the dyonic result.
 Recall from Sec.~\ref{dyonicflows} that the filling fraction $\nu = a/c$, so $\sigma_{xy} = \nu$ as expected for a quantum Hall state.  In addition, the longitudinal conductivity vanishes as required.

Similar results where found for dyonic dilatonic black holes in \cite{Kachru2}, with the difference that their dyonic quantum hall states were not gapped. On the contrary, as we will see in Sec.~\ref{sec:Gap}, our system has a mass gap $\Delta$, so we expect that at low temperature $T \ll \Delta$ the conductivity is supressed as
\be
\sigma_{xx} \sim e^{-\Delta/T}.
\ee
This should be contrasted with the power law behavior $\sigma_{xx} \sim T^{2/z}$ ($z$ being the dynamical critical exponent) found in \cite{Kachru2,Burgess} where there was no hard gap.

\subsubsection{Holographic DC conductivity at finite magnetic field}

We will now sketch the derivation of (\ref{sh}) directly from the magnetic frame. First, we show that
\begin{align} \label{azzb1}
a_z^{(1)}&=O(\left(h_t^z\right)^{(0)})+\omega L \tau_2^{UV} \frac{q-h \tau_1^{UV}}{h}a_{ z}^{(0)}+O(\omega^2), \,\,\, \notag \\
a_{\bar z}^{(1)}&=O(\left(h_t^{\bar z}\right)^{(0)})-\omega L \tau_2^{UV} \frac{q-h \tau_1^{UV}}{h}a_{\bar z}^{(0)}+O(\omega^2).
\end{align}
The detailed derivation of these equations can be found in App.~\ref{az1}.  Substituting (\ref{azzb1}) into (\ref{szzb}) and taking into account that the functional differentiation of the metric field sources with respect to those of the gauge field vanish, we obtain
\be \label{szz}
\sigma^{z\bar z}\big|_{\omega=0}=-2i\frac{q}{h}=-\sigma^{\bar z z} \big|_{\omega=0}, \,\,\, \sigma^{ z z}\big|_{\omega=0}=\sigma^{\bar z \bar z}\big|_{\omega=0}=0.
\ee
Finally, substituting equations (\ref{szz}) in (\ref{sxy}) reproduces precisely equation (\ref{sh}).


\section{Mass gap in the dyonic frame}\label{sec:Gap}

In order to complete our identification of the dyonic solutions as quantum Hall states, we need to show the spectrum of charged fluctuations is gapped not only for the pure electric solutions but for dyonic solutions as well.  Note that although we engineered our Schr\"odinger potential in Sec.~\ref{constraints} to yield a gap for purely electric solutions and to diverge in the IR for dyonic solutions, and although the spectrum can be expected to be discrete and gapped in any frame due to $SL(2,\mathbbm{Z})$ invariance of the constraint \eqref{TDunstable}, it is \textit{a priori} possible that an $SL(2,\mathbbm{Z})$ transformation maps a gapped spectrum into a ungapped one and vice versa. The resulting Schr\"odinger potential in the dyonic frame could for example go to zero in the UV, allowing for scattering states. In fact, starting from a continuous spectrum in the electric frame one can easily show that a gapped spectrum can be reached by $SL(2,\mathbbm{Z})$ if no other conditions such as the ones in Sec.~\ref{constraints} are imposed. In order to arrive at this result, we first establish the following facts about the fluctuation problem, as well as its Schr\"odinger potential:
\begin{itemize}

\item We discuss in Sec.~\ref{Univ} the behavior of the Schr\"odinger potential for the charged excitation problem in the IR and in the UV. We find that it is universal in both cases, in the sense that it is independent of the $SL(2,\mathbb{Z})$ frame, the charge density or the frequency. In particular, the IR and the UV behavior  is such that it allows only for discrete and gapped states in all $SL(2,\mathbb{Z})$ frames. 

\item We then show in Sec.~\ref{Norm} that the vector fluctuations in any dyonic frame have unique IR boundary conditions.  Under the conditions given in Sec.~\ref{constraints}, the modes which are regular in the IR also have a finite IR on-shell action (i.e. are normalizable), in contrast to irregular modes, whose on-shell action is diverging (i.e.  are nonnormalizable). We also show that regularity and normalizability in the electric frame implies regularity and normalizability in the generic dyonic frame, and vice versa.

\end{itemize}
We finally discuss in section \ref{spec} the important qualitative behavior of the Schr\"odinger potential $V_{schr}$ in (\ref{psi}) as frequency and magnetic field are varied.\footnote{For instance, one could work along the lines of \cite{Papadimitriou:2013jca} in order to extract how the transport coefficients depend on the several parameters. The techniques used in  \cite{Papadimitriou:2013jca} are useful for more complicated set-ups as the present one.} We furthermore compute the low-lying spectrum in the dyonic frame and comment on its behavior under variation of the magnetic field induced by the $SL(2,\mathbb{Z})$ transformations. We find that the spectrum as a whole is invariant under $SL(2,\mathbbm{Z})$ transformations and has simple scaling properties under changes of the magnetic field induced by changes of the charge density in the electric frame. We leave a full analysis of all aspects of the spectrum to a future work. On a technical side note, our numerical results for the wavefunctions confirm that the singularity of the fluctuation equations found in Sec.~\ref{access} is accessory in the sense of differential equations, and hence fully accessible for both of the independent solutions.


\subsection{The fluctuation problem as a Schr\"odinger problem and universal behavior in the Schr\"odinger potential}\label{Univ}

Our analysis in this section is based on the Schr\"odinger potential for the decoupling variable $E_z$, as defined in (\ref{psi}). 
The resulting Schr\"odinger problem is slightly non-standard, as the potential itself depends not only on external parameters such as the charge density and the magnetic field, but also on the frequency $\omega$ itself. One should note that the potential given in \eqref{psi} does not rely on any symmetry and in particular not on $SL(2,\mathbb{Z})$, but only on the form of the action \eqref{totalaction} and the Ansatz \eqref{generalmetric}.\footnote{For example, relaxing $SL(2,\mathbb{Z})$ will allow general functions of the scalars in front of the $F^2$ and $F\tilde F$, and this will affect the fluctuation equations.} On a technical level we generate the Schr\"odinger potential numerically by first transforming the numerically obtained electric frame RG flows  Fig.~\ref{bagel} into the dyonic frame via equations (\ref{e1}) and (\ref{calGFtrafo}), and then numerically evaluate the Schr\"odinger potential on the so-obtained dyonic RG flows.


\subsubsection{Universal IR behavior of the Schr\"odinger potential and wave function regularity}\label{IR}

We first show, by analyzing the asymptotic UV and IR behavior, that the Schr\"odinger potential $V_{schr}$ admits only discrete normalizable modes, implying that the spectrum of charged fluctuations is discrete and gapped. Using the IR background scaling solutions (\ref{CDBHsolnA}, \ref{putsaf}, \ref{CDBHsolnf}, \ref{CDBHsolnAt}) in \eqref{gop}, \eqref{g1p} and  \eqref{psi}, we find that in the IR\footnote{Here the IR limit is $r\rightarrow r_0$, where $r_0$ is the IR end point of the radial coordinate after proper normalisation of the charge density by \eqref{restr3}. For the general analysis about the IR parameters, see App.~\ref{RGdetails}.} 
limit the Schr\"odinger potential behaves as
\begin{eqnarray} \label{g01IR}
g_0 &\sim& -h^2\left(\frac{B\tau_2}{C^2}\right)^{IR} \sim \frac{C_0}{(r-r_0)^2},
\,\,\,
g_1\sim \frac{C_1}{(r-r_0)}, \,\,\,\,\ r\sim r_0,\,\,\, \mbox{where} \notag \\
C_0&=& \frac{- w v}{4} = \frac{(4+3\gamma^2-2\gamma\delta-\delta^2)(-2-\gamma\delta+\delta^2)}{4}, \,\,\,C_1=\frac{3(\gamma-\delta)^2}{4}\,,\notag\\\label{VsIR}
 V_{schr}^{IR} &=& \frac{3(3(\g-\d)^2-8)(\g-\d)^2+16 w v}{64(r-r_0)^2}.
\end{eqnarray}
with $w$ and $v$ defined in \eqref{wvdef}. 
Remarkably, the IR behavior of  $g_0$ and $g_1$ does not depend on the quantities $\omega$, $h$ or $q$, but only on the parameters $\gamma$ and $\delta$ determining the IR behavior of the scalar potential and the gauge kinetic coupling.  
While extracting the leading behavior of $g_1$ is straightforward, the expression for $g_0$ is more involved: One first needs to extract the IR behavior of the scalars in the dyonic frame,
\begin{align} \label{t12IRp}
\tau_2^{IR} \sim \frac{1}{\tau_2^{el; \, IR}} \,\,, \,\,\,\,\,\,\,\,  q-h\tau_1^{IR} & \sim \frac{1}{(\tau_2^{el; \, IR})^2}  \,\,, \,\,\,\,\,\,\,\,  \tau_1^{IR} \,'  \sim \frac{1}{r}\times\frac{1}{\tau_2^{el; \, IR}}.
\end{align}
After inserting the IR geometry in a general dyonic frame, i.e. (\ref{CDBHsolnA}, \ref{putsaf}, \ref{CDBHsolnf}, \ref{CDBHsolnAt}) with (as explained in Sec.~\ref{dyonicsolutions}) $\tau_2 \rightarrow 1/\tau_2$, $q\rightarrow h$, and $\tau_1=q/h$, the denominator of \eqref{gop} has two contributions, one $\propto h^2$ and one $\propto \omega^2$.  It is the first contribution which dominates as a consequence of the thermodynamic constraint \eqref{TDunstable}. The latter contribution would dominate for black holes which settle down smoothly to their extremal ground state, i.e. if the opposite of \eqref{TDunstable} holds. For the particular values used in this work \eqref{gammas} we find
\be\label{g0g1Vsparticularvalues}
C_0 = -0.159\,,\quad C_1 = 2.623\,,\quad V_{schr}^{IR} = \frac{0.567}{(r-r_0)^2}\,.
\ee
In particular the numerator of (\ref{VsIR}) is positive for our choice of parameters \eqref{gammas}, creating an infinite potential barrier in the IR and implying that $\psi(r)$ admits one regular and one irregular (exponentially growing) solution which, in terms of the decoupling variable \eqref{Ez}, \eqref{ExEy} are given by \eqref{measure},
\be \label{EzIR}
E_z ^{(IR)}=  C_1 (r-r_0)^{v}+C_1 (r-r_0)^{-\frac{w}{4}} = C_1 (r-r_0)^{0.926}+ C_2 (r-r_0)^{-1.715}\,,
\ee
where we have used \eqref{gammas} in the second equality. Due to \eqref{Gubser} the first mode is always regular, while the second one is always irregular. An interesting observation from equations (\ref{EzIR}) and \eqref{hIR} is that both modes have the same power law as the metric fluctuation in the electric frame \eqref{hIR}. Due to $SL(2,\mathbb{Z})$ invariance of the metric the IR behavior of the metric fluctuations must be the same in any $SL(2,\mathbbm{Z})$ frame, implying that either both terms in \eqref{ExEy} have the same scaling in any frame, or that the gauge field fluctuations must always be subleading to the metric fluctuations in \eqref{ExEy}. We will show in Sec.~\ref{GC} that both possibilities exist even after imposing the constraints of Sec.~\ref{constraints}, but for our parameter choice \eqref{gammas} only the latter possibility is realized. We will see in sec~\ref{Norm} that the irregular fluctuation is also nonnormalizable. In summary, it is reassuring to see that the conditions we imposed in Sec.~\ref{constraints} automatically singles out a natural boundary condition for the fluctuations in the dyonic frame as well. In the following (in particular in section \ref{spec}), we employ the regular boundary conditions (i.e. set the irregular mode to zero in the IR) in numerically obtaining the fluctuations and their spectrum.


\subsubsection{Universal behavior in the UV} \label{UV}

In the UV the $AdS_4$ solution is given in the coordinate system (\ref{generalmetric}) by $B = 1$, $C = D = e^{2r/L}$ and constant $\tau_1$ and $\tau_2$.  With this background, the $r\to\infty$ behavior of \eqref{go} and \eqref{g1} is
 \begin{eqnarray}\label{Vuv}
g^{UV}_0 \sim O(\exp(-2r/L)),
\,\,\,
g^{UV}_1\sim \frac{1}{L} \ ,\,\, \Rightarrow 
V_{schr}^{UV} \approx \frac{1}{(2L)^2} \ ,\,\, r\gg L.
\end{eqnarray}
This is a general result in the sense that it applies to any Einstein-Maxwell-axio-dilaton background with constant magnetic field that asymptotes to $AdS_4$.\footnote{In the special case where $\omega=0$ the $V_{schr}$ behaves as $V_{schr}^{(UV)} \sim 4/(3L)^2$. Hence the $\omega\rightarrow 0$ limit does not commute with the UV limit. However, since we are searching for bound states with nonzero $\omega$, this is not of importance to us.}
The potential is flat in the UV, but since we include the conventional $\omega^2$ term in \eqref{psi} into the frequency-dependent Schr\"odinger potential we are searching for zero energy solutions which have to tunnel through that infinitely long and finitely high barrier, which turn out to be
 \bea\label{EzUV}
\psi^{UV} \sim a^{(0)} e^{ \frac{r}{2L}} +a_1^{(0)} e^{-\frac{r}{2L}},  \, \,\, \Rightarrow \,\,\, E_z ^{(UV)}  \sim a^{(0)} +a^{(1)} e^{-\frac{r}{L}} \ ,\,\, r\gg L.
 \eea
where $a^{(0)} $ and $a^{(1)} $ are arbitrary constants. The UV asymptotics of the two solutions behave like gauge fields in $AdS_4$, and hence the coefficients $a^{(0)}$ and $a^{(1)}$ must be identified, respectively, with the source and the VEV of the dual conserved current. In particular the energy-momentum VEV, which enters the metric fluctuation at higher order in $e^{-r/L}$, does not mix with the conserved current directly. In Sec.~\ref{spec} we will find $\omega_n$'s for vanishing source, whose wave-functions $\psi_n$'s are hence normalizable. These wave functions describe charged meson-like excitations around the QH state.


\subsection{Normalizability in the presence of a background magnetic field}\label{Norm}

In section \ref{EvsB} we derive some useful results on how the fluctuations transform as we change $SL(2,\mathbb{Z})$ frames, which we then use in Sec.~\ref{2fl} to prove that normalizability is preserved under $SL(2,\mathbb{Z})$. The latter fact is important in establishing that the regular IR boundary conditions actually give rise to a well-defined generating functional in the dual field theory.


\subsubsection{Relating the gauge IR fluctuations of electric and dyonic frames}\label{EvsB}

We first express the IR behavior of the gauge field in the dyonic frame using our knowledge of the analytic expressions of the fluctuations in the electric frame (i.e. $h=0$). Using (\ref{azIR}) and (\ref{hIR})\footnote{We suppressed here the IR integration constant $A_{IR}$ appearing in \eqref{hIR} for simplicity.}  in (\ref{newaxy}) along with the the background (\ref{CDBHsolnA})-(\ref{qeq}) and $h=0$ we obtain  
\begin{subequations} \label{apeIR}
\begin{align}
\tilde{a}_x^{IR}&= -\frac{i}{\omega} \Bigg[i\omega \left(C^1_x (d+c\tau_1^{IR}) r^{\frac{w}{4}}+...\right) \notag\\ &
\hspace{1in}-c \left(  \frac{C_1^y
}{v}\underbrace{ \left(q^2- 4\frac{v
\sqrt{\Lambda^2} }{u}  \right)}_{\mbox{zero by (\ref{qeq})}} r^v+...\right)   \Bigg]  \label{apregIR},\\
\tilde{a}_x^{IR}&= -\frac{i}{\omega} \Bigg[i\omega \left(C^2_x (d+c\tau_1^{IR}) r^{-v}+...\right) \notag\\ &
\hspace{1in}-c \left(4  \frac{C_2^y
}{w}\underbrace{ \left(q^2- 4\frac{v
\sqrt{\Lambda^2} }{u}  \right)}_{\mbox{zero by (\ref{qeq})}} r^{-\frac{w}{4}}+...\right)   \Bigg] 
\label{apiregIR}
\end{align}
\end{subequations}
where $\tau_1^{IR}=a/c$ and $(a,b,c,d)$  are the $SL(2,\mathbb{Z})$ parameters, fulfilling $ad-bc=1$. \eqref{apregIR} is for the regular solution in \eqref{azIR}, \eqref{hIR}, while \eqref{apiregIR} is for the irregular case. An analogous pair of equations exists for the $y$ component. The dots represent sub-leading corrections and it is \textit{a priori} not clear whether the dots next to the power $r^{ \pm w/4}$ (power $r^{ \pm v}$) in equation (\ref{apregIR}) can be more important than  $r^{\pm v}$ ($r^{\pm w/4}$). However, since the leading power associated with $\tilde{F}_{\mu \nu}$ of the transformation (\ref{Ftrafo}), i.e. the terms $\sim r^v$ and  $\sim r^{-w/4}$, vanish, we find that   if $w/4<v$ then the leading power is given by $w/4$ in (\ref{apregIR}), while if $w/4>v$ then the leading term can either be given by the sub-leadings corrections to $r^v$ or by $w/4$.  As we will see in Sec.~\ref{Norm} and App.~\ref{puts}, there are regions of the $\gamma-\delta$ plane where either possibility is realized. Analogous facts apply for (\ref{apiregIR}). 

In summary, given that $w>0$ and $v>0$ (see (\ref{Gubser})), equation (\ref{apeIR}) shows the following not \textit{a priori} expected result:

{\it $SL(2,\mathbb{Z})$ maps the regular solutions onto regular solutions and the irregular solutions onto irregular ones.}


\subsubsection{On shell action to second order in the fluctuations}\label{2fl}

We now turn to the question whether the regular IR boundary conditions from \eqref{EzIR} indeed correspond to normalizable IR perturbations, {\it i.e.}~whether they lead to a finite on-shell action. 
Expanding the on shell action (\ref{totalaction}) to second order in the fluctuations schematically yields terms $\propto a_i^2$, $\propto h_{ti}^2$, or $\propto a_i h_t^i$, with or without radial derivatives acting on them. The prefactors are functions of the background geometry, $q$, $h$, and the scalars. Then, a sufficient condition for normalizability is if all these terms are separately integrable in the IR. 
The action (\ref{totalaction}) has three terms given by (\ref{gaugeaction}), (\ref{potentialaction}) and (\ref{gravtauaction}). Since we are interested in vector perturbations only, the scalars don't fluctuate, and the metric fluctuations and the background metric are $SL(2,\mathbb{Z})$ invariant. This implies that $S^{(2)}_{\hat g}$ in \eqref{e3} and $S^{(2)}_V$ in \eqref{potentialaction} contain exactly the same terms in any $SL(2,\mathbb{Z})$ frame and in particular the same terms as those in the electric frame.

We begin with $S^{(2)}_{\hat g}$, \eqref{e3}. Expanding to second order in  fluctuations yields
\begin{align}\label{S2gr}
 S_{\hat g}^{(2)} \sim \int (\sqrt{-g})^{(2)} \,R_B dr&+\int \sqrt{-g_B} R^{(2)} dr \sim \notag\\
\int \frac{(B C D)''}{ B^2 C D} \sqrt{\frac{B}{D}} C^2 \left((h_t^x)^2+(h_t^y)^2 \right)dr &+ \int  \sqrt{BDC^2} \frac{\left(B C D ((h_t^x)^2+(h_t^y)^2) \right)''}{ (B  D)^2}dr \sim \notag\\
 &\sim \int  \frac{C^2}{r^2 \sqrt{BD}}((h_t^x)^2+(h_t^y)^2)dr
\end{align}
where the superscript B denotes background fields. The first integral in (\ref{S2gr}) contains the $(\sqrt{-g})^{(2)}$ fluctuations and the second integral the $R^{(2)}$ ones. The last integral gives the leading power of all the terms in  $S_{\hat g}^{(2)}$. 
 
Next, we proceed to the kinetic scalar terms in \eqref{gravtauaction}, which yield
\begin{align}\label{S2sc}
 S_{scal}^{(2)} &\sim \int (\sqrt{-g})^{(2)} \left(V-\frac{1}{2 \gamma^2} \frac{(\tau_1')^2+(\tau_2')^2}{\tau_2^2} \right) dr\notag\\
& =  \int  \sqrt{\frac{B}{D}} C^2 \left((h_t^x)^2+(h_t^y)^2 \right) \left(V-\frac{1}{2 \gamma^2} \frac{(\tau_1')^2+(\tau_2')^2}{\tau_2^2} \right)dr.
\end{align}

Finally we consider the gauge piece of the action, eq.~\eqref{gaugeaction}. To second order in fluctuations and assuming $a_i(r) e^{i \omega t}$ and $h_{ti}(r) e^{i \omega t}$ with $i=x,y$, we can organize the expansion conveniently as
\be \label{S2B1}
S_F^{(2)} \sim \int \tau_2 ( \sqrt{-g})^{(2)} F_B^2   dr +  \int \tau_2 \sqrt{-g_B} (F^2)^{(2)}   dr +\int \frac{\tau_1}{2}\tilde \epsilon^{\m\n\r\s}(F_{\m\n}F_{\r\s})^{(2)} dr\,,
\ee 
where we dropped the $d^3x$ integration for simplicity of notation. The three pieces of (\ref{S2B1}) separately read
\begin{subequations}\label{S2F}
\begin{align}
S_{F_1}^{(2)} &\equiv \int \tau_2 ( \sqrt{-g})^{(2)} F_B^2 dr \sim -\int  \sqrt{\frac{B}{D}} \frac{((q-h \tau_1)^2-h^2 \tau_2^2)}{ \tau_2} \left((h_t^x)^2+(h_t^y)^2 \right)dr\label{SF1}\\
S_{F_2}^{(2)} &\equiv \int \tau_2 \sqrt{-g_B} (F^2)^{(2)}   dr \sim 2 \int  \sqrt{\frac{B}{D}} \frac{((q-h \tau_1)^2-h^2 \tau_2^2)}{ \tau_2} \left((h_t^x)^2+(h_t^y)^2 \right)dr\notag\\
&-4\int (q-h \tau_1) \left( a_x' h_t^x+a_y' h_t^y \right)dr+ 2\int \sqrt{\frac{D}{B}}\tau_2 \left(a_x'^2+a_y'^2 \right) dr\notag\\
&+4i h\omega \int  \tau_2 \sqrt{\frac{B}{D}} \left(a_y h_t^x-a_x h_t^y \right)dr+2\omega^2 \int \tau_2  \sqrt{\frac{B}{D}}  \left(a_x^2+a_y^2 \right)dr
 \label{SF2}\\
S_{F_3}^{(2)} &\equiv \int \frac{\tau_1}{2}\tilde \epsilon^{\m\n\r\s}(F_{\m\n}F_{\r\s})^{(2)} dr \sim i\omega \int (a_x a_y'+a_x' a_y)dr \label{SF3}
\end{align}
\end{subequations}
where the $\omega$-dependence of the fluctuations and the $d\omega dx dy$ integrations have been suppressed.\footnote{For example $a_x^2$ is a shorthand for $V_2\int a_x(r;\omega)a_x(r;-\omega) d\omega$ where $V_2$ is the spatial volume}

The next step is to substitute the IR behavior of the background fields and of the fluctuations in the integrands of (\ref{SF2}) and check whether all terms are separately integrable. 
We start by arguing that this is the case for the particular values of $\gamma$ and $\delta$ given by (\ref{gammas}). In the next section, we discuss general values of $(\gamma,\delta)$, and show that the existence of a  gap is an $SL(2,\mathbb{Z})$ invariant property. The leading IR behavior of the background geometry was described in Sec.~\ref{dyonicsolutions}. The leading IR behavior of the gauge field and metric fluctuations is given by \eqref{IRfluc}. Using these pieces of data in (\ref{S2gr}), (\ref{S2sc}), and (\ref{S2F}) one can readily verify that all the terms are separately integrable for the first case in \eqref{IRfluc}, which is the regular IR mode in the case \eqref{gammas}. This shows that the particular choice for $\gamma$ and $\delta$ of (\ref{gammas}) the regular IR modes are normalizable in any dyonic frame, once normalizable boundary conditions in the electric frame are chosen. In other words, normalizability in the electric frame implies normalizability in any other generic $SL(2,\mathbb{Z})$ (dyonic) frame, once the constraints of Sec.~\ref{constraints} are obeyed. In the next section we will argue the same for general $(\gamma,\delta)$. Finally, if we had chosen the irregular solution in (\ref{EzIR}), the terms $S_{F_3}^{(2)}$ from (\ref{SF3}) would have been nonnormalizable. The regular mode in \eqref{EzIR} is hence normalizable, and the irregular mode in \eqref{EzIR} is nonnormalizable.

\subsection{$SL(2,\mathbb{Z})$ invariance of the existence of a gap}\label{GC}

The purpose of this section is to prove that

\vspace{0.1in}

{\it  For $SL(2,\mathbb{Z})$ covariant background geometries obeying the constraints in Sec.~\ref{constraints} (in particular \eqref{t2infcond}) , the existence of a gap is an $SL(2,\mathbb{Z})$ invariant property. Equivalently, for such geometries, a gap exists in the electric frame if and only if it exists in any other $SL(2,\mathbb{Z})$ (dyonic) frame.}

In the following we outline the proof while we omit intermediate steps for appendix \ref{puts} and the remarks 2 below. \vspace{0.1in}

We begin by pointing out that a sufficient, but not generally necessary, condition for normalizability is that all of the terms of the on-shell action, that is the terms (\ref{S2gr}), (\ref{S2sc}) and (\ref{S2F}) are separately  finite. We proceed to show that such a sufficient condition does apply in our-set-up for the allowed region in the $(\gamma,s)$ plane discussed in Sec.~\ref{constraints}.

We want to show that existence of a gap in the electric frame implies existence of a gap in any other $SL(2,\mathbb{Z})$ frame. Given the constraints imposed in Sec.~\ref{constraints}, the classification of IR geometries  in \cite{Charmousis:2010zz} show that the current-current correlator has a discrete and gapped spectrum. Since the Schr\"odinger potential goes to a universal constant \eqref{Vuv} in the UV, we will momentarily forget about the UV constraint \#5 of Sec.~\ref{constraints}, and focus on the IR constraints. The allowed region in the $\gamma-\delta$ plane is mostly\footnote{Momentarily forgetting about the UV constraint \#5 in Sec.~\ref{constraints}, the constraint which restricts the region allowed by \eqref{Gubser} strongest is the thermodynamic instability constraint \eqref{TDunstable}. The other constraints then restrict this region further only by a little bit.} defined by the intersection of the constrains (\ref{Gubser}) and  (\ref{TDunstable})\footnote{The constraint (\ref{discreteIR}), when superimposed with the constraints \eqref{Gubser}, \eqref{TDunstable} and \eqref{t2infcond}, does not restrict the allowed region further and hence it is omitted from the discussion. This point is further analyzed in appendix \ref{puts}.}. This yields the two disconnected regions (related by $(\gamma,\delta) \leftrightarrow (-\gamma,-\delta)$) between the purple, green and orange curves of the left panel in Fig.~\ref{reLR}. These two disconnected regions are further restricted to $\gamma<0$ for the case $\delta>\gamma$ and $\gamma>0$ for the opposite case $\delta<\gamma$ by the requirement \eqref{t2infcond}.


The next step is to extract the leading behavior of the fluctuations in the just defined regions. From now on, due the symmetry $(\gamma,\delta) \leftrightarrow (-\gamma,-\delta)$, we will assume $\delta>0$ and refer to the right panel of Fig.~\ref{reLR}. In this upper triangle, the leading behavior of $a_z^{IR}$ in the generic magnetic frame exhibits two different behaviors, depending on whether the point $(\gamma,\delta)$ is on the left (i.e. red dot) or on the right (i.e. black dot) of the blue curve defined by the exponent (\ref{AF2}). If $(\gamma,\delta)$ is on the left (right) of the blue curve then $a_z^{IR}$ behaves according to (\ref{AF3b}) (according to (\ref{AF5b})). Hence, for the fluctuations in a generic dyonic frame we have the following two regular possibilities:
\be \label{IRfluc}
\left(a_i, h_t^i \right)^I_{IR} \sim \left(r^{\frac{1}{2}(\delta^2-\gamma^2)},r^v\right)\,\,\,\, \mbox{or} \,\,\,\,\,     \left(a_i, h_t^i \right)_{IR}^{II} \sim \left(r^{\frac{w}{4}},r^v\right).
\ee

The final step is to substitute the fluctuations for case I of (\ref{IRfluc}) and the IR background (\ref{CDBHsolnA})-(\ref{qeq}) into eq.s~(\ref{S2gr}), (\ref{S2sc}), and (\ref{S2F}), and extract the conditions on $(\gamma,\delta)$ for which these fluctuations are integrable. The intersection of these conditions defines a region $S_I$ of the $(\gamma,\delta)$ plane. The same analyis for case II in \eqref{IRfluc} yields a region $S_{II}$.  $S_I \cup S_{II}$ is then intersected with the allowed (triangle) region defined above.\footnote{In principle we would have to impose all constraints of Sec.~\ref{constraints}, and intersect the allowed region of Fig.~\ref{fig:constraints} with $S_I \cup S_{II}$. It turns however out that $S_I \cup S_{II}$ has complete overlap with the region allowed by \eqref{Gubser}, \eqref{TDunstable} and \eqref{t2infcond} already, hence this analysis suffices.} It turns out that $S_I \cup S_{II}$  has maximal overlap with the allowed region. Taking further into account that the allowed region used above is defined through $SL(2,\mathbb{Z})$ invariant constraints only, namely \eqref{Gubser}, \eqref{TDunstable} and \eqref{t2infcond}, we conclude that normalizability in the electric frame implies normalizability in any dyonic frame and vice versa.

\paragraph{Remarks 1: Clarifications and consistency checks}

\begin{itemize}

\item There are two physical reasons to demand \eqref{t2infcond} in the electric frame. First, in top-down constructions $\tau_2\sim g_s^{-1}$, which means that the IR of the electric frame will be a controlled string theory background at weak string coupling. Secondly, only if \eqref{t2infcond} holds will the $SL(2,\mathbb{Z})$ transformation (\ref{e1}) imply $\tau_1^{IR}=a/c$ in the generic dyonic frame, and hence will the IR value of the axion coincide with the fractionalized Hall conductivty $\sigma_{xy} = q/h=a/c$ (see (\ref{sxyh})). 

\item Since $E_z=\omega a_z+ h h_t^z$ (see (\ref{ExEy}) and (\ref{Ez})), the leading behavior of $E_z$ should be the same as the dominant term of either  $a_z$ or $h_t^z$, for both the regular and irregular mode (see \ref{EzIR}). From the constraints $w>0$, $v>0$, $w/4>v$ (see left panel in Fig.~\ref{reLR} and caption) and (\ref{hIR}), (\ref{apeIR}) and (\ref{avsE}) one finds that the leading IR behavior of the combination $\omega a_z+ h h_t^z$ is $r^v$ for the regular and $r^{-w/4}$ for the irregular solution, exactly as in (\ref{EzIR}). Hence \eqref{IRfluc} and its counterpart for the irregular mode is compatible with \eqref{EzIR}.

\item The allowed triangular region defined above has maximal overlap with the curve $w/4-v=3 /4(\gamma - \delta)^2-1>0$ (above the dashed line in the right panel of Fig.~\ref{reLR}). Thus, according to (\ref{apregIR}), the leading power of the gauge field fluctuation can, in principle, either be $r^{w/4}$ or a sub-leading power to $r^v$. Which power is the leading one is not  obvious \textit{a priori}, and needs a detailed investigation. In fact, according to the right panel of Fig.~\ref{reLR} and equation (\ref{AF3b}) (equation (\ref{AF5b})), in the parts of the allowed triangle which lie below (above) the blue curve, eq.~(\ref{apregIR}) is dominated by the sub-leadings of $r^v$ (by $r^{w/4}$), respectively. Thus, as shown in more detail in App.~\ref{puts}, within the allowed triangle region either behavior is possible. 

\item As explained in Sec.~\ref{constraints}, normalizability does not impose any additional constraints in the electric frame. Our sufficient but not necessary normalizability requirement of Sec.~\ref{2fl} must reproduce this result. Taking the limit of vanishing magnetic field $h\rightarrow 0$ in (\ref{S2gr}), (\ref{S2sc}) and (\ref{SF2}) and using the electric frame IR background (see (\ref{CDBHsolnA})-(\ref{qeq})) and (regular) fluctuations  (\ref{azIR}) and (\ref{hIR}), we note that  the terms in (\ref{S2gr}) and (\ref{S2sc}) are $SL(2,\mathbb{Z})$ invariant and yields conditions $w>0$ and $w/4+v>0$, which hold due to Gubser's constraint \eqref{Gubser}.  Integrating the rest of the terms in (\ref{SF3}) yields either the same ($w/4+v>0)$ or trivially fulfilled ($1+1/4 (\gamma - \delta)^2>0$) conditions. Hence, in any case, there is no additional normalizability constraint on the electric frame vector fluctuations.

\end{itemize}

\paragraph{Remarks 2: Details in checking normalizability}

\begin{itemize}

\item The terms (\ref{S2gr}) and (\ref{S2sc}) do not transform under $SL(2,\mathbb{Z})$ and hence suffice to be checked in the electric frame. It is thus expected that these terms to be finite by definition of the allowed region above. Indeed, (\ref{S2gr}) integrates to $\sim r^{v+w/4}$ which is IR finite due to \eqref{Gubser}, as is (\ref{S2sc}), which  integrates to $\sim r^{w/4+v}+r^{w/4}$.

\item Although the first term in (\ref{SF1}) and in (\ref{SF2}) transform under $SL(2,\mathbb{Z})$, their IR behavior, according to (\ref{t12IRp}), is $SL(2,\mathbb{Z})$ invariant. Hence, it again suffices to check these terms in the electric frame yielding $r^{w/4+v}$, which again is finite due to \eqref{Gubser}. 

\item Furthermore, since $h_t^i \sim r^v \sim E_z \gg a_i$ in the IR, the last two terms in (\ref{SF2}) are less important compared to the first term\footnote{According to (\ref{t12IRp}), the scalar dependence of all three terms is the same, namely $1/\tau_2^{el;IR}$.}. Given that the first term in \eqref{SF2} is integrable by the previous bullet point, these terms will be as well.

\item Given that the fluctuations vanish in the IR and that $(q-h\tau_1)_{IR} \rightarrow 0$, the second term of (\ref{SF2}) and (\ref{SF3}) are integrable in the IR.

\item The last term that needs to checked is the third term of (\ref{SF2}). For the two cases I and II of (\ref{IRfluc}), this term evaluates to either $\sim r^{1 + 1/4 (\gamma - \delta)^2}$ or $\sim r^{w/4+v-2\gamma (\delta-\gamma )}$, respectively. The first exponent is clearly positive. The second exponent is also positive in the allowed region because of the constraints $w>0$, $v>0$ (i.e. Gubser's constraint \eqref{Gubser}) and $\gamma (\gamma -\delta)>0$ (i.e. \eqref{t2infcond}) (c.f. Sec.~\ref{constraints} and App.~\ref{puts} for a detailed explanation of these constraints). This completes the investigation.

\end{itemize}


\subsection{Spectrum analysis: Dependence on frequency, charge density and magnetic field}\label{spec}

In this section we analyze how the fluctuations behave as the charge density $q$, the magnetic field $h$ (i.e. the $SL(2,\mathbbm{Z})$ frame) and the frequency $\omega$ is varied, by studying the behavior of the Schr\"odinger potential under these changes. This will provide us with some intuition about how the spectrum changes.

\begin{figure}[htb]
\includegraphics[scale=0.8]{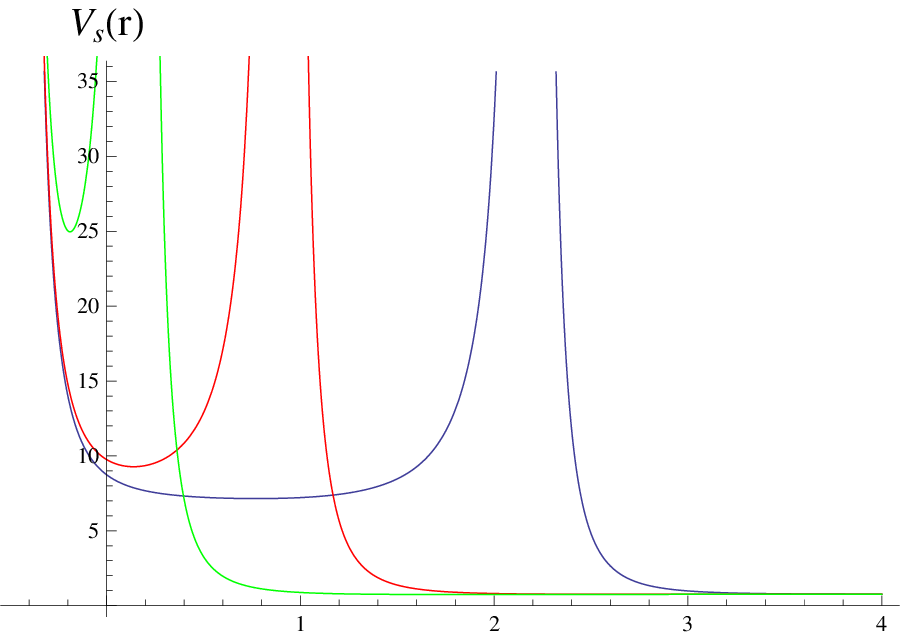}\hspace{0.2in}
\includegraphics[scale=0.8]{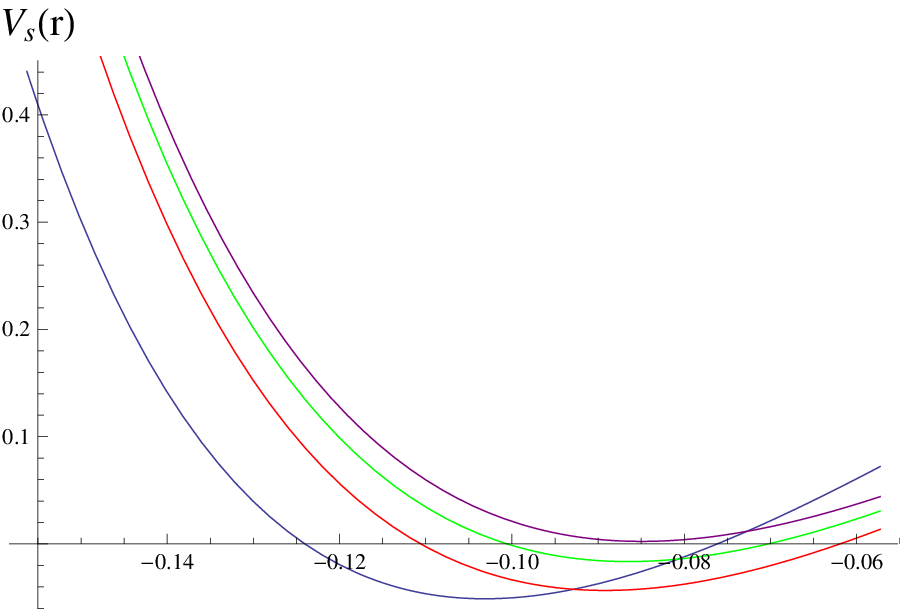}
%
	\caption{Left panel: The Schr\"odinger potential $V_{schr}(r;a,b,c,d;\omega)$ as a function of $r$ for different frequencies $\omega=0.01$ (blue), $\omega=0.1$ (red), $\omega=0.4$ (green). The IR end point is at $r_0=-0.457$. The initial background in the electric frame has $q=1$ and $\tau_1^{(IR)}=a_0=0$ (non running axion). The new charge and magnetic field in the dyonic frame are given by $q'=a q$ and $h'=c q$, for $SL(2,\mathbbm{Z})$ parameters $a=1,\,b=1,\,c=1,\,d=2$. This plot shows that the accessory singularity of Sec.~\protect\ref{access} moves towards the IR as the frequency is raised. Furthermore, the valley between the divergence in the deep IR and the singularity becomes higher with increasing frequency. 
Right panel: Change of $V_{schr}$ with increasing magnetic field for fixed $\omega=0.68$ and for fixed charge density ($a=1$).  The $SL(2,\mathbb{Z})$ parameters are:  $b=1,\,c=1,\,d=2$ (blue, $h=q=1$),  $b=1,\,c=4,\,d=5$ (red, $h=4\,, q=1$),  $b=1,\,c=10,\,d=11$ (green, $h=10\,, q=1$),  $b=1,\,c=1000,\,d=3001$ (purple, $h=1000\,, q=1$). Large magnetic fields lift the negative valleys in which bound states are supported up to positive valleys, i.e. increase the gap.
}		 
	\label{Vs1}
\end{figure}
%

\begin{figure}[htb]
\begin{center}
	\includegraphics[scale=0.85]{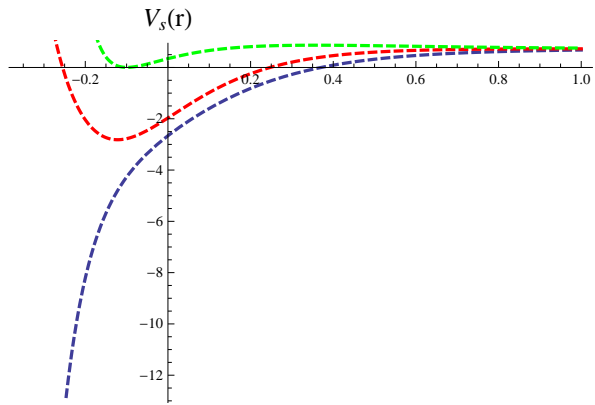}
\end{center}
	\caption{
Change in the Schr\"odinger potential for a fixed $SL(2,\mathbb{Z})$ frame with $a=b=c=1$ and varying frequencies $\omega=0.68$ (dashed, green), $\omega=1$ (dashed, red), $\omega=2$ (dashed, blue). For fixed magnetic field, there is a minimum $\omega_{min}$ (here $\omega_{min} \sim 0.68$) such that $V_{schr}$ develops a negative valley and allows the possibility for normalizable modes as those of Fig.~{\protect\ref{w10}}. The valley gets deeper and wider for increasing $\omega$, allowing excited states with more and more nodes to form.
}
	\label{Val}
\end{figure}


\begin{figure}[htb]
	\includegraphics[scale=0.57]{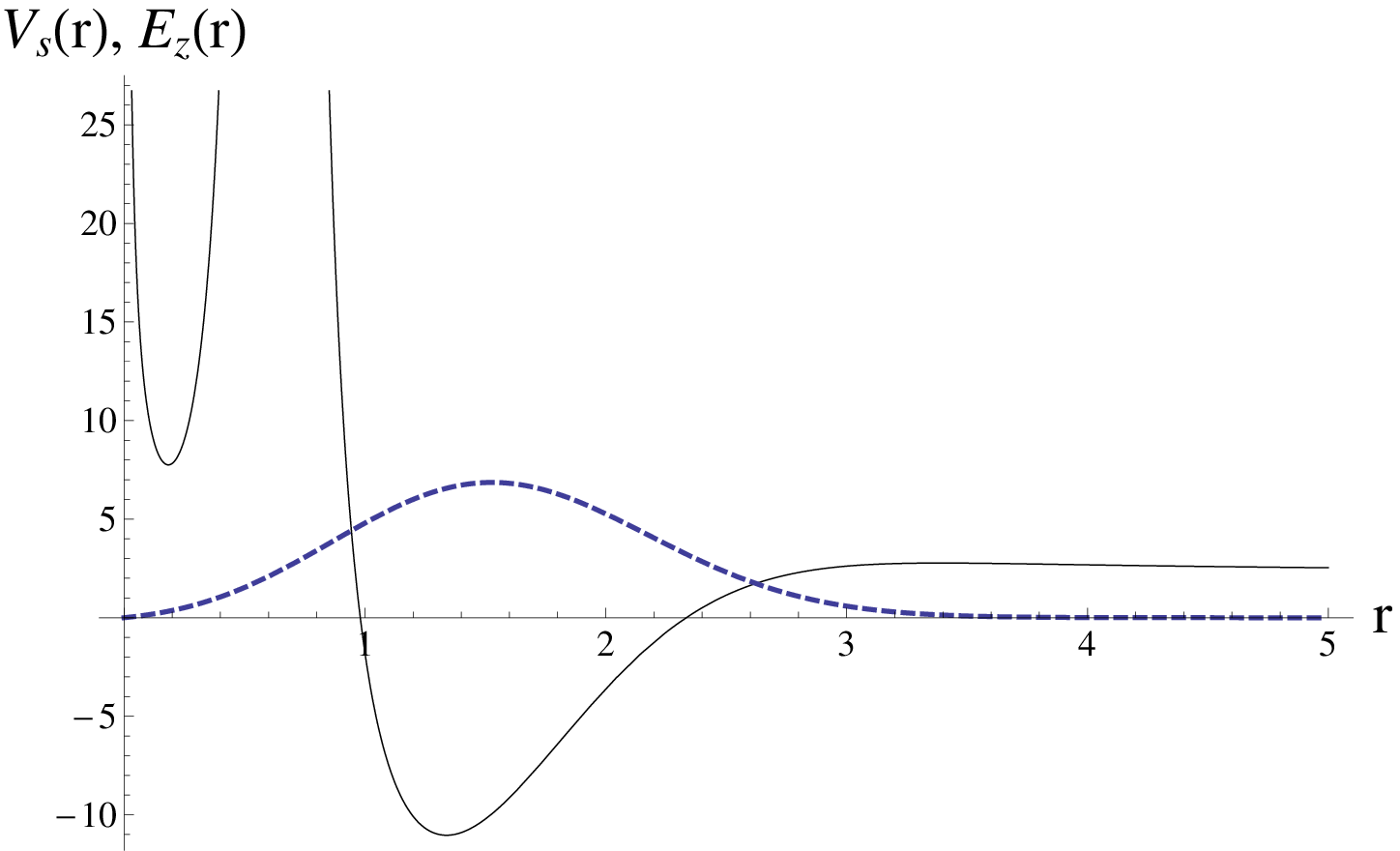} \hspace{0.2in}
	\includegraphics[scale=0.83]{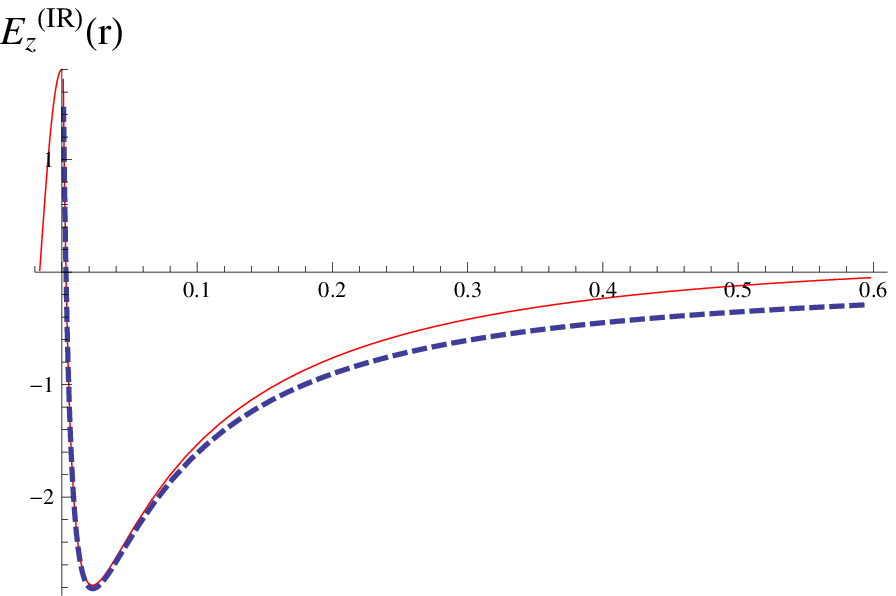}
	\caption{Left panel: The ground state wave-function (blue dashed) in the corresponding Schr\"odinger potential (black solid). This plot is not numerical but a sketch  Right panel: The first excited state $\omega_1=1.9793$ wavefunction as obtained from numerics (blue dashed) and the analytic IR approximation {\protect\eqref{JIR}} (red solid). In both cases the $SL(2,\mathbbm{Z})$ parameters were $a=b=c=d=1$. 
}
	\label{w10}
\end{figure}


\subsubsection{Constant magnetic field and charge density, varying frequency}\label{varyw}

Having constructed the dyonic RG flows, we can easily evaluate the Schr\"odinger potential $V_{schr}$ numerically. Keeping the dyonic background fixed (by considering a fixed $SL(2,\mathbb{Z})$ transformation from the electric frame), $V_{schr}(\omega;r)$ shows the following properties:

\begin{itemize}
\item  We checked that the IR and the UV asymptotics of $V_{schr}$ are the ones predicted in sections \ref{IR} and \ref{UV} respectively, i.e. IR behaves as in (\ref{VsIR}) and the UV as in (\ref{Vuv}) with the correct numerical value for the AdS Radius $L =L_{(0)}\approx 0.76$, which is its value for a flow to the $\tau=i$ fixed point in the electric frame (see (\ref{L})).

\item Fig.~\ref{Vs1}: As $\omega$ is increased, the intermediate accessory singularity in the Schr\"odinger potential moves to the IR.

\item For small $\omega$ compared to $h$ the potential is positive, $V_{schr}>0 \,\,, \forall \,\, r$. Since our Schr\"odinger equation is written without the customary $\omega^2 \Psi$ term on the RHS, we are actually searching for zero energy states in the given potential. Positivity of $V_{schr}$ for small frequencies means that no zero energy states exist for small frequencies, i.e. the spectrum is gapped. Fig.~\ref{Val}: As $\omega$ is increased, negative valleys of $V_{schr}$ begin to appear towards the IR\footnote{The green curve represent the curve for the limiting frequency below which the negative valleys disappear.}. For any $\omega$, the deepest region of the negative valleys is mostly located far in the IR. As we increase $\omega$ further, the negative valleys become deeper in the IR and wider towards the UV, allowing more and more bound states with higher numbers of nodes for increasing frequency. The spectrum is hence discrete and gapped, there are no scattering states. The complete potential in the whole range of $r$ look like the blue curve of the right panel in Fig.~\ref{w10}.

\item Due to the behavior of the potential explained in the last point above, we find normalizable bound state wave functions for discrete values $\omega_n$. A typical wave function is depicted in the  blue curves in Fig.~\ref{w10} (left panel: ground state, right panel: first excited state). 

The increasing depth and width of the valleys as $\omega$ grows suggests that one can fit more and more eigenstates inside. Vanishing of the source then implies normalizability and an infinite transport coefficient associated with the field $E_z$. Hence there exists a frequency-pole in the current-current 2-point function. In other words, the corresponding frequencies $\omega_n$ should be identified with the discrete spectrum of the theory.

\vspace{0.1in}
$\bullet$  We have computed the first few $\omega_n$'s numerically. We found for the choice of electric background input parameters $a=b=c=d/2=1$, $q=1$, $a_0=0$\footnote{See discussion in the beginning of section \ref{Univ}. In particular, $a_0=0$ correspond to a non-running axion in the electric frame. Although $a_0$ is not physical (see Sec. \ref{UVfixedpoints}), we use it as a toy model as the present analysis is preliminary.} the following eigenfrequencies: $\omega_0 \approx 1.7$, $\omega_1 \approx 2$,  $\omega_2 \approx 2.4$. Some numerical details on how these eigen-modes are obtained are reserved for appendix \ref{Ndi}. What is further explained there is why the number of nodes increases (see Fig.~\ref{w10}) and hence the subsequent states are obtained as the frequency increases.
\vspace{0.1in}

\end{itemize}


\subsubsection{Constant frequency and charge density, varying magnetic field}

The spectrum in principle also depends on the $SL(2,\mathbb{Z})$ parameters $(a,b,c,d)$, as well as the magnetic field $h$ and charge density $q$ in the dyonic frame, which in turn are both scaled by the electric frame charge density. This section investigates the change of the spectrum as the magnetic field is varied by changing $SL(2,\mathbbm{Z})$ frame. 
More precisely, we fix an $\omega$ and a $q$ and we plot $V_{schr}$ for (generally) fixed $q'=a q$, i.e. for a fixed $SL(2,\mathbb{Z})$ parameter $a$ as the $SL(2,\mathbb{Z})$ varies and hence as $h'=c q$ changes.
\begin{itemize}
\item As expected from the analytic limits, we checked that the IR and the UV asymptotics are $SL(2,\mathbb{Z})$ invariant. Furthermore, although not shown in the figures, the intermediate (regular) singular point moves to the UV as $h$ increases, as expected  from \eqref{bo} (the position of the singular point depends only on the dimensionless combination $\omega^2/h^2$).
\item Right panel in Fig.~\ref{Vs1}: As $h$ increases the negative valleys of $V_{schr}$ shrink and at some point they disappear. Increasing magnetic field hence generates a larger gap: generally, one needs to increase $\omega$ sufficiently in order to create a negative valley for $V_{schr}$ that may accommodate $\omega_0$.
\item   We found numerically that $V_{schr}$ and hence the whole spectrum is nearly $SL(2,\mathbbm{Z})$ invariant for small filling fractions $\nu=q/h$ (or equivalently large magnetic fields $h\gg q$).
\item For high enough frequencies\footnote{Actually the numerical evidence is already suggestive for $\omega > \omega_n, \,\,n \geq 1$.}, we observe numerically that the spectrum does not change more than a few percent as the magnetic field changes many orders of magnitude.
\end{itemize}

Given that $E_z$ is a linear combination of metric and gauge fluctuation (see (\ref{ExEy})) implies that the 2-point functions of $E_z$ include 2-point functions of the gauge field, the metric tensor and mixed ones.  Hence, those first few $\omega_n$'s that we have computed, is not a-priori clear whether they correspond to the metric or the gauge field. Therefore, it is possible that the $\omega_n$'s that we have computed in this regime of parameters corresponds to metric-metric fluctuations which should be $SL(2,\mathbb{Z})$ invariant. This is supported by the observation of Sec.~\ref{GC} that $E_z$ scales in the UV in the same way as the metric fluctuation. We hope to clarify these issues and make a precise statement about how exactly the spectrum transforms under $SL(2,\mathbb{Z})$ in generic dyonic frames in a future work.


\subsubsection{Constant frequency and filling fraction, scaling charge density and magnetic field}

The charge $q$ in the dyonic frame can be firstly changed directly in the electric frame followed by an $SL(2,\mathbb{Z})$ transformation, which will induce a magnetic field. Starting from an initial charge density $q_0$ in the electric frame  (see (\ref{cd}), (\ref{b}) and (\ref{bq})) and changing it via a rescaling with a factor of $k$ we obtain the new charge density $q_0'=k q_0$. Performing then, an  $SL(2,\mathbb{Z})$ transformation, which induces a magnetic field, we obtain $q'=a q_0'=k a  q_0$ and $h= c q_0'=k c  q_0$. Thus, assuming the same $SL(2,\mathbb{Z})$ transformation for either $q_0$ or $q_0'=k q_0$, the new charges are $k$ times the old ones. Then, as one can show using equation (\ref{famsol}), the quantities $g_0$ and $g_1$ and hence the fluctuation equation (\ref {Ezz}) remain invariant if the frequency is also rescaled according to $\omega'=\omega {\sqrt k}$. This is an exact scaling of the whole spectrum and can be summarized as

\be
\mbox{Fixed $SL(2,\mathbb{Z})$ and $q_0'=kq_0$ in el. frame}\,\, =>\,\, \nu \equiv q/h=\nu'\,\,\,  \mbox{and}\,\,\, \omega_n \sim \sqrt{h}\,\, .
\ee


\section{Conclusions and future directions}\label{conclusions}

In this paper, we have taken a major step toward a realistic holographic model of the fractional quantum Hall effect, extending the work of \cite{Kachru2} and \cite{Burgess}.  Based on strong evidence from condensed matter physics, real-world quantum Hall states seem to be governed by a modular group action on a two-dimensional subspace of couplings at low temperatures.  We implemented this modular group action by starting with an $SL(2,{\mathbbm Z})$-invariant Einstein-Maxwell-axio-dilaton theory and chosing a suitable $SL(2,{\mathbbm Z})$-invariant scalar potential satisfying several constraints required to yield physically sensible results.  In our model, quantum Hall states correspond to holographic RG flows from $AdS_4$ in the UV to dyonic dilatonic black holes in the IR, where the filling fraction is equal to the IR value of the axion. Although our solutions are UV complete, the specific details of the UV completion are not important for the quantum Hall state, which only depends on the leading exponential behavior of the scalar potential in the IR. In practice, we have constructed explicit $SL(2,{\mathbbm Z})$-covariant backgrounds and studied their fluctuations. The backgrounds, although constructed numerically, involve a set of conceptual issues, such as the number of independent input parameters and their physical meanings, which this paper clarifies.

We, in particular, engineer the scalar potential to ensure that the dyonic dilatonic black holes, like all good QH states, exhibit a hard mass gap. To achieve this, we needed to adjust the two parameters $\gamma$ and $\delta$ such that the dyonic dilatonic black holes undergo a Hawking-Page like confinement-deconfinement transition as the temperature is lowered. The low-temperature geometry then is confining, and the excitations around it are automatically  discrete and gapped, which we confirm by analyzing the spectrum of vector fluctuations around the ground state.  We found that the size of the gap scales, at fixed filling fraction, with the square root of the magnetic field. The gap ensures that the longitudinal DC conductivity $\sigma_{xx}$ vanishes exponentially at low temperatures. 

On a technical level, we first computed the AC conductivity of the purely electric solutions, corresponding to states at finite charge density and vanishing magnetic field, and found a delta function contribution to the AC conductivity, $\sigma_{xx}(\omega) \sim \delta(\omega)$, as expected for a translationally invariant system.  The longitudinal conductivity scales  with the charge density as $\sigma_{xx} \sim \sqrt{q}$. We also found a universal formula for the anomalous Hall conductivity in Einstein-Maxwell-axio-dilaton systems, which, in particular, does not depend on the $SL(2,\mathbbm{Z})$ structure.
 
We then mapped the purely electric solutions via $SL(2,{\mathbbm Z})$ to the dyonic dilatonic black holes, our QH states. The $SL(2,\mathbbm{Z})$ transformation of the conductivity automatically led, by virtue of the delta function contribution in the electric frame, to the correct Hall conductivity for the QH states given by the filling fraction, $\sigma_{xy} = \nu$. The filling fraction then turned out to be related to the $SL(2,\mathbbm{Z})$ transformation parameters $a$ and $c$, as well as to the IR value of the axion, by $\nu = \frac{Q}{H}=\frac{a}{c} = \tau_{1}^{IR}$. 

To compute the conductivity and spectrum of these QH states, we analyzed the vector fluctuations around these dyonic solutions in detail. We then went on to calculate the excitation spectrum in the QH states numerically. In order to do so, we derived in Sec.~\ref{Norm} a general sufficient criterion for the finiteness of the generating functional, and in particular showed in Sec.~\ref{GC} that IR regular and normalizable fluctuation modes are mapped onto themselves by $SL(2,\mathbbm{Z})$. 

This analysis demonstrates that the existence of a mass gap and the discreteness of the spectrum are both preserved under $SL(2,\mathbbm{Z})$ mappings.  In particular, because our initial electric solution was engineered to have a gap, the generic dyonic solutions, which represent QH states, are also gapped. This is the most important result of our paper, which improves upon the previous gapless models QH \cite{Kachru2,Burgess}.

We also investigated how the size of the gap changes under $SL(2,\mathbbm{Z})$.  
Focussing on transformations which vary the magnetic field for a fixed charge density and therefore varying $\nu$, we saw that the gap changes very little for smaller values of the magnetic field,  while for larger values, the gap becomes field independent.  In summary, the size of the gap is approximately $SL(2,\mathbbm{Z})$ invariant.  We emphasize that such an indication is preliminary and still needs to be studied in detail.

Along the way, we observed that the potential of the corresponding  Schr\"odinger problem has a singularity at a finite radial position.  Such a singularity is actually quite general, appearing in any Einstein-Maxwell system with a constant magnetic field \cite{Edalati:2009bi}, and in particular, it does not rely on the $SL(2,\mathbbm{Z})$ symmetry.
However, we showed that this singularity is accessible and the fluctuations are consequently regular. In particular, one is not restricted to the $\omega \ll h$ limit of \cite{Edalati:2009bi} when computing the conductivities. Any calculations may proceed smoothly for any values of the frequency $\omega$ and the magnetic field $h$.

We would like to conclude with a couple of open questions and directions for future research.  The fluctuation analysis of Sec.~\ref{sec:Gap} showed that the AC conductivity  for both the pure electric and dyonic black holes implies a discrete and gapped charged excitation spectrum, {\it i.e.}~the AC conductivity is a sum of poles and associated delta functions. How a discrete excitation spectrum, which seems to be preserved under $SL(2,\mathbbm{Z})$ as our direct fluctuation analysis suggests, is actually compatible with the $SL(2,\mathbbm{Z})$ transformation of the AC conductivity tensor, remains an open problem.

We have some indication that the set of fixed points for our chosen $SL(2,{\mathbbm Z})$-invariant model is actually quite rich:  In addition to the $AdS_4$ UV fixed points and dilatonic dyonic IR scaling solutions, there are gapless $AdS_2 \times R^2$ fixed points, as well as solutions with running axion, exhibited in App.~\ref{app:Ansatz}. A full analysis of all the possible IR fixed points and RG flows and the associated phase space is quite involved and is beyond the scope of this paper.  One could hope that metallic transitions states between QH plateaux would be mediated holographically by, for example, one of these ungapped IR solutions.  We hope to return to this problem the future.

Finally, as discussed in Sec.~\ref{dyonicsolutions}, the exact $SL(2,{\mathbbm Z})$ duality of this model yields an infinite number of QH states, with the filling fraction $\nu$ taking every rational value.  Only a finite set of these states is observed in real experimental systems, presumably because impurities break the $SL(2,{\mathbbm Z})$ duality.  One might try to model such effects by introducing disorder into the holographic model.  However, a simpler approach might be to explicitly break the $SL(2,{\mathbbm Z})$ duality by introducing a scalar potential with only approximate $SL(2,{\mathbbm Z})$ invariance.  For example, one could truncate the infinite Eisenstein series \eqref{aa14} to yield a potential which is approximately invariant under a finite subset of $SL(2,{\mathbbm Z})$ transformations.  The goal would be to obtain a more realistic holographic model in which the number of QH states increased as the disorder was reduced and $SL(2,{\mathbbm Z})$ duality was more fully realized. We plan to pursue this line of thought in the future as well.


\bigskip
\noindent
{\bf \large Acknowledgments}

We owe a large debt to Elias Kiritsis, who helped initiate this project, contributed many useful insights, and provided extensive comments on the draft.  In addition, would like to thank S. Hartnoll, C. Herzog, J. McGreevy, I. Papadimitriou, D. Son, A. Stern and R. Tiwari.  M.L. is supported by funding from the European Research Council under the European Union's Seventh Framework Programme (FP7/2007-2013) / ERC Grant agreement no.~268088-EMERGRAV.  
This work is part of the $\Delta$-ITP consortium, a program of the Netherlands Organisation for Scientific Research (NWO) that is funded by the Dutch Ministry of Education, Culture and Science (OCW). The work of R.M. was supported by the World Premier International Research Center Initiative (WPI), MEXT, Japan, as well as by the European Union grant FP7-REGPOT-2008-1-CreteHEPCosmo-228644 and by the EU program ``Thales'' ESF/NSRF 2007-2013. The work of A.T. is supported in part by the Belgian Federal Science Policy Office through the Interuniversity Attraction Pole P7/37, by FWO-Vlaanderen through project G020714N, by the Vrije Universiteit Brussel through the Strategic Research Program ``High-Energy Physics'', and by the Vrije Universiteit Brussel Research Council. Finally, we would like to thank the organizers of the conference "Quantum Field Theory, String Theory and Condensed Matter Physics" (held in Kolymbari, Crete), where this work was finalized, for their kind hospitality.


\appendix

\renewcommand{\theequation}{\thesection.\arabic{equation}}

\section{Equations of Motions in Several Coordinate Systems}
\label{ABCtodomainwall}

Here we present the equations of motion for the Einstein-Maxwell-axio-dilaton system using a more general metric ansatz and show how these specialize to the domain wall parametrization used in the main text.

The general Einstein-Maxwell-axio-dilaton action is
 \be
 S=M^{2}\int d^{4}x\sqrt{-g}\left[R-{1\over 2}(\partial\phi)^2-{W\over 2}(\pa \tau_1)^2+V(\phi,\tau_1)-{Z\over 4}F^2-{\tau_1\over 8} \e^{\m\n\r\s}F_{\m\n}F_{\r\s}  \right] \ ,
\label{generalaction}
\ee
and we have chosen the functions $W$ and $Z$ to be
\be
 W(\phi)={e^{-2\gamma \phi}\over \gamma^2}\sp Z(\phi)=e^{\gamma\phi} \ .
\label{generalWandZ}
\ee
We adopt the following ansatz for the metric:
\be
\label{generalmetric}
ds^2 = -D(r) dt^2 + B(r) dr^2 + C(r) \left(dx^2 + dy^2 \right) \ .
\ee
Note that $B(r)$ is not dynamical, but encodes the choice of radial coordinate. 
For the gauge field, we consider a radial electric gauge field $A_t(r)$ along with a constant magnetic field
\be \label{Fxy}
F_{xy} = h \ .
\ee
We further assume the scalars $\phi$ and $\tau_1$ also only depend on $r$.

Using primes to denote radial derivatives, the gauge field equation is
\be
\partial_{\m}\left(Z\sqrt{g}F^{\m\n}\right)={1\over 2} \tilde{\e}^{\n\m\r\s}\pa_{\m}\left(\tau_1 F_{\r\s}\right)
\label{generalMaxwell}
\ee
with solution
\be
A_t'={\sqrt{BD}(q-h\tau_1)\over ZC}
\label{generalgauge}\ee
where $q$ is a constant of integration.

The scalar curvature of the metric \eqref{generalmetric} is
\be
R={1\over B}\left[{1\over 2}\left({C'\over C}-{D'\over D}\right)^2+{B'\over B}\left({C'\over C}+{D'\over 2D}\right)-2{C''\over C}-{D''\over D}\right]
\label{generalcurvature}
\ee
and Einstein's equations are
\be
{D'\over D}{C'\over C}+{1\over 2}{C'^2\over C^2}+{B(q-h\tau_1)^2\over 2ZC^2}+{h^2BZ\over 2C^2}-BV-{1\over 2}\phi'^2-{W\over 2}\tau_1'^2=0
\label{generalTrr}
\ee
\be
{C''\over C}+{1\over 2}\phi'^2+{W\over 2}\tau_1'^2-{1\over 2}\left({B'\over B}+{D'\over D}\right){C'\over C}
-{1\over 2}{C'^2\over C^2}=0
\label{generalTrr+Ttt}
\ee
\be
{D''\over D}-{C''\over C}-{1\over 2}\left({D'\over D}-{C'\over C}\right)\left({D'\over D}+{B'\over B}\right)={B(q-h\tau_1)^2\over ZC^2}+{h^2BZ\over C^2} \ .
\label{generalTrr+T33}
\ee
The dilaton equation of motion is
\be
\sqrt{1\over DBC^2}\partial_r\left(\sqrt{DC^2\over B}\phi'\right)+{\pa V\over \pa \phi}+{(q-h\tau_1)^2\over 2Z^2C^2}\frac{\partial Z}{\partial \phi} - {h^2\over 2C^2}\frac{\partial Z}{\partial \phi} - {\tau_1'^2\over 2B}\frac{\partial W}{\partial \phi}=0 \ ,
\label{generaldilaton}
\ee
and the axion equation is
\be
W\sqrt{1\over DBC^2}\partial_r\left(\sqrt{DC^2\over B}\tau_1'\right)+{\pa V\over \pa \tau_1}+{\tau_1'\phi'\over B}\frac{\partial W}{\partial \phi}+{h(q-h\tau_1)\over ZC^2}=0 \ .
\label{generalaxion}
\ee

To specialize to the domain wall ansatz, we set
\bea
D &\to& fe^{2A}  \\
B &\to& {1\over f} \\
C &\to&  e^{2A} \ .
\eea
The equations of motion then become
\be
\tau_2'^2+4 \gamma^2 \tau_2^2A'' + {\tau_1'}^2 = 0
\label{dwEOMAApp}\ee
\be
f'' + 3A' f' - e^{-4A} \left(  \frac{(q-h\tau_1)^2}{\tau_2} -  h^2\tau_2 \right)= 0
\label{dwEOMfApp}\ee
\be
\tau_1''+\left[ 3A' + \frac{f'}{f} - 2\gamma \frac{\tau_2'}{\tau_2}  \right] \tau_1' + \frac{\gamma^2 h \tau_2 }{f}  (q-h\tau_1) e^{-4A} +  \frac{\gamma^2 \tau_2^2}{f}  \frac{\partial V}{\partial \tau_1}  = 0
\label{dwEOMaxionApp}\ee
\be
(\log \tau_2)''+ \left[ 3A' + \frac{f'}{f} \right](\log \tau_2)'+ \frac{\tau_1'^2}{\tau_2^2}  + \frac{\gamma^2\tau_2}{f}\left[ \frac{\partial V}{\partial \tau_2} + \frac{1}{2}e^{-4A} \left( \frac{(q-h\tau_1)^2}{\tau_2^2} - h^2\right) \right]=0
\label{dwEOMdilatonApp}\ee
\be
-\frac{1}{2}\left(\frac{\tau_2'}{\gamma\tau_2}\right)^2 +6{A'}^2 + 2A'\frac{f'}{f} - \frac{V}{f} +\frac{1}{2f}e^{-4A}\left( \frac{(q-h\tau_1)^2}{\tau_2}  + h^2 \tau_2 \right) = 0
\label{dwEOMphiApp}\ee
\be
F_{rt} = A_t'= \frac{(q-h \tau_1)}{\tau_2} e^{-A} \ .
\label{dwEOMgaugeApp}\ee
To obtain the pure electric equations (\ref{electricEOMA}, \ref{electricEOMf}, \ref{electricEOMphi}, \ref{electricEOMaxion}, \ref{electricEOMgauge}), set $h=0$. The axion-free equations (\ref{CDBHeomA}, \ref{CDBHeomf}, \ref{CDBHeomphi}) are found by also setting $\tau_1 = 0$.

To obtain the hyperscaling-violating Lifshitz form of the metric, one can set
\bea
D &\to& r^{\theta - 2z}  \\
B &\to& L^2 r^{\theta-2} \\
C &\to& r^{\theta-2} \ .
\eea
Note that when we write hyperscaling-violating Lifschytz metric in \eqref{HSV}, we relabel the radial coordinate $\rho$, which is related the the radial coordinate $r$ in the domain wall parametrization by \eqref{r_in_terms_of_rho}.


\section{Computational details of the RG flow }\label{RGdetails}

This section discusses in detail the construction of the numerical solutions in the absence of a background magnetic field but in the presence of a charge density. We show how the solutions are constructed, how the UV and the IR parameters involved match the two-parameter family of solutions that we expect, and how these solutions can be properly normalized in the UV by using appropriate scaling symmetries of the equations of motion. We also discuss which are the scales involved. For concreteness, instructive examples of flows are constructed and numerical evidence of our analytical results, in particular for the scaling symmetry that allows us to scale out one dimensionful field theory quantity (which we choose to be the charge density), are provided. In the following we will work with a convention in which both the gauge field and the scalar potential are dimensionless.

\subsection{Parameter Counting at UV Fixed Points}
\label{UVperturbations}

We will choose a domain wall ansatz for the metric
\be
\label{domainwallmetric1}
ds^2 = e^{2A(r)} \left(-f(r) dt^2 + dx^2 + dy^2\right) + \frac{dr^2}{f(r)} \ ,
\ee
and assuming the only $r$-dependent component of the gauge field is $A_{t}$, with a solution to Maxwell's equations given by \eqref{dyonicansatz} with $h=0$.
The action \eqref{totalaction} admits, inside the fundamental domain, two $AdS_4$ fixed points given e.g. by $\tau^{(0)}$ and $\tau^{(1)}$ in \eqref{AdS4FundDomain}. The other cusp point $\tau^{(2)}$ is then related to $\tau^{(1)}$ by T-duality. The AdS radii of the two fixed points for our choice of $(\gamma,s)$ (c.f.~eq.~\eqref{gammas}) are given by
\be\label{L}
L_{(0,1)} = \sqrt{6/V(\tau^{(0,1)},{\bar\tau}^{(0,1)})}\,,\quad L_{(0)} = 0.572\,,\quad L_{(1)}= 0.573\,.
\ee

To understand the RG flow from the UV, we analyze perturbations of the $AdS_4$ fixed point solutions.  For generic values of the UV dimensions $\Delta_\phi$, $\Delta_{\tau_1}$, the expansion of the five fields $A$, $f$, $\tau_1$, $\phi=\log \tau_2/\gamma$, and $A_t$ of the domain wall metric is\footnote{For special values of $\Delta_\phi$ and $\Delta_{\tau_1}$, logarithms connected to anomalies will appear in the expansion.}
\begin{eqnarray}
\label{UV1}
A &=& A_1^{UV} \frac{r}{L} + A_0^{UV} + \dots \\
\label{UV2}
f &=& f_0^{UV} + f_1^{UV} e^{-3 A} + \dots \\
\label{UV3}
\phi&=& \phi^{UV} + J_\phi e^{-(3-\Delta_\phi) A} + O_1 e^{-\Delta_\phi A} +\dots \\
\label{UV4}
\tau_1 &=& \tau_1^{UV} + J_{\tau_1} e^{-(3-\Delta_{\tau_1}) A} + O_1 e^{-\Delta_{\tau_1} A} +\dots \\
\label{UV5}
A_t &=& \mu - \frac{L}{A_1^{UV} \tau_2}(q -h \tau_1^{UV} ) e^{-A} + \dots\,,
\end{eqnarray}
which contains ten integration constants: $A_0^{UV}$, $A_1^{UV}$, $f_0^{UV}$, $f_1^{UV}$, $J_\phi$, $O_\phi$, $J_{\tau_1}$, $O_{\tau_1}$, the chemical potential $\mu$ and charge density $q$.\footnote{Note that $O_\phi$ and $O_{\tau_1}$ are related, but not equivalent, to the VEVs of the corresponding operators ${\cal O}_\phi$ and ${\cal O}_{\tau_1}$. The exact relation between VEVs and sources is has to be determined by holographic renormalization along the lines of \cite{Papadimitriou:2011qb,Papadimitriou:2010as,Papadimitriou:2004ap}.}

We can use the residual gauge freedom of the domain wall metric \eqref{domainwallmetric1} to fix some of these constants by properly adjusting the UV asymptotics of the corresponding RG flows. The equations of motion are left invariant under the following three residual gauge transformations:
\begin{subequations}\label{rester}
\begin{align}\label{restr1}
r &\rightarrow c r,\,t\rightarrow t/c\,\, \mbox{which induce}\,\,  f\rightarrow c^2 f \,,\\\label{restr2}
 (t,\vec{x})&  \rightarrow b (t,\vec{x})\,\, \mbox{which induces} \,\,e^A \rightarrow b e^A\,\,\mbox{and}\,\, q \rightarrow  b^2 q, \\\label{restr3}
r & \rightarrow r-r_0\,\,  \mbox{which induces}\,\, A \rightarrow A-r_0/L\quad \text{and}\quad q\rightarrow q e^{2 r_0/L}\,.
\end{align}
\end{subequations}
These three transformations allow us to remove three integration constants in the UV, while enforcing the in the UV by varying the IR length scale will allow us to remove a fourth one (c.f.~Sec.~\ref{IRparameters} for the discussion from the IR point of view). Two constants are removed easily: Applying \eqref{restr2} sets $A_0^{UV}=0$, while \eqref{restr1} can be used to fix the asymptotic blackening factor to $f_0^{UV}=1$. The gravitational constraint \eqref{dwEOMphi} constrains the expansion \eqref{UV1}-\eqref{UV5} via (after using \eqref{L})
\be\label{UVconstraint}
{A_1^{UV}}^2 f_0^{UV} = 1\,,
\ee
fixing $A_1^{UV}=1$ once $f_0^{UV} = 1$ by \eqref{restr1}. We stress that the constraint \eqref{UVconstraint} is not automatic, but needs to be imposed dynamically by varying the parameters of the IR geometry. We will discuss this further in Sec.~\ref{numericalstrategy}. This finally reduces the above UV expansion (\ref{UV1}, \ref{UV2},
\ref{UV3}, \ref{UV4}, \ref{UV5}) to the more familiar form
\begin{eqnarray}\label{UV1a}
A &=& \frac{r}{L} + \dots\\\label{UV2a}
f &=& 1 + f_1^{UV} e^{-3 r/L} + \dots\,\\\label{UV3a}
\phi &=& \phi^{UV} + J_\phi e^{-(3-\Delta_\phi) r/L} + O_1 e^{-\Delta_\phi r/L} +\dots\,\\\label{UV4a}
\tau_1 &=& \tau_1^{UV} + J_{\tau_1} e^{-(3-\Delta_{\tau_1}) r/L} + O_2 e^{-\Delta_{\tau_1} r/L} +\dots \,\\\label{UV5a}
A_t &=& \mu - \frac{L(q - h \tau_1^{UV})}{\tau_2} e^{-r/L} + \dots\,.
\end{eqnarray}
The four VEVs out of the seven remaining constants, which we take to be $f_1^{UV}$, $O_\phi$,  $O_{\tau_1}$, and $\mu$, are fixed in terms of the other three, which we take to be\footnote{Since in quantum Hall samples usually the charge density is fixed by the doping of the semiconductor we choose to keep $q$ fixed and work in the canonical ensemble.} $q$, $J_{\phi}$, $J_{\tau_1}$, by demanding regularity in the IR. Finally, as we will prove in detail in Sec.~\ref{familyq}, we can now use the dimensional scaling of dimensionfull field theory quantities induced by the third residual symmetry \eqref{restr3} to scale out one of the left dimensionful parameters. In Sec.~\ref{familyq} we choose this to be $q$ and give sufficient numerical evidence that \eqref{restr3} indeed scales all quantities with the proper powers fixed by dimensional analysis. These powers are fixed by looking at how \eqref{restr3} acts on the expansion \eqref{UV1a}-\eqref{UV5a}. We are therefore left with a two-parameter family of solutions labelled by the ratios
\be\label{UVratios}
\frac{J_\phi}{q^{3-\Delta_\phi}}\,,\quad \frac{J_{\tau_1}}{q^{3-\Delta_{\tau_1}}}\,.
\ee
As we will see in the next section, the IR scaling solutions are also parametrized by two integration constants, $q$ and $\tau_{1IR}$.  The RG flows provide a mapping between the UV parameters $(J_{\phi}, J_{\tau_1})$  and the IR parameters $(q, a_0)$. Note that since $q$ is a integration constant in the $r$ direction, it shows up both in the UV and the IR data.


\subsection{Counting of IR parameters \& Numerical Strategy}
\label{IRparameters}

As seen in section \ref{IRscalingsolutions}, in the IR $\tau_2 \to \infty$ while the leading behavior of the potential for large $\tau_2$ is $V(\phi) = -2 \Lambda \tau_2^s$ (see (\ref{aa15})) where $\Lambda=-\zeta(2s)$, $Z\equiv \tau_2=e^{\gamma \phi}$. In particular $V$ is independent of $\tau_1$, and hence solutions with any constant $\tau_1$ are allowed. 
In this situation the equations (\ref{dwEOMA})-\eqref{dwEOMphi} reduce to those in \cite{Charmousis:2010zz}) if the magnetic field is zero as well. Hence the IR solutions are the same as those in \cite{Charmousis:2010zz}) with $\tau_1=$constant. Using the residual gauge freedom \eqref{restr1}-\eqref{restr3} we can generalize the scaling solutions of Sec.~8 of \cite{Charmousis:2010zz} by including three additional parameters $c$ (via \eqref{restr1})
\footnote{More precisely, in the way \eqref{CDBHsolnA1}-\eqref{CDBHsolnAt1} are written we replaced $p/c\rightarrow p$, such that $c$ only shows up in \eqref{CDBHsolnf1}.}
, $A_{IR}$ (via \eqref{restr2}), and $r_0$ (via \eqref{restr3})
\begin{subequations}\label{IRsol}
\begin{align}
A(r) &=A_{IR} + \frac{(\gamma - \delta)^2}{4} \log \left(p(r-r_0)\right), \label{CDBHsolnA1}\\
f(r) &= \frac{-16 \Lambda}{wu c^2} e^{-\delta \phi_{IR}} \left(p(r-r_0)\right)^{v},\label{CDBHsolnf1}  \\
\phi(r) &= \phi_{IR} + (\delta-\gamma) \log\left(p(r-r_0)\right),\label{CDBHsolnphi1} \\
\tau_1(r)&=a_0=\mbox{constant}, \label{CDBHsolnax} \\
A_t(r) &= \frac{8}{w} \sqrt{\frac{v\Lambda}{u}} e^{-\frac{(\gamma+\delta)}{2}\phi_{IR}}\left(p(r-r_0)\right)^{\frac{w}{4}}\label{CDBHsolnAt1}
\end{align}
\end{subequations}
where $w = 3\gamma^2 - \delta^2 - 2\gamma\delta + 4$, $u = \gamma^2 -\gamma\delta + 2$, and $v = -\delta^2 + \gamma\delta + 2$. The constraint equation (\ref{CDBHeomphi}) implies the following algebraic relation between the IR parameters
\be \label{res1}
q^2 =e^{-4 A_{IR}}\frac{-4 \Lambda e^{(\gamma- \delta) \phi_{IR}} (2 +
\gamma \delta - \delta^2)}
{(2 + \gamma^2 - \gamma \delta)}
\ee

As a result, only two out of the three constants $q$, $A_{IR}$ and $\phi_{IR}$ are independent. The constant $p$ can be taken arbitrary for now, but is important to note that it has dimension of mass and hence renders the argument of the logarithm dimensionless, and which is the parameter rescaling $r$. The scale $p$ is associated with an effective IR scale generated along the flow from the UV.

\paragraph{Fixing the effective IR scale} So far we are left with five independent IR parameters, $A_{IR}$, $p$, $q$, $a_0$, $r_0$. From experience with the flows in Sec.~\ref{electricsolutions} we can anticipate that $q$ and $a_0$ are the proper IR parameters, as $q$ is a integration constant, and $a_0$ labels the flows that foliate the (e.g. right hand side) of the fundamental domain when integrating up from the IR. The IR parameter $A_{IR}$  will be determined by demanding the validity of
\be\label{res2}
6A'^2f=V(\tau_{UV},{\bar\tau}_{UV})\,\,\,\mbox{at the UV}\,.
\ee
 The constant $c$, which leaves (\ref{res2}) invariant, will rescale $f$ such that $f_{UV}=1$, corresponding to the use of the scaling symmetry \eqref{restr1}. Using\eqref{restr3} the constants $r_0$ must be chosen such that $A_{UV}$ has the form $A_{UV}=r/L+A^0_{UV}$ with $A^0_{UV}=0$. The constant $\phi_{IR}$ will be fixed by inverting the constraint (\ref{res1}) to be
\be \label{res3}
\phi_{IR}=\frac{1}{\gamma- \delta} \left(4 A_{IR}-\log\left[ \frac{-4 \Lambda    (2 +
\gamma \delta - \delta^2)}
{q^2(2 + \gamma^2 - \gamma \delta)} \right]\right)
\ee
and by demanding constantly its validity during the procedure that the rest of the parameters are found. 
Equation \eqref{restr3} can be used to scale out one dimensionfull UV quantity. The fact that (\ref{res2}) is invariant under \eqref{restr1}-\eqref{restr3} implies that $p$, $r_0$, and the pair $A_{IR}$ and $\phi_{IR}$ can be  obtained simultaneously. That is the directions of $p$, $r_0$, and of the pair $A_{IR}$ and $\phi_{IR}$, for fixed $a_0$, are orthogonal in parameter space. This fact simplifies the numerical approach significantly. Hence, we are  left with two physical parameters in the IR, $q$ and $a_0$. The RG flow will relate them dynamically to the UV data discussed in Sec.~\ref{UVperturbations}.


\subsubsection{Numerical Strategy}\label{numericalstrategy}

Before we outline the steps of the numerical construction of the RG flows, we make two important remarks:
\begin{enumerate}
\item The potential that we will use is a run-away potential at the IR, consisting of a single exponential with corrections. Therefore, the scaling solutions \eqref{IRsol} will only be approximate solutions to the full equations. Using them as initial conditions, there is hence no need in perturbing the IR solutions, in order to numerically evolve towards the UV. We however list the radial perturbations in Sec.~\ref{numericsappendix} for completeness. The flows of this section correspond to choosing $d_1=d_2=d_3=d_4=0$ there.
\item The smallest value of $\tau_2$ inside the fundamental domain is attained at the lowest UV fixed points (see (\ref{AdS4FundDomain})). In the large $\tau_2$ expansion (\ref{aa15}) the contributions of the Bessel functions with larger values of $n,m$ can be neglected, as they decay exponentially in $\tau_2$ towards larger $\tau_2$. We thus can truncate the sum in \eqref{aa15} to low values. In the numerics for this paper we truncated to $|m|,|n|\leq 3$. Although in this way we loose $SL(2,{\mathbb Z})$ invariance in the full $\tau$ plane, the truncated potential still well-approximates the full potential inside the fundamental domain. We will then use $SL(2,\mathbb{Z})$ transformations to map the fundamental domain to its images.
\end{enumerate}

The steps for the numerical approach are summarized as follows:
\begin{enumerate}
\item Although strictly not necessary, it is convenient to choose a dimensionless variable for the numerical evolution. Since we are shooting from the IR, the natural variable is
\begin{align}\label{n}
n-n_0\equiv p(r-r_0)=r-r_0\,.
\end{align}
Since we will use the trial value $p=1$ (in units of the scalar potential) below in order to construct flows, this is equivalent to the use of $r$ (which is dimensionless in units of $d$ already).

\item The initial shooting point $n_I$ must be chosen deep enough in the IR such that the sub-leading terms in the scalar potential are much smaller than the leading exponential. On the other hand, because of the Bessel functions involved in the scalar potential, $n_I$ cannot be too small or there will be numerical problems. A good choice turns out to be $n_I = O(10^{-4})$ different than $n_0$. For such an $n_I$, the sub-leading exponential in the potential being $10^{-8}$ times smaller compared to the leading one (the Bessel functions are $O(10^{-10^8})$ suppressed).

\item We pick a value for $\tau_1^{(0)}=a_0$, and a initial value for the charge $q$. The charge $q$ will change during application of the scaling symmetries \eqref{restr1}-\eqref{restr3}. There are two different cases that yield to two different UV AdS$_4$ geometries which, are not related by $SL(2,{\mathbb Z})$: $a_0=0$ flows to the $\tau^{(0)}=i$ fixed point, and $0<|a_0| \leq \frac{1}{2}$ flows to $\tau^{(1)} = e^{\frac{2\pi i}{3}}$ if $a_0>0$, and to $\tau^{(2)} = e^{\frac{\pi i}{3}}$ if $a_0<0$. As mentioned above, $\tau^{(0)}$ has a different AdS radius than $\tau^{(1,2)}$, but the radii of $\tau^{(1)}$ and $\tau^{(2)}$ are the same, as they are related by T-duality. The values of $a_0$ may hence be restricted to the right part of the  fundamental domain by $a_0\in[0,1/2]$. Hence, in what follows we will only consider the two distinct fixed points $\tau^{(0)}$ and $\tau^{(1)}$.

\item We start out with trial values $p=1$, $A_{IR}=0$, $n_0$=0 for the IR parameters. With this data we compute IR initial conditions from \eqref{IRsol} and (\ref{res1}) (functional values and first derivatives), and run {\it NDSolve} up to a large value of $r \sim O(10^3)$.

\item {\bf Determination of the IR scale:} This is the main step in order to generate a viable solution.
As already described, $c$ sets the effective IR length scale of the solutions \eqref{IRsol}. We determine $p$ in a two stage process. Firstly, by varying $A_{IR}$ in the IR, aiming at fulfilling the constraint \eqref{res2} in the UV\footnote{Where the UV fixed point in (\ref{res2}) can be either of the two, $\tau^{(0)}$ or $\tau^{(1)}$ mentioned in step 3.}. Practically we used Newton's method and required \eqref{res2} to be fulfilled to $O(10^{-6})$. While tuning $A_{IR}$ to fulfill (\ref{res2}), equation (\ref{res1}) was valid and hence for each trial $A_{IR}$ value, the value of $\phi_{IR}$ was determined.

\item In the second stage we act on the resulting IR parameters with the residual gauge transformations \eqref{restr1}-\eqref{restr2}, in order to adjust to the above discussed UV asymptotics, $f\rightarrow 1$, $A \simeq r/L$ as $r\rightarrow \infty$ by an appropriate choice of $p$ and $n_0$ respectively. This completes the determination of the IR scale $p$. At the end we scale out $q$ by \eqref{restr3}, setting it effectively to $q=1$. The result is the properly normalized RG flow from a particular UV $AdS_4$ fixed point to the IR scaling geometries \eqref{IRsol}.
At the end we check by plotting that the asymptotics indeed are as expected (see Fig.~\ref{bagel}).

\end{enumerate}


\subsection{Numerical checks of the scaling of {\protect\eqref{restr3}} and {\protect\eqref{famsol}}} \label{familyq}

We now turn to the numerical evidence for the residual gauge transformation \eqref{restr3} that we used to scale out the charge density. The main point of this section is to show that for a fixed IR axion value $a_0$ and assuming that the numerical solution for a given $q_0$ has been found, then the numerical solutions are known for any other $q$ without any further numerics by application of the UV scale transformation \eqref{restr3}. Note that in principle \eqref{restr3} could have been broken by matter conformal anomalies. The fact that the scaling \eqref{restr3} works tells us that for our choice of scalar dimensions there is no matter conformal anomaly in the system.

Although in principle it is not necessary, it is convenient to define a scaling parameter $k$ which tunes the the charge density via
\be\label{b}
k= \frac{q}{q_0}\,,
\ee
i.e. after changing the charge, its new value will be $q = q_0 k$. Then the claim is that for any $k$, keeping $a_0$ fixed, the transformation
\be \label{bq}
q_0 \rightarrow k q_0
\ee
acts on the IR parameters as:
\be\label{fampar}
p_0\rightarrow p_0 \hspace{0.1in}, A_{IR}^0 \rightarrow A_{IR} ^0-1/2\log(k), \hspace{0.1in} n_0 \rightarrow n_0+L/2 \log(k)\,
\ee
This transformation is nothing but \eqref{restr3}. As a consequence of (\ref{res1}) the IR value of the dilaton is unchanged,
\be\label{fampar2}
\phi_{IR}^0\rightarrow \phi_{IR}^0.
\ee
That these transformed values solve the equations of motion successfully with the correct asymptotics fixed once, i.e.~with (\ref{UVconstraint}), $A \sim r/L$, and  $f\sim 1$, has been checked numerically for a large number of $k$'s that differ by six orders of magnitude. We will describe this evidence below. Concluding, for fixed $a_0$ there is one index family of solutions labeled by k. Including the axion, there is a two indexed family of solutions given by the relations
\begin{subequations}\label{famsol}
\begin{align}
a(r;q,a_0)&=a(r-L/2 \log(k);q_0,a_0), \notag\\
 \phi(r;q,a_0)&=\phi(r-L/2\log(k);q_0,a_0),\label{famsolsc}\\
f(r;q,a_0)&=f(r-L/2 \log(k);q_0,a_0), \notag\\
A(r;q,a_0)&=A(r-L/2 \log(k);q_0,a_0)+1/2\log(k) \label{famsolgeo}.
\end{align}
\end{subequations}
where (\ref{famsolsc}) and (\ref{famsolgeo}) refers to the way the scalars and the geometry respectively transforms under (\ref{bq}).
In particular, once the solutions for a given pair of values $q_0,a_0$ is (numerically) obtained and properly normalized, then the solutions for any $q,a_0$ can be constructed without further numerics. The ratios vev/source for the fields $f$, $\phi$ and $a$ satisfy usual dimensional scaling
\be \label{scaleO}
\frac{vev}{source}\big|_{q=k}= k^{3/2-O_i} \frac{vev}{source}\big|_{q=1}\,\,\, \mbox{where} \,\,\, i=f,\phi,\tau_1 \,\,\, \mbox{with} \,\,\, O_f=0
\ee
while for the $A_t$ field
\be
\frac{vev}{source}\big|_{q=k}= \sqrt{k} \frac{vev}{source}\big|_{q=1}.
\ee


\subsubsection{Numerical confirmation of equations {\protect\eqref{scaleO}} }

\begin{figure}[htb]
	\includegraphics[scale=0.8]{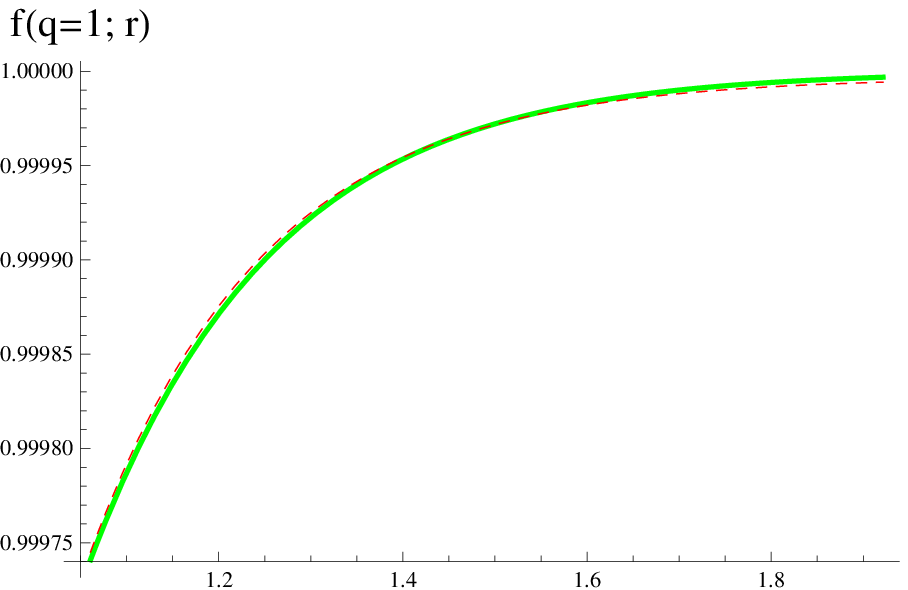}\hspace{0.05cm}
	\includegraphics[scale=0.8]{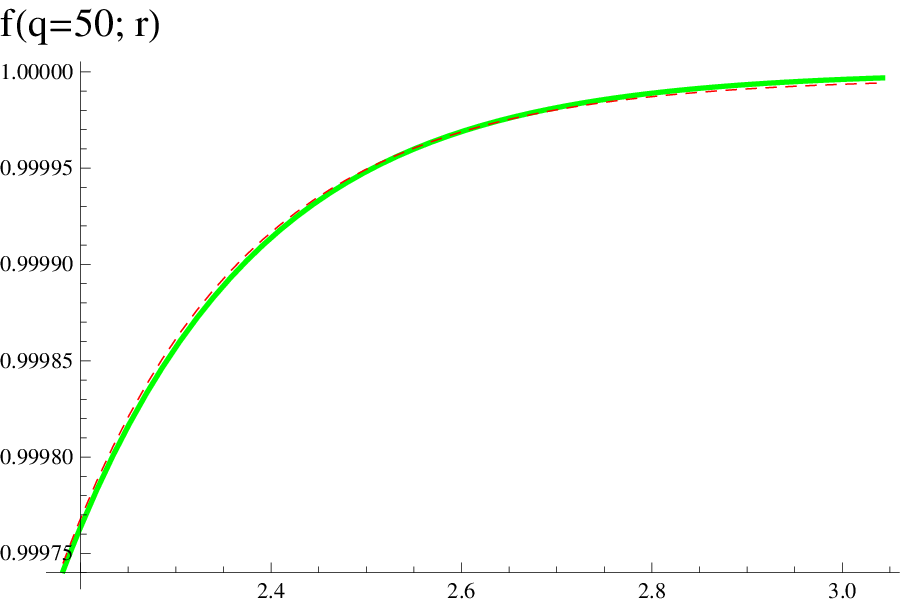}\hspace{0.05cm}
	\caption{Both panels concern the field $f(r)$ as a function of r (in units where $d=1$) for $a_0=0.2$ which results in a running axion. 
	Left panel: concerns $q=1$. The green curve is the numerical solution. The red dashed curve is the fitting of the numerical curve at larger arguments. In the present case the fitting is achieved by $1-0.065\exp{(-r/L)}$ where the value $L=L_1=\sqrt{6/V(\tau^{(1)})} \approx 1.745$ has been used. Right panel: similar things as in left panel apply but with  $q=50$ this time. The fitting is now achieved by $1-22.981\exp{(-r/L)}$. The only visible difference between the left and the right panel is a shift which is (very accurately) equal to $L/2\log(50)$ as expected. Observing that $22.981/0.065=353.553$ and furthermore that $353.553\approx 50^{3/2}$ suggests that the claim of {\protect\eqref{scaleO}} is correct.}
\label{scale_vev_sou}
\end{figure}

The plots in Fig.~\ref{scale_vev_sou} are in striking agreement with the scalings of the vevs over the sources as the charge density varies. We only show the details for the field $f$.
The scaling of $f$ is as dictated by equations (\ref{scaleO}). The caption of Fig.~\ref{scale_vev_sou}  provides further explanations and supports our claims.


\subsubsection{Numerical confirmation of equations {\protect\eqref{famsol}} }
\begin{figure}[htb]
	\includegraphics[scale=0.8]{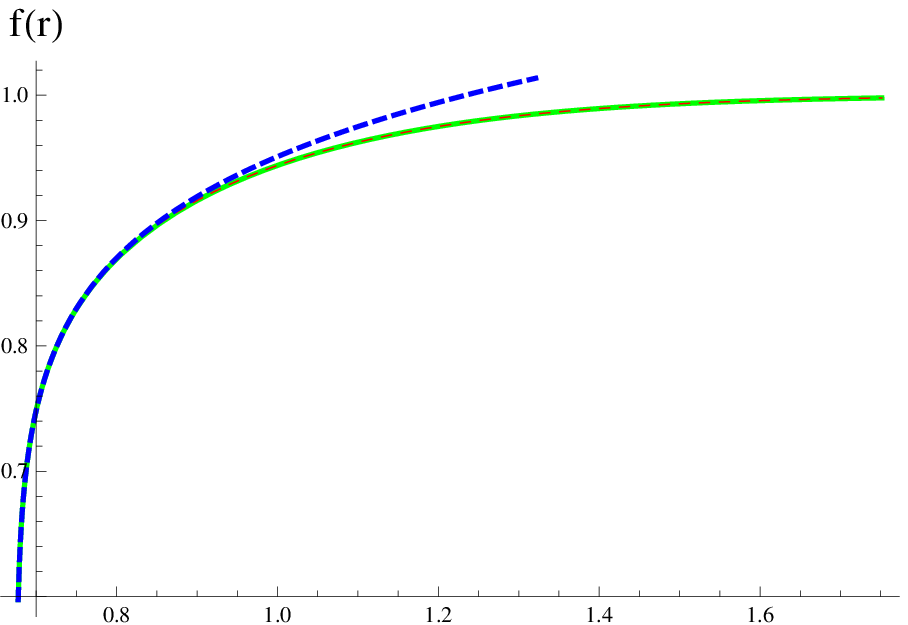} \hspace{0.2in}
	\includegraphics[scale=0.8]{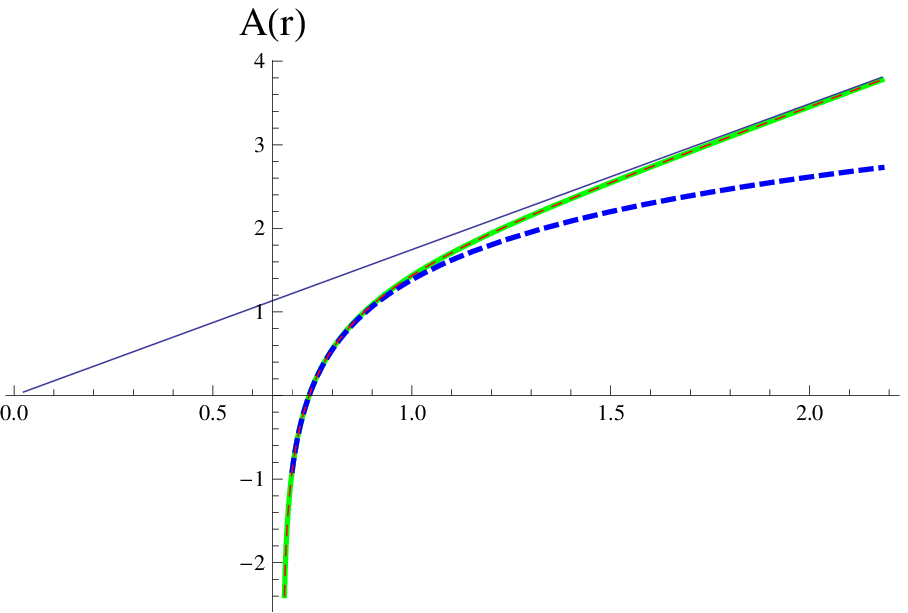}\\
	\includegraphics[scale=0.8]{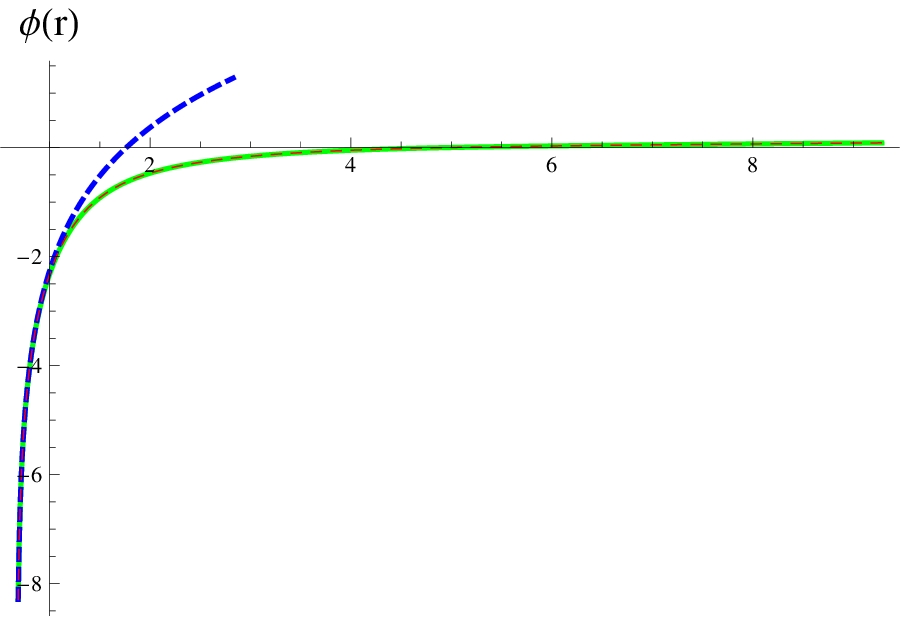}\hspace{0.2in}
	\includegraphics[scale=0.8]{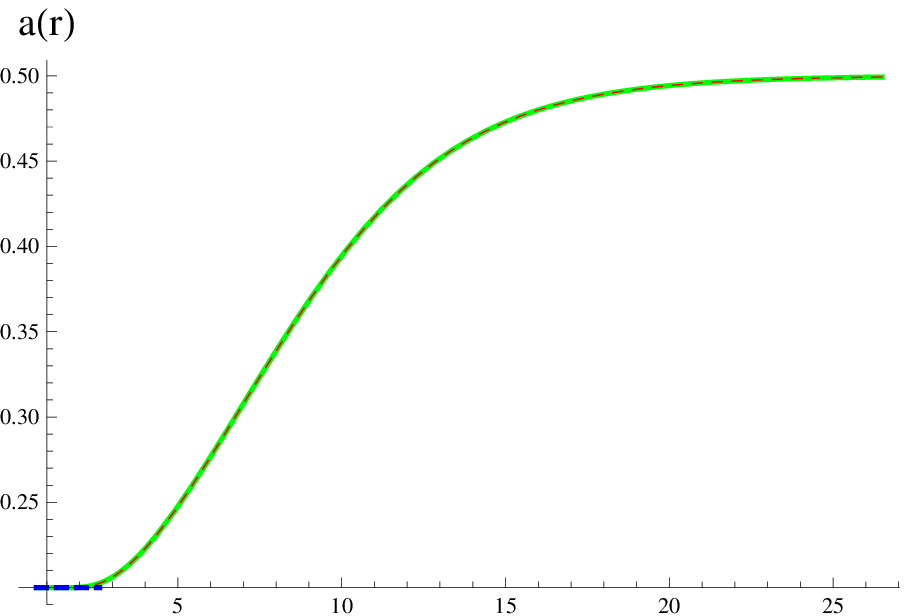}
	\caption{Check of the scaling law {\protect\eqref{famsol}} and of  the IR and the UV asymptotics.  Upper Left: Blackening Factor $f(r)$, Upper Right: Scale Factor $A(r)$, Lower Left: Dilaton $\phi(r)$, Lower Right: Axion $\tau_1(r)$. Red dashed curves are obtained for $q=1$ ($k=q_0=1$), green solid curves are obtained by applying the scaling {\protect\eqref{famsol}} on the numerical solutions obtained for $q=50$ ($k=50$). The blue dashed curves are the analytical solutions at the IR {\protect\eqref{IRsol}} (our initial condition for numerical evolution) for $q=50$. The UV behavior of the fields is the expected one. In particular, the scalars and the blackening factor asymptote to their expected constant values. The
	field $A(r)$ becomes linear with slope $1/L$ to a great accuracy as it is supported by the purple straight line, which has equation $r/L$.}
	\label{bagel}
\end{figure}

We check scaling (\ref{famsol}) by superimposing, as in Fig.~\ref{bagel}, the background fields $\phi$, $\tau_1$, $A$ and $f$ for a small value of $q=1$ (red dashed  line)
, as well as for a scaled-up value $q=50$ (green solid line). The red dashed curves have been scaled according to \eqref{famsol}. More specifically, we used a flow with varying axion, with starting values $p=23.237$ and $r_0=(-0.777+0.5\log(k)) L$ with $L=L_1=\sqrt{6/V(\tau^{(1)})} \approx 1.745$ ($k=1$ for $q=1$ flow and $k=50$ for the scaled flow). $\phi_{IR}$ is given by (\ref{res1}).  Included in that plot are also the IR asymptotics (\ref{IRsol}) (blue dashed line), as well as the UV asymptotics for $A(r)$ \eqref{UV1a} (black solid line). 

The plots confirm the following. (i) The small argument asymptotic are the hyperscaling solutions, equations (\ref{IRsol}). (ii) The solutions solve the Einstein's equations including the constraint equation by construction. This has been checked numerically for any of the solutions we have constructed. (iii) The geometry asymptotes properly to AdS$_4$. This implies $A\rightarrow r/L$, $f \rightarrow 1$ and the scalars asymptote to their UV values for $r\gg L$. (iv) The two families of solutions as $q$ takes two different values behave as dictated by equations (\ref{famsol}).

\subsection{Holographic Renormalization in the Charged Sector}\label{sec:holoRG}

\subsubsection{Charge Density}\label{chargedensity}


In this section we will derive the boundary charge density from holographic renormalization of the gauge boson part of the action \eqref{SF1}. In the extremal case we need to impose a regularity condition on the $t$-component of the gauge field. The usual argument that $A_t$ has to vanish at the horizon only applies to nonextremal horizons. Our geometry is well-defined in the IR, and can in principle support a nonnormalizable leading mode there, which would correspond to another integration constant $C$ in the integral of (\ref{dyonicansatz}) if integrated around \eqref{IRsol},
\be\label{IRAt}
A_t= C + \int_{r_0}^r  (q - h \tau_1)e^{-\gamma \phi(r) -A(r)} dr\,
\ee
where $r_0$ is defined in (\ref{IRsol}). In what follows we set $r_0$ equal to zero for simplicity as its precise value does not affect any of the subsequent arguments. The fact that the integral converges at $r=0$ follows from (\ref{CDBHsolnAt1}) and the Gubser bound \eqref{Gubser}. The nonnormalizable mode is however not allowed by the following consideration: In the UV the chemical potential $\mu$ is given by the value of $A_t$ at the boundary, i.e. the leading term in the large $r$ limit. The charge density is related to the sub-leading term in the large $r$ expansion,
\be\label{AtUV}
A_t = \mu-(q-h\tau_1^{UV})Le^{-\gamma \phi_{UV}-r/L}+\mbox{subleading as }r\rightarrow \infty
\ee
where
\be\label{mu}
\mu= A_t(\infty) = A_t^{(0)} = \int_{0}^{\infty} (q-h\tau_1(r))e^{-\gamma \phi(r) -A(r)} dr \,.
\ee
Comparing \eqref{IRAt} with \eqref{AtUV} we see that the IR constant $C$ would shift the UV chemical potential, making it ill-defined. Since the charge density however is well-defined, this would destroy the thermodynamic relationship of the chemical potential being the conjugate variable to the charge in the system. Hence, the natural choice is to fix $C=0$, in which case the chemical potential is given by the definition \eqref{mu}. Also note that we could have introduced the integration constant as a lower limit in the integral of \eqref{IRAt}. Since $r$ (or better $e^A$) has the interpretation of renormalization group scale, this would mean that we do not calculate the chemical potential in \eqref{mu} by including all the charged modes down to zero energy, but only up to a finite IR cutoff. Since we do not know what sets this IR cutoff, the procedure is clearly ill-defined. We need to integrate all the way into the deep IR.

We now proceed to evaluate the charge density, defined from the holographic effective action as
\be\label{Jt}
\rho \equiv<J^t>= \frac{\delta S_{eff}}{\delta (A_t^{(0) QFT})} =\frac{\delta S_{eff}}{\delta \mu}.
\ee
where the derivative is understood to be taken with respect to the gauge field \textit{after} sending the cutoff to infinity. In order to calculate the charge density we need to evaluate the on-shell part of the action $S_F$ in \eqref{totalaction}. In what follows we will write it as a total derivative. Working with an inverse radial coordinate $u = L e^{-r/L}$, for which the asymptotic UV metric becomes
\be \label{pp}
\ ds^2=\frac{L^2}{u^2}\left(-dt^2+dx^2+dy^2+du^2\right)
\ee
and, for generality including a background magnetic field (see (\ref{Fxy})), we find
\begin{align}
\label{SFbdry}
S_F&=-\frac{M^2}{4 } 2\times \int \ud^{4}x\left[\sqrt{-g}~\tau_2 ( \nabla_{\mu}A_{\nu})F^{\mu \nu}+\frac{\tau_1}{2}\tilde{\epsilon}^{\m\n\r\s}(\partial_{\m}A_{\n})F_{\r\s}\right] \notag\\
&=-\frac{M^2}{2 }  \int \ud^{4}x \Bigg\{ \left[\sqrt{-g}~\nabla_{\mu}(A_{\nu}\tau_2 F^{\mu \nu})+\partial_{\m}(\frac{\tau_1}{2}\tilde{\epsilon}^{\m\n\r\s}A_{\n}F_{\r\s})\right]\notag\\
&-A_{\nu}\underbrace{\left[ \nabla_{\mu}(\sqrt{-g}\tau_2 F^{\mu \nu})+\partial_{\m}(\frac{\tau_1}{2})\tilde{\epsilon}^{\m\n\r\s}F_{\r\s}\right]}\limits_{=0 \text{ by {\protect\eqref{Maxwell_eq}}}} \Bigg \}\notag\\
&=-\frac{M^2}{2 }  \int \ud^{4}x \,\,\, \partial_{\mu}\left[ \sqrt{-g} \left(A_{\nu}\tau_2 F^{\mu \nu}+\frac{\tau_1}{2}\frac{\tilde{\epsilon}^{\m\n\r\s}}{\sqrt{-g}}A_{\n}F_{\r\s}\right)\right]
\notag\\
&=\frac{M^2}{2 }  \int \ud^{3}x  \,\,\, n_{u} \sqrt{-\gamma} \left[ A_{\nu}\tau_2 F^{u\nu}+\frac{\tau_1}{2}\frac{\tilde{\epsilon}^{u \n\r\s}}{\sqrt{-g}}A_{\n}F_{\r\s}\right]_{u=0} \notag\\
&=-\frac{M^2}{2 }  \int \ud^{3}x \left[ \,\,\, \frac{1}{\sqrt{g^{uu}}} \frac{L^3}{u^3}A_t^{(0)}\tau_2^{(UV)} F_{ut}g^{uu}g^{tt} + 2\times \frac{\tau_1}{2}  \tilde{\epsilon}^{utxy}A_t F_{xy}\right]
\Big|_{u=0} \notag\\
&=\frac{M^2}{2 }  \int \ud^{3}x  \,\,\,A_t^{(0)}\left[\tau_2^{(UV)} F_{ut} +\tau_1^{(UV)}h
\right] \Big|_{u=0} \notag\\
&=\frac{M^2}{2 }  \int \ud^{3}x  \,\,\,A_t^{(0)} \left[ \tau_2^{(UV)} \frac{q- \tau_1^{(UV)}h} {\tau_2^{(UV)}}+\tau_1^{(UV)}h \right] =
\frac{M^2 q}{2}  \int \ud^{3}x  \,\,\,A_t^{(0)}
\end{align}
where a few explanations are in order. First, the Maxwell part of the on shell action is finite and no counter-terms are required. We can hence remove the cutoff right from the start. In the first equality we used  $\partial_ {\m}A_{\n}-\partial_{\n}A_{\m}=\nabla_ {\m}A_{\n}-\nabla_{\n}A_{\m}$ and the antisymmetry of $F_{\m \n}$ to pull out a factor of two. In the second equality we used $\tilde{\epsilon}^{\m\n\r\s}\partial_{\m}F_{\r\s}=0$. In the third equality we used (\ref{Maxwell_eq}) and the fact that for any antisymmetric tensor $\nabla_{\mu}F^{\mu \nu}=\partial_{\mu}({\sqrt -g} F^{\mu \nu})/{\sqrt -g}$ while we multiplied and divided the second term in the first line by ${\sqrt -g}$ in order to apply the divergence theorem in the next step. In the fourth equality $\gamma_{ab}$ is the induced metric on the boundary of the AdS$_4$, $n_u=1/{\sqrt{g^{uu}}}(0,0,0,-1)$ is a time-like outward pointing unit vector on the boundary and $\sqrt{-\gamma/g^{uu}}=\sqrt{-g}$. In the fifth equality the second term does not vanish because we include a background magnetic field (see (\ref{Fxy})), as we want to analyze fluctuations about the dyonic background later on.  We then use $\tilde{\epsilon}^{txyu}=+1$ (see section \ref{axiodilaton}),  (\ref{dyonicansatz}) and (\ref{pp}). Finally we replaced $F_{ut}$ by equation (\ref{generalgauge}) evaluated on the Poincare patch  \eqref{pp}.

When the last line of (\ref{SFbdry}) gets differentiated with respect to $A_t^{(0)}$ one must pull out an additional factor of two.\footnote{{We note that the variation here is with respect to a field that is constant on-shell, and hence the $\delta$ function is to be understood as a menomic that we somehow need to cancel the volume integral, which otherwise would be divergent. We also could have formally divided by the volume of the 2+1-dimensional space-time in \eqref{Jt} from the beginning. The correct way however to deal with the volume factors is to use the Euclidean path integral and observe that the  length of the time circle is $\propto T^{-1}$, and that the variation of the partition function with respect to $\mu$ gives the total charge, i.e. charge density integrated over 2-dimensional space (c.f.~e.g.~Eq.~(7.4)~in~\cite{Jensen:2011xb}). In that case the definition of charge density would be $\rho = \frac{T}{Vol_2}  \frac{dZ}{d\mu} = \frac{T}{Vol_2} \frac{d S_{ren}}{d\mu}$, where $Z$ is the partition function. This relation should then replace \eqref{Jt}. The rest of the calculation would proceed accordingly.}} The reason is because the original Lagranian of the gauge field appears as $F^2$. Hence, differentiating the $F^2$ with respect to $A_t^{(0)}$ and retracing the steps of the calculation of equation (\ref{SFbdry}), effectively replaces the integral of the last equality with the factor two. Thus, combining equations (\ref{Jt}) and (\ref{SFbdry}) we obtain
\be\label{cd}
\rho \equiv<J^t>=\frac{\delta S}{M\delta A_t^{(0)}}=
q M .
\ee
The charge density $\rho$ has dimensions of mass square because $q$ has dimensions of mass and hence it has the right dimensions. We observe that the final result for the on shell action and the charge density is independent on the magnetic field. The result implies that even if we tune $q/h$ such that the equality $q/h=\tau_1$ applies, as for example at our quantum Hall plateaux states at $\tau_2=0$, the naive expectation that the magnetic charge would screen the electric one only applies in the bulk deep in the IR: The charge density as measured at the UV $AdS_4$ fixed point is  always given by (\ref{cd}) regardless from the value of $h$.


\subsubsection{Quadratic Fluctuations and Conductivity Formula}

We begin by introducing the complex coordinates
\be \label{zzb}
z=x+i y, \,\,\,{\bar z}=x-iy.
\ee
The reason is because, as one could verify, the linear combination of the components $(x)+ i(y)$ of Maxwell equations and $(tx)+ i(ty)$ of Einstein equations decouple from the ones with $i \leftrightarrow -i$. This implies that any differentiation of of field with a ${\bar z}$ label with respect to a source with $z$ label is zero. In fact, this is the utility of intruding such a coordinate transformation.  
 
The transformation (\ref{zzb}) induces the following transformation on the boundary (Minkowskian) metric $\eta_{\mu \nu}$
\be
\eta_{zz}=\eta_{{\bar z} {\bar z}}=\eta^{zz}=\eta^{{\bar z} {\bar z}}=0, \,\, \eta_{{\bar z} z}=\frac{1}{2},\,\,  \eta^{{\bar z} z}=2.
\ee

In the domain wall parametrization we have that at the UV, the solution to the fluctuation equation (\ref{gaugeperturbations}) behaves as
\begin{align}\label{aoa1}
a_z=a_z^{(0)}+a_z^{(1)}e^{-\frac{r}{L}}+...,r\gg L\,\,\, \mbox{where}\,\,\, a_z^{(0)}=source\,\,\, and\,\,\,a_z^{(1)}=vev
\end{align}
where $L$ is the radius of the AdS$_4$. Equation (\ref{gaugeperturbations}) is the one we used\footnote{In particular, we took a linear combination of the $x$-component plus the $i y$-component of (\ref{gaugeperturbations}).} in performing the numerical evaluations, as those appearing in Fig.~\ref{axflalarger}, for the source/vev. On the other hand, in evaluating the on shell action its more useful to work in inverse radial coordinates in which case equation (\ref{aoa1}) becomes
\begin{align}\label{aoa1z}
a_z=a_z^{(0)}+a_z^{(1)}\frac{u}{L}+...,u \rightarrow 0\,\,\, \mbox{where}\,\,\, a_z^{(0)}=source\,\,\, and\,\,\,a_z^{(1)}=vev.
\end{align}

Starting from the forth equality in (\ref{SFbdry}) we can expand the Maxwell part of the action to second order in the fluctuations obtaining
\begin{align} 
S_F^{(2)}+S_{GR}^{(2)}&=-\frac{M^2}{2} \int \ud^{3}x  \,\,\, \left[\sqrt{-\gamma}  n_{u} a_{x}\tau_2 f^{ux}+\tau_1 \tilde{\epsilon}^{u x t y}a_{x}f_{ty}+(x\leftrightarrow y)\right]_{u=0}
 \notag\\ &+M^2\int d^4x \,\,\mbox{gravity}
\end{align}
where $f_{\mu \nu}=\partial_{\mu} a_{\nu}-\partial_{\nu} a_{\mu}$. Last equation then yields
\begin{align}\label{isa}
S_F^{(2)}+S_{GR}^{(2)}&=-\frac{M^2}{2} \int \ud^{3}x  \,\,\, \left[~ \tau_2 a_x a_x'+
  \tau_1 \tilde{\epsilon}^{u x t y}a_{x}{\dot a_{y}}+(x\leftrightarrow y)\right]_{u=0} +M^2\int d^4x \,\,\mbox{gravity} \notag\\
&= -\frac{M^2}{2} \int \ud^{3}x \left[ \,\,\, \tau_2^{(UV)} (a_x a_x'+a_y a_y')+\tau_1^{UV}(a_{x}{\dot a_{y}}-{\dot a_{x}}{a_{y}}) \right]_{u=0}  \notag\\
&+M^2\int d^4x \,\,\mbox{gravity}
\end{align}
where primes and dots denote differentiation with respect to the radial and time direction respectively.

The gravity fluctuations come either from expanding the determinant and they have the form $h_{tx}^2$ and $h_{ty}^2$ or they come from the $F^2$ term. The ones coming from $F^2$ have the form $h_{tx}^2$, $h_{tx}a_x$ and $h_{tx}a'_x$ and their y-parts, where prime denotes a $u$-derivative. The $h_{tx}^2$ are pure gravity terms while the $h_{tx}a_x$ terms can be written as $h_{tx} (h_t^x)'$ by using the (rx) component of Einsteins equations (see ((\ref{ahry}) with $h=0$). Likewise, the last terms $h_{tx}a'_x$ can be expressed in terms of $h_{tx} (h_t^x)'$ and $h_{tx} (h_t^x)''$ by differentiating the (ux) components of Einstein's equations used in the previous step. Finally, $h_t^{''x}$ can be expressed in terms of $h_t^x$ and $h_t^{'x}$ using the second order equation of motion for $h_t^x$ and which, can be obtained as follows. By first contracting (\ref{Einstein_eq}) with $g^{\mu \nu}$ and expanding to first order in the fluctuations. Since the scalars and hence the scalar potential do not fluctuate, this implies the equation $R^{(1)}=0$ where $R^{(1)}$ is the first order correction of the scalar curvature due to the metric fluctuations. Equation $R^{(1)}=0$  provides the second order differential equation for $h_t^x$ (decoupled from $a_x$). In all, the second order part $S_F$ of the on shell action can be written as part that involves only the gauge field and as a part that involves the metric fluctuations. Both fields appear with zeroth and first derivatives only. We again note, as in deriving (\ref{SFbdry}), that the Maxwell part of the second order contribution in the fluctuations of the on shell action is finite. Hence no counter-terms have been required.

Writing $S_F^{(2)}$ in the coordinate system (\ref{zzb}) and going to Fourier space
\be
a_z(t)=\int_{-\infty}^{\infty} \frac{d \omega}{2 \pi} a_z(\omega)
\ee
and performing the integrations of the resulting $\delta(\omega+\tilde\omega) d\tilde\omega'$ yields
\be
S_F^{(2)}=-\frac{M^2}{2} \int \frac{d\omega}{2\pi}\frac{dz d{\bar z}}{2}  \,\,\, \left[2 \tau_2\left (a_z(\omega) a_{\bar z}'(-\omega)+a_z' (\omega)a_{\bar z}(-\omega)\right)+ 4 \omega \tau_1
a_{\bar z}(\omega) a_z(-\omega) \right]_{u=0}. 
\ee
where we dropped terms of the form $\omega a_z(\omega) a_z(-\omega) $ and $\omega a_{\bar z}(\omega) a_{\bar z}(-\omega) $ from symmetry-antisymmetry considerations and where a factor of $1/2$ comes from the measure. Last equation is written as
\begin{align} \label{SF}
S_F^{(2)}&=-\frac{M^2}{2} \int \frac{d\omega}{2\pi}\frac{dz d{\bar z}}{2}  \,\,\, \Bigg \{   4 \omega \tau_1^{UV}a^{(0)}_{\bar z}(\omega) a^{(0)}_z(-\omega) +\notag \\ &
2  \frac{\tau_2^{(UV)}}{L}  \left [a^{(0)}_z(\omega) a_{\bar z}^{(1)}\left(-\omega;a_{\bar z}^{(0)},(h_t^{\bar z})^{(0)}\right) +a^{(1)}_z \left(\omega;a_z^{(0)},(h_t^z)^{(0)}\right) a^{(0)}_{\bar z}(-\omega)\right] \Bigg\}
\end{align}
where we show the explicit (linear) dependence of the source $a_z^{(1)}$ on $a_{ z}^{(0)}$ and on the sources $(h_t^{ z})^{(0)}$ but not on $a_{\bar z}^{(0)}$ and on $(h_t^{\bar z})^{(0)}$ and similarly for $a_{\bar z}^{(1)}$. This implies that $\sigma^{zz}$ and $\sigma^{\bar z \bar z}$ vanish. In particular, the conductivities are found by 
\footnote{A factor of M is accompanies $a_i^{(0)}$ because in our conventions we have that $A^{\mu}_{CFT}=M A^{\mu}_{gravity}$, and $A^{\mu}_{gravity}$ is dimensionless. In the following expressions we first differentiate w.r.t $A_{CFT}$, and then translate this into $A_{gravity}$.}
\be \label{fac2}
\sigma^{ij}=\frac{i}{\omega} \frac{\delta^2 S_F} { \delta (M a_i^{(0)}(\omega)) \delta (M a_j^{(0)}(\omega'))} \rightarrow 2 \times \frac{i}{\omega} \frac{\delta^2 S_F} { \delta (M a_i^{(0)}(\omega)) \delta (M a_j^{(0)}(\omega'))} 
\ee
where the expression after on the right of the arrow has an additional factor of two in order to match the standard normalizations conventions in the literature (i.e. see \cite{Kachru2,Hartnoll:2007ai} etc) of the Maxwell term in the action.
Due to the $\delta$ functions that appear from the functional differentiation, one has $\omega'=-\omega$. Hence, the conductivity formula eventually yields
\begin{align}  \label{szzb}
&\sigma^{z{\bar z}}=-\frac{i}{\omega}\left[  \frac{\tau_2^{(UV)}}{L}  \left( \frac{\delta(a_{\bar z}^{(1)}(-\omega))}{\delta(a^{(0)}_{\bar z}(-\omega))} +  \frac{\delta(a_{ z}^{(1)}(\omega))}{\delta(a^{(0)}_{ z}(\omega))}\right)+ 2\omega \tau_1^{UV} \right], \notag\\
&\sigma^{{\bar z}z}=-\frac{i}{\omega}\left[  \frac{\tau_2^{(UV)}}{L}  \left( \frac{\delta(a_{\bar z}^{(1)}(\omega))}{\delta(a^{(0)}_{\bar z}(\omega))} +  \frac{\delta(a_{ z}^{(1)}(-\omega))}{\delta(a^{(0)}_{ z}(-\omega))}\right)- 2\omega \tau_1^{UV}\right], \,\,\,
\sigma^{zz}=\sigma^{\bar z \bar z}=0.
\end{align}
where in the functional differentiations, the metric sources are kept constant and then set equal to zero. 

(iii) Finally, the conductivities in $x,y$ coordinates are given by
\begin{subequations} \label{sxy}
\begin{align}  
&\sigma_{xy}=i \left(\sigma_{\bar z z}-\sigma_{z\bar z }\right)=\frac{ i}{4} \left(\sigma^{z\bar z }-\sigma^{\bar z z}\right)=-\sigma_{yx},\\
&\sigma_{xx}= \sigma_{\bar z z}+\sigma_{z\bar z }=\frac{ 1}{4} \left(\sigma^{z\bar z }+\sigma^{\bar z z}\right)=\sigma_{yy}.
\end{align}
\end{subequations}




\section{Radial perturbations of the IR scaling solutions {\protect\eqref{IRsol}}}
\label{numericsappendix}

As mentioned in Sec.~\ref{numericalstrategy}, for general $\gamma,\delta$ we do not need to perturb \eqref{IRsol} in order to numerically generate our flows.  They are however necessary, for example, to assess possible condensation instabilities of the IR fixed points as the parameters of the system are varied. We calculated the first-order correction to the charged dilatonic scaling solutions (\ref{CDBHsolnA1}, \ref{CDBHsolnf1}, \ref{CDBHsolnphi1}, \ref{CDBHsolnAt1}),
\begin{eqnarray}
A(r) &=& A_{IR} + \frac{(\gamma-\delta)^2}{4} \log \frac{r}{l} + d_1 \left(\frac{r}{l}\right)^{\nu} \\
f(r) &=& - \frac{16 (\tau_2^{IR})^{-\frac{\delta}{\gamma}} \left(\frac{r}{l}\right)^{1- \frac{3}{4} (\gamma-\delta)^2 + \frac{w u}{4}}\Lambda}{w u^2} + d_2 \left(\frac{r}{l}\right)^{\nu - v}\\
\tau_1(r) &=& \tau_1^{IR} + d_4\left(\frac{r}{l}\right)^{\xi}  \\
\tau_2 (r) &=& \tau_2^{IR} \left(\frac{r}{l}\right)^{\gamma (\delta-\gamma)} e^{d_3\left(\frac{r}{l}\right)^\nu}
\eea
where, as before,
$u  =  \gamma^2 - \gamma \delta + 2$,
$v = \delta^2-\gamma\delta +2$,
$w u = 3 \gamma^2 - \delta^2 - 2 \gamma \delta +4$, and we traded the IR mass scale $p$ in the last sections for a IR length scale $l=1/p$, and set the IR end point of the radial coordinate in (\ref{CDBHsolnA1}-\ref{CDBHsolnAt1}) to $r_0=0$, for simplicity. It can easily be reinstated by a shift \eqref{restr3}. The scaling coefficients $\nu$ and $\xi$ are given by
\bea
\nu &=& \frac{1}{8}\left( -w + \sqrt{w (36+(\gamma-\delta)(17 \delta +\gamma(19+8(\gamma-\delta)\delta)))} \right)\\
\xi &=& \frac{\delta^2 + 10 \gamma\delta - 11\gamma^2 - 4}{4}
\eea
Note that the constraints 1-4 in Sec.~\ref{constraints} immediately imply the reality of $\nu$. 
The equations of motion relate the amplitudes of the $A$, $f$, and $\tau_2$ perturbations, such that ($\Lambda = -\zeta(2s)$)
\bea
\frac{d_3}{d_1} &=&  \frac{2 \gamma (\nu-1)}{\gamma-\delta}\\
\frac{d_1}{d_2} &=& \frac{ (\tau_2^{IR})^{\frac{\delta}{\gamma}} (\delta-\gamma)(\delta^2 -\gamma\delta - \nu -2)(\delta^2 + 2 \gamma\delta - 3 \gamma^2 - 4)}{32\Lambda(\delta^2 - \gamma\delta - 2)}\\
&& \times \frac{(\delta^2 + 2 \gamma\delta - 3 \gamma^2 - 4 - 4 \nu )(\gamma(\delta-\gamma)-2)}{2(2+5\nu)\gamma +3 ( 1-\nu)\gamma^3 -2 (4 + 3\nu + (4+\nu)\gamma^2)\delta - (\nu-3)\gamma \delta^2 + 2 \delta^3 } \, . \nonumber
\eea
There are therefore only two independent perturbation strengths, which we can take to be $d_1$ and $d_4$. The power $\xi$ is always real, but $\nu$ can acquire complex values, signalling an instability. For our values $(\gamma,\delta=-\gamma s)$, \eqref{gammas}, it is real.


\section{The missing steps in the proof of equation ({\protect \ref{kmk}})} \label{pf}

In this appendix we prove equation (\ref{kmk}). The proof follows as a corollary of the following theorem of second order differential equations.

\vspace{0.1in}
\underline{Theorem:} Let $x_0$ be a regular singular point of the second-order linear differential equation $y''(x)+p(x)y'(x)+q(x)y(x)=0$. Let $d_1$ and $d_2$ (Re($d_1) \geq $Re($d_2$)) be the roots of the indicial equation associated with $x_0$. If $d_1 - d_2 \in N$ (positive integer), then the differential equation has two non-trivial linearly independent solutions of the form
\begin{subequations}
\begin{align}
y_1(x)&=|x-x_0|^{d_1} \sum_{n=0}^{\infty} a_n(x-x_0)^n,\,\,\, {a_0\neq 0} \\
y_2(x)&=|x-x_0|^{d_2} \sum_{n=0}^{\infty} c_n(x-x_0)^n+cy_1(x)\ln|x-x_0|,\,\,\, {c_0\neq 0}\label{2ndsol}
\end{align}
\end{subequations}
where the constant $c$ may or may not be equal to zero.

The proof of the theorem along with the Frobenius method in solving these kind of equations can be found in any standard textbook for differential equations. We hence omit the proof and we instead prove the following corollary and hence the condition (\ref{kmk}).

\vspace{0.1in}
\underline{Cor.:} Let $x_0$ be a regular singular point for the differential equation of the previous theorem and let $d_1=2$ and $d_1=0$ be the roots of the corresponding  indicial equation. If in the neighborhood of the singular point $p(x)=-1/(x-x_0)+m_1+O(x-x_0)$ and $q(x)=k_0/(x-x_0)+k_1+O(x-x_0)$ then
\be \label{apsin}
k_0^2+m_1k_0+k_1=0\,\,\, \mbox{if and only if}\,\,\, c=0
\ee
where $c$ is the coefficient of the logarithmic singularity of the second linearly independent solution (see \ref{2ndsol}). Equivalently, a regular singular point of a differential equation in this class of differential equations is completely regular if and only if equation (\ref{apsin}) applies.

\vspace{0.1in}
{\it Proof:}
\vspace{0.1in}

Without loss of generality we assume $x_0=0$. Then from the theorem we have that $y_1(x)=x^2(a_0+a_1x+a_2x^2+....)$ and $y_2(x)=(c_0+c_1x+c_2x^2+....)+c\log(x)y_1(x)$. Substituting $y_2$ in the differential equation and taking into account that the terms proportional to the logarithms cancel because $y_1$ is a solution, we obtain $y''+py'+qy=(c_0k_0-c_1)/x+(2ca_0+c_0k_1+c_1k_0+c_1m_1)+O(x)=0$. This implies that $c_1=c_0 k_0$. Focusing on the second term and substituting $c_0$ yields $2 a_0 c+c_0(k_0^2+m_1k_0+k_1)=0$ where $a_0\neq0$ and $c_0\neq0$ by the theorem. Hence $c=0<=>k_0^2+m_1k_0+k_1$=0 $\Box$.


\section{The missing steps in the proof of equation ({\protect \ref{azzb1}})}  \label{az1}

The first step is to consider the constraint equation (\ref{ahry}) and along with the $(ry)$-component part multiplied by i and take their sum. This yields
\be \label{eo}
a_z'= \frac{1}{ hD\tau_2}\left( \sqrt{DB}(\omega a_z-h h_t^z) \left( q -h \tau_1  \right) +\omega C^2 h'{_t}^{z}  \right)
\ee 
where $h_t^z$ at any order in $\omega$ in the UV at Poincare coordinates expands as
\be \label{e2}
h_t^z=(h_t^z)^{(0)}+\frac{z^3}{L^3} (h_t^z)^{(1)}+... ,\, \mbox{UV as $z\rightarrow 0$.}
\ee
Expanding the right hand side of (\ref{eo}) to order $\omega$ and to lowest order in $z$ using (\ref{e2}) and (\ref{aoa1z}) yields
\begin{align} \label{e33}
\frac{1}{L}a_z^{(1)}(\omega)&=\frac{1}{L}\Big(a_z^{(0,1)}+\omega a_z^{(1,1)}\Big) +O(\omega^2) =-\frac{ q- h \tau_1^{UV}    }{ \tau_2^{UV}} (h_t^z)^{(0,0)}  \notag\\ 
 &+ \omega \left(\frac{ q- h \tau_1^{UV}    }{h \tau_2^{UV}}\left(a_z^{(0,0)}-h (h_t^z)^{(1,0)} \right) + \frac{3}{hL \tau_2^{UV}} (h_t^z)^{(0,1)}   \right)     +O(\omega^2) 
\end{align}
where the upper left superscript indicates the order in $\omega$ while the right one shows either the source (0) or the VEV (1). Last equation implies
\be \label{e4}
a_z^{(1)}=L\left(-\frac{ q- h \tau_1^{UV}    }{ \tau_2^{UV}} (h_t^z)^{(0)}+\omega \frac{ q- h \tau_1^{UV}    }{ h\tau_2^{UV}} a_z^{(0)}      \right)+\ \frac{3\omega}{h \tau_2^{UV}} (h_t^z)^{(0,1)}
+O(\omega^2) 
\ee
and it looks almost as (\ref{azzb1}). The proof will be completed if we show that the last term in (\ref{e4}) does not depend on $a_z^{(0)}$ which is equivalent in showing that the VEV of $h_t^z$ to zeroth order in $\omega$ does not depend on the source $a_z^{(0)}$ of the gauge field. In order to show it we begin by expanding the right hand side of (\ref{eo}) to order $\omega^0$ yielding
\be \label{e5}
\omega^0 a_z'(z)= -\omega^0 \frac{\sqrt{DB} \left( q -h \tau_1  \right)  }{ D\tau_2} h_t^z(z)
\ee
where the superscripts that indicate the order in $\omega$ have been suppressed.

It can be seen by direct substitution that (\ref{e2}) solves (the $z$-version of) equation (\ref{axh}) to leading order in $\omega$ automatically. Hence, in order to specify the solution $h_t^z$ to leading order in $\omega$, one must employ the $(tz)=(tx)+i(ty)$-component of the Einstein's equation. As it can be checked, the $(tz)$-component to leading order in $\omega$ contains $h_t^z,\, h'{_t}^{z}, \, h''{_t}^{z}$ and $a_z'$ terms. This second order differential equation can be decoupled completely in terms of $h_t^z$ using (\ref{e5}). The solution, as usual, can be fixed by IR regularity leaving the VEV $(h_t^z)^{(0,1)}$ completely determined in terms of the source $(h_t^z)^{(0,0)}$ but not in terms of the source $a_z^{(0,0)}$. The precise solution of $h_t^z$ to leading order in $\omega$ is not important. What is emphasized however is that this solution is independent on $a_z$ and in particular independent  on $a_z^{(0)}$. In more detail, to leading order in $\omega$, the VEV term of $h_t^z$ is proportional to $(h_t^z)^{(0)}$ alone but it has no contribution from $a_z^{(0)}$. Hence, equation (\ref{e4}) has the form of the top equation (\ref{azzb1}). 

In a similar way, one can show that the $a_{\bar z}$ has the form of the lower equation (\ref{azzb1}) and this completes the proof.


\section{The IR solutions of the vector fluctuations}\label{IRfluctsol}

\subsection{IR solutions in the electric frame}\label{IRsolelectric}

In specifying the ratios $a_{ z}^{(1)}/a_{ z}^{(0)}$ required in (\ref{selw}) one needs to solve (numerically) the fluctuation equations (\ref{azzb}). The necessary boundary conditions are taken from the IR by demanding regularity of the solution. Fortunately, as we will see, the IR solutions can be found analytically.

Writing the leading behavior of (\ref{azzb}) in the IR taking into account that the frequency terms are subdominant \footnote {The $\omega^2$ term in the fluctuation equation is suppressed compared to the $A_t^2 \sim q^2$ term by virtue of the constraint \eqref{TDunstable}.
In the deep IR the $\omega \, \tau_1'$ term is also negligible compared to the $\sim q^2$ term, because in the $\tau_1$ begins to flow due to terms in the large $\tau_2$ expansion of the potential (\ref{Eisensteinlargetau2}) which are exponential suppressed by virtue of the Bessel functions at large arguments, implies that $\tau_1'$ is very small. In summary, the $q^2$ contribution dominates.} 


we have
\be \label{IRazDE}
a_z^{el;IR}\,''+\frac{8-3(\gamma-\delta)^2}{4r}a_z^{el;IR}\,'-\frac{(2 + (\gamma- \delta) \delta)  (4 + (\gamma - \delta) (3 \
\gamma + \delta)) }{4r^2} a_z^{IR}=0
\ee
where the superscript $el;IR$ emphasizes the IR behavior in the absence of a magnetic field (electric frame). In arriving in (\ref{IRazDE}) we used the background equations (\ref{CDBHsolnA})-(\ref{qeq}). The solutions to (\ref{IRazDE}) are given by
\be 
a_z^{el;IR}= C^1_z r^{\frac{w}{4}}+ C^2_z r^{-v} = a_z^{el;IR}=C^1_z r^{1.715}+ C^2_z r^{-0.926}\,,
\ee
where $C^1_z$ and $C^2_z$ are arbitrary constants, $w$ and $v$ are defined in \eqref{wvdef}, and we used our choice of parameters \eqref{gammas} in the second equality. Analogous equations apply for the ${\bar z}$ components. 
The positive power provides the regular solution and hence the boundary condition for the subsequent numerical evaluation.

For completeness, we compute the IR behavior of the metric fluctuation by employing equation (\ref{Einsteinpert}). Substituting the IR behavior of all the fields we have
\be \label{hIR}
(h_t^z)^{IR} =-\frac{q}{v}e^{3 A_{IR}}C^1_z r^{v}+ \frac{4 q}{w}e^{3 A_{IR}}C^2_z r^{-\frac{w}{4}}
= -\frac{q}{v}e^{3 A_{IR}}C^1_z r^{0.926}+ \frac{4 q}{w}e^{3 A_{IR}}C^2_z r^{-1.715}.
\ee
where $A_{IR}$ is the integration constants in the IR solution of the metric world-factor (see (\ref{CDBHsolnA})), and we used our choice of parameters (\ref{gammas}) in the second equality.


\subsection{The IR solution of the decoupling variable $E_z$}\label{Ndi}

Due to numerical constraints, the starting point we begin the numerics in the IR may not be exactly the point $r=r_0$ and hence we are forced to consider a nearby point $r=r_{in}$. We choose $r_{in}$ to be as close to $r_0$ as possible such that the numerical process does not become dysfunctional\footnote{Here $r_{in}$ should be identified with $n_I$ in step 2, below equation (\ref{n}).}. However, such a shift in the initial point of beginning the numerics requires considering sub-leading corrections in $1/(r-r_0)$. In particular, instead of 
\be \label{IRgs}
E_z''+\frac{3(\ga-\delta)^2}{4(r-r_0)} E_z'-\frac{wv} {4(r-r_0)^2}E_z=0,
\ee
which describes $E_z$ at the IR (see (\ref{Ezz}) with $g_0$ and $g_1$ given by (\ref{g01IR})), we need to consider the following equation
\bea\label{EzzIRG}
E_z''+\frac{3(\ga-\delta)^2}{4(r-r_0)} E_z'+ \left( \frac{e^{2(A_{IR}+\d \phi_{IR}} w^2 u^2 \omega^2}{256 \Lambda^2 (p(r-r_0))^{\frac{8+(\g-\d)(\g+3\d)}{2}}} -\frac{wv} {4(r-r_0)^2}\right)E_z=0.
\eea
The quantities $p$, $A_{IR}$ and $\phi_0$ are IR parameters that are discussed in appendix \ref{RGdetails}. Deep inside the IR, the regular solution to equation (\ref{EzzIRG}) is given by the regular solution of (\ref{EzIR}). Evidently, equation (\ref{EzzIRG}) corrects equation (\ref{IRgs}) when we move away from the end-IR point $r_0$. It is notable that equation (\ref{EzzIRG}) does not depend on any $SL(2,\mathbb{Z})$ parameter . Hence this equation can be used as a boundary condition on the fluctuation, for any dyonic background obtained via (any) $SL(2,\mathbb{Z})$ transformation from the electric one.

According to appendix \ref{RGdetails}, for input parameters $q$ and $a_0$ in the electric-frame background, all the rest IR parameters are determined. In particular, for $q=1$ and $a_0=0$ (non-running axion; see appendix \ref{RGdetails}) equation (\ref{EzzIRG}) becomes
\bea\label{EzzIR}
E_z''+\frac{2.623}{r-r_0} E_z'+ \left( \frac{0.393 \omega^2}{(r-r_0)^{1.937}} -\frac{0.159} {(r-r_0)^2}\right)E_z=0, \,\, r_0\approx -0.457.
\eea
The closest possible starting point for the numerics in constructing the underlying background we were able to choose in the shooting method we applied is approximately the point $r=r_{in}=-0.457$. Yet, according to (\ref{EzzIR}), the sub-leading contribution for $r\geq r_{in}$ is equally important to the leading one and this is the reason that such a correction must be included in the boundary conditions. The sub-leading corrections depend on $\omega$ and the analytic solution in the neighborhood of $r_{in}$ including these corrections for any other charge\footnote{Than $q=1$ that we considered in arriving to equation (\ref{EzzIR}).} is
\be \label{JIR}
E_z=\frac{1}{(r-r_0)^{0.812}} J\left (\pm 27.255, \frac{\omega}{\sqrt{k}}(r-r_0)^{0.033} \right), \,\, k=\frac{q}{q_0}=q.
\ee
In the $r\rightarrow r_0$ limit, the solution with the positive sign in the Bessel function, which corresponds to the regular solution that vanishes in the IR, maps onto the regular solution (\ref{EzIR}) as should. Equation (\ref{JIR}) provides the boundary conditions for the fluctuation $E_z$, for any initial charge (density) $q$ but for fixed $a_0=0$\footnote{A different choice in $a_0$ would change $A_{IR}$ and $\phi_0$ (see appendix \ref{RGdetails}) in (\ref{EzzIRG}) and hence the coefficient 0.393 in (\ref{EzzIR})).} in the electric frame and any $SL(2,\mathbb{Z})$ transformation to the dyonic frame.

\begin{figure}[h!]
	\begin{center}
	\includegraphics[scale=1]{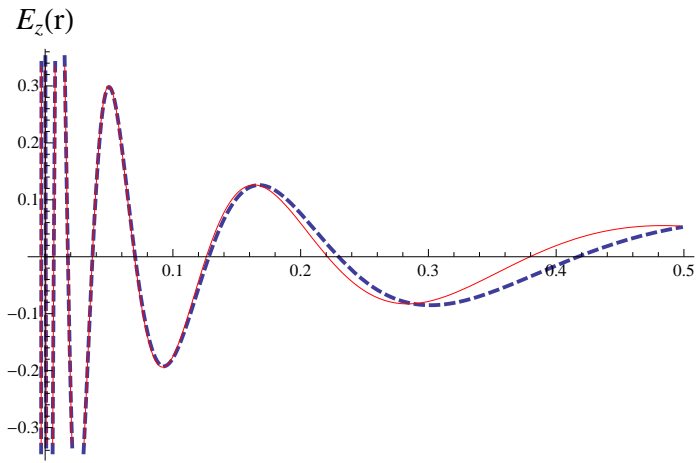}
	\end{center}
	\caption{The numerical curve (blue) superimposed with equation {\protect\eqref{JIR}} in the IR region. Here, $a=b=c=1$ and $\omega=10$. This plot provides a more intricate example for the Bessel-like behavior of the fluctuation in the IR and it shows how the number of nodes and hence the excitation increases as $\omega$ increases. Towards the UV, the two curves diverge from each other. }
	\label{w10_IR}
\end{figure}

Now, things become clearer: due to the sine-like behavior of the Bessel function, as $ \omega $ increases, there are more oscillations in the IR and hence more nodes. The oscillations occur in the region where $V_{schr}$ attains the negative values as discussed previously in section \ref{varyw}. As the number of oscillations increases, the label $n$ on $\omega_n$ increases. The number of nodes specifies the eigenvalue; i.e. $n$ nodes correspond to $\omega_n$. For instance, the left (right) panel of Fig.~\ref{Vs1} has zero (one) nodes and hence it corresponds to $\omega_0\approx 1.7$ ($\omega_1 \approx 2$).  Fig.~\ref{w10_IR} corresponds to a larger value of $\omega$, $\omega=10$ ($\omega=10$ does not correspond to an eigenvalue). Evidently, according to Fig.~\ref{w10_IR}, as $\omega$ increases, the number of nodes increases. From the right panel of Fig.~\ref{Vs1} and from Fig.~\ref{w10_IR} we see that the IR solution \eqref{JIR} seems to give a good approximation to the numerically obtained wave-function, for a relatively large $r$-interval.



\section{Reduced allowed region for general $\gamma$ and $\delta$ and corresponding leading behavior of $a_z^{IR}$ \label{puts} }

The allowed region in the $\gamma-\delta$ plane for general backgrounds that belong in the class that we use here, is defined through the Gubser bound, equation (\ref{Gubser}), through the thermodynamical instability, equation  (\ref{TDunstable}) and through (\ref{discreteIR}). Furthermore, the set-up of this problem, requires that $\tau_2^{el} \rightarrow \infty$ in the IR where  $\tau_2^{el}$ the scaling solution (\ref{putsaf}) \footnote{A quick way to see this is to begin from equation (\ref{szz}) that states that $\sigma_{xy}c\sim q/h+O(\omega)$. On the other hand, the ratio $q/h$ is given by the IR behavior of $\tau_1$, equation (\ref{qoh}), in the presence of $h$ field. Requiring that $q/h$ is fractional and taking into account (\ref{e1}) implies that in the IR, $\tau_2^{el} \rightarrow \infty$.}. Therefore, $\tau_2^{IR}=e^{\gamma \phi} \rightarrow \infty$ and (\ref{putsaf}) imply the additional constrain $\gamma(\delta-\gamma)<0$ or equivalently
\be \label{gdsign}
\tau_2^{IR} \equiv e^{\gamma \phi^{IR}} \rightarrow \infty=> \,\,\,\, \gamma<0, \,\,\,\, \delta>\gamma  \,\,\,\,  \mbox{or}  \,\,\,\, \gamma>0, \,\,\,\, \delta<\gamma .
\ee
As we show graphically in the left panel of Fig.~\ref{reLR}, both possibilities of (\ref{gdsign}) have overlap with the allowed region defined by (\ref{Gubser}) and (\ref{TDunstable})\footnote{One of the Gubser constrains, $u>0$, and the constrain (\ref{discreteIR}), do not restrict the allowed region further and hence we avoid including them in Fig.~\ref{reLR} keeping the plots as simple as possible.}.

Using (\ref{axp}), we extract the leading behavior of $a_z$ \footnote{At this level of investigation, $a_x \sim a_y \sim a_z$ and $E_x \sim E_y \sim E_z$ and hence any $x$, $y$ indices are treated as equivalently.} and we show that the leading IR behavior of $E_z$ coincides with the leading IR behavior of $h_t^i$, i.e. that the inequality 
\be\label{avsE}
E_z^{IR} \gg a_z^{IR}. 
\ee
applies. The starting point is to note that the $h^2$ term in the denominator of \eqref{axp} dominates over the $\omega^2$ term because of the constraint (\ref{TDunstable}). This implies that
\be \label{AF1}
(C^2)^{IR} \ll (D \tau_2)^{IR} \sim \frac{D^{IR}}{\tau_2^{el;IR}}.
\ee
The numerator has also two terms. Taking their ratio, transforming the scalars in the $h$-frame where appropriate and using the IR behavior of the backgrounds fields yields
\be \label{AF2}
\left( \frac{\omega C^2 E_z'}{-ihE_z\sqrt{BD} (q-h \tau_1)}\right)^{IR} \sim r^{-1-\frac{1}{4} (\gamma-\delta)(5\gamma+3\delta)}
\ee
where in addition, the middle equation in (\ref{t12IRp}) has been used. There are two possibilities here, $-1-\frac{1}{4} (\gamma-\delta)(5\gamma+3\delta)<0$ and $-1-\frac{1}{4} (\gamma-\delta)(5\gamma+3\delta)>0$. 
\begin{figure}[htb]
	\includegraphics[scale=0.87]{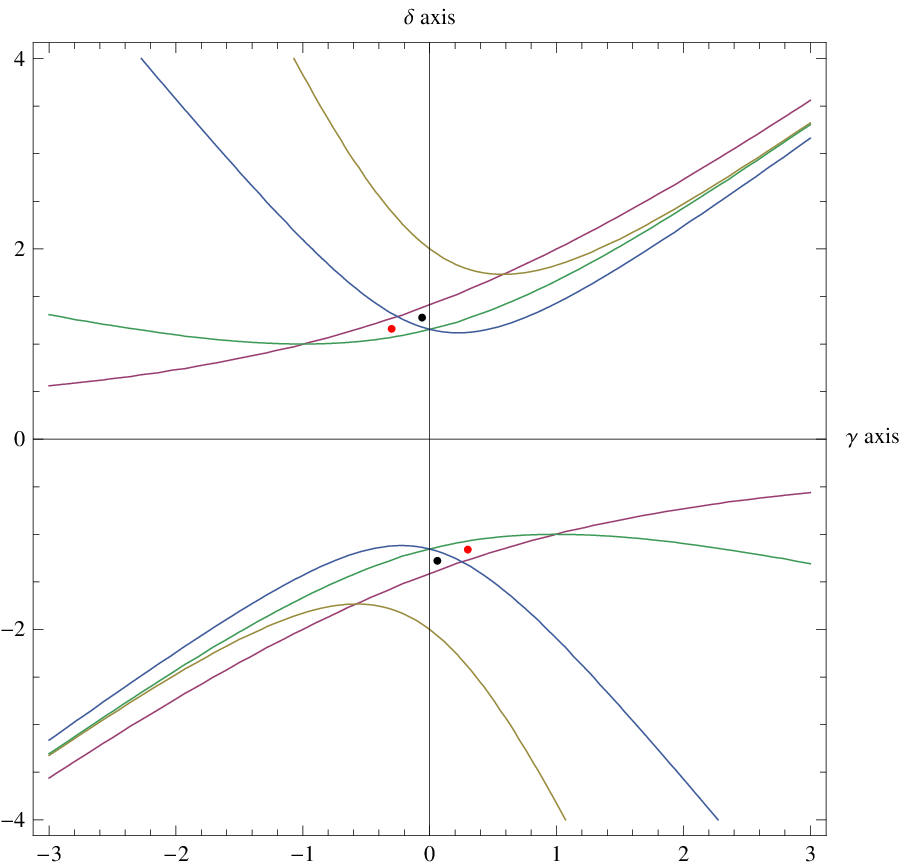} 
	\includegraphics[scale=0.78]{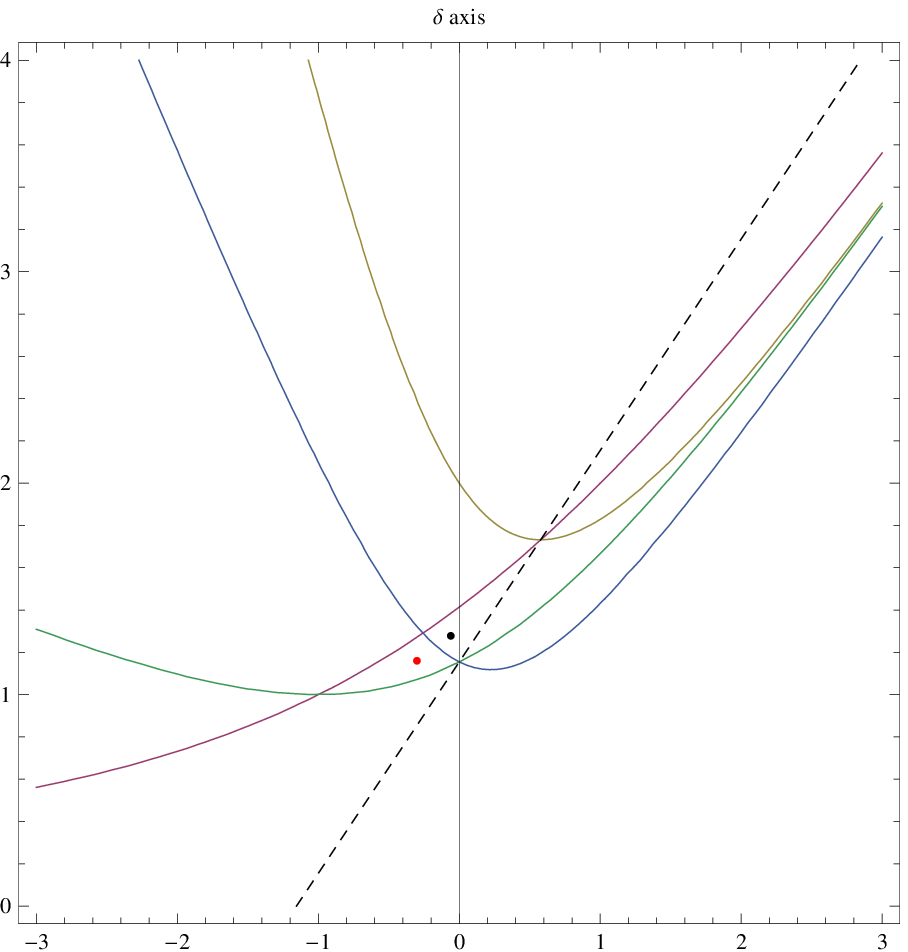}\\
	\caption{Horizontal (vertical) axis: $\gamma$ ($\delta$). {\bf Left panel}: shows the allowed region. It exists either in the upper left box or the lower right one (see \protect\eqref{gdsign}) and is located inside the closed curve(s) defined by the purple, green and the vertical ($\delta$) axis. In particular, the orange curve defined by the equation $v\equiv -2 - \gamma \delta + \delta^2$ does not participate in the definition of the allowed region. The blue curve is given by equation $-1-\frac{1}{4} (\gamma-\delta)(5\gamma+3\delta)=0$. This expression is negative (positive) outside (inside) the curves and this defines case 1 (case 2). Evidently, both cases have overlap with the allowed region (see red (black) points for case 1 (case 2)).
	{\bf Right panel}: It zooms in the $\delta>0$ region for a cleaner illustration (similar results apply for the $\delta<0$ region). The panel contains the same information as the left panel with one additional curve (dashed straight line). The equation of this line is given by the negative sign of $\gamma-\delta=\pm 2/\sqrt{3}$ (the line with the plus sign is not displayed). The points above this line satisfy $w/4-v=\frac{3}{4} (\gamma-\delta)^2-1>0$ and hence, according to  \protect\eqref{AF5}, in this region we have $a_z^{IR}  \ll E_z^{IR} $. Moreover, above the dashed line we have $w/4>v$.}
	\label{reLR}
\end{figure}

\vspace{0.1in}
Case 1: $-1-\frac{1}{4} (\gamma-\delta)(5\gamma+3\delta)<0$. An example for parameters that belong in this region is provided by the red points in the left panel of Fig.~\ref{reLR}. In such a case, in equation (\ref{axp}), the $\sim \omega E_z'$ dominates over the $\sim hE_z$ one. Hence, from (\ref{axp}), (\ref{AF1}) and (\ref{AF2}) it is deduced that
\be \label{AF3}
a_z^{IR} \,' \sim  \left(\frac{C^2}{D \tau_2}\right)^{IR} E_z^{IR} \,'=> a_z^{IR} \sim  r^{-2-\frac{1}{2} (\gamma-\delta)(\gamma+3\delta)}E_z^{IR}  \sim  r^{2 (\gamma-\delta)^2-4 w}E_z^{IR} 
\ee
where $E_z^{IR}\sim r^v$ is given by (\ref{EzIR}). Therefore, using (\ref{wvdef}) yields
\be \label{AF3b}
 a_z^{IR} \sim  r^{(2 (\gamma-\delta)^2-4 w)+v} \sim r^{\frac{1}{2} (\delta^2-\gamma^2)}.   
\ee

Taking into account the constraint (\ref{TDunstable}), ones sees that the power of $r$ multiplying $E_z^{IR}$ in (\ref{AF3}) is positive. Hence, we have shown that in the region defined by the inequality of case 1, the inequality (\ref{avsE}) does apply. Furthermore, since $E_z^{IR}$ is regular, $ a_z^{IR}$ from (\ref{AF3b}) should also be regular implying that $\delta^2-\gamma^2$ must be positive. After all, according to (\ref{apregIR}) and the text that follows, regular solutions are mapped onto regular solutions. Positivity of $\delta^2-\gamma^2>0$ can be shown using that $2(\delta-\gamma)^2-w=-4 - \gamma^2 - 2 \gamma \delta + 3 \delta^2>0$ and $v=2 + \gamma \delta - \delta^2>0$ (see ({\ref{wvdef}}), (\ref{Gubser}) and (\ref{TDunstable})). Then, the quantity $2(\delta-\gamma)^2-w+2v=\delta^2-\gamma^2>0$.

\vspace{0.1in}
Case 2: $-1-\frac{1}{4} (\gamma-\delta)(5\gamma+3\delta)>0$. An example for parameters that belong in this region is provided by the black points in the left panel of Fig.~\ref{reLR}. In such a case, the $h E_z$ term in (\ref{axp}) dominates over the $\omega E_z'$ one. Hence, from (\ref{axp}), (\ref{AF1}) and (\ref{AF2}) it is deduced that
\be \label{AF5}
a_z^{IR} \,' \sim  \left(\frac{\sqrt{BD} (q-h \tau_1)}{D \tau_2}\right)^{IR} E_z^{IR}  \sim \sqrt{ \frac{B^{IR}}{D^{IR}}} \frac{E_z^{IR} }{\tau_2^{el;IR}} 
    => a_z^{IR} \sim  r^{\frac{3}{4} (\gamma-\delta)^2-1}E_z^{IR} 
 \ee
where $E_z^{IR}\sim r^v$ is given by (\ref{EzIR}). Therefore, using (\ref{wvdef}) yields
\be \label{AF5b}
 a_z^{IR} \sim  r^{\frac{w}{4}}.
\ee
Equation (\ref{AF5b}) is consistent with (\ref{apregIR}), which shows that regular solutions are mapped onto regular solutions. 

There is not a-priori a constraint on the sign of $\frac{3}{4} (\gamma-\delta)^2-1$, coming from some physical requirement, as in case 1. This sign can generally be positive in which case $a_z^{IR} \ll E_z^{IR}$ or it an be negative in which case $a_z^{IR} \gg E_z^{IR}$. $a_z^{IR} \gg E_z^{IR}$ would lead a contradiction as $a_z^{IR}$, can not always be more important than $E_z^{IR} \equiv \omega a_z^{IR} + i h (h_t^z)^{IR} $ for arbitrary $\omega$. Fortunately, the sign of $\frac{3}{4} (\gamma-\delta)^2-1$ when the $\gamma$ and $\delta$ are restricted
in the reduced allowed region due to the additional constrain (\ref{gdsign}), 
is always positive and hence $a_z^{IR} \ll E_z^{IR}$. This argument is completed in the caption of the right panel of Fig.~\ref{reLR}.

Therefore, we have shown that in either case, inequality (\ref{avsE}) applies.

\vspace{0.1in}
Remark: Although we have shown that $a_z^{IR} \ll E_z^{IR}$ applies in the allowed region in the parameter plane, there is not a definite leading IR behavior of $a_z^{IR}$.
If the pair of $(\gamma,\delta)$ belongs in region defined in case 1 above, then $a_z^{IR}$ behaves according to (\ref{AF3}). On the other hand, If the pair of $(\gamma,\delta)$ belongs in region defined in case 2, then $a_z^{IR}$ behaves according to (\ref{AF5}). Therefore, it seems that any possible investigation that treats the pair of $(\gamma,\delta)$
generically, must be done with care and it must be partitioned in these two cases.



\section{Alternate IR solutions\label{app:Ansatz}}

In this appendix we investigate the possible infrared asymptotics of our system.  There are two kinds of possible IR ground states, solutions with fixed scalars and solutions in which the scalars run. We first classify the former, and find that generically $AdS_2$ ground states exist for all values of the magnetic field and charge density. We then classify all solutions with running scalars and find two classes, namely solutions which flow to the dyonic dilatonic black holes which we identified as our fractional quantum Hall states in Sec.~\ref{dyonicsolutions}, as well as solutions in which both the axion and the dilaton diverge in the IR. We conclude by describing how the $AdS_2$ fixed points may play a crucial role in the transitions between the fractional quantum Hall states, which we plan to investigate in future work.

\subsection{Solutions with fixed scalars}

We are looking for solutions with $\phi=\phi_*$, $\tau_1=\tau_{1*}$.
From (\ref{generaldilaton}) and (\ref{generalaxion}) we obtain that $C$ is constant and we scale it to one from now on,\footnote{By replacing $q\rightarrow q/C$, $h\rightarrow h/C$.} and
\be
{\pa V\over \pa \tau_1}\Big|_*+{h(q-h\tau_{1*})\over Z_*}=0\sp {\pa V\over \pa \phi}\Big|_*+{Z'_*\over 2Z_*}\left({(q-h\tau_{1*})^2\over Z_*}-{Z_*~h^2}\right)=0
\label{e13}\ee
The two equations above can be written as $SL(2,{\mathbb Z})$ covariant complex equations
\footnote{Note that from the fact that $E_s$ is an eigenfunction of the Laplacian on the upper half plane, $\tau_2^2\pa_{\tau}\pa_{\bar\tau}E_s={s(s-1)\over 4}E_s$, we can constrain the difference in operator dimension in the axion and dilaton directions in field space.}
\be
\tau_2\partial_{\tau}V\Big|_{*}-i{(q-h\bar\tau_*)^2\over 2\tau_2^*}=0
\label{e18}\ee
The rest of the equations for $B,D$ become
\be
2V_*={(q-h\tau_{1*})^2\over Z_*}+h^2Z_*={|q-h\tau_*|^2\over \tau_2^*}
\label{e14}\ee
\be
{1\over 2B}\left(2{D''\over D}-{D'\over D}\left({D'\over D}+{B'\over B}\right)\right)={(q-h\tau_{1*})^2\over Z_*}+{h^2Z_*}
\label{e15}\ee
They have the $AdS_2\times R^2$ solution with
\be
B=D={\ell_*^2\over r^2}\sp {1\over \ell_*^2}=V_{*}
\label{e16}\ee
We define the complex number
\be
Q=q-h\tau
\label{e19}\ee
that controls the charge density and magnetic field.
Then if $(\tau_*,Q)$ is a fixed point solution, so are all its $SL(2,{\mathbb Z})$ copies.
What this implies is that Q is a torus (abelian) variety, due to $SL(2,{\mathbb Z})$ covariance.

The condition \eqref{e18} fixes either the position $\tau_\ast$ of the $AdS_2$ fixed point in terms of the external charges $(q,h)$, or
vice versa. It turns out that it is easy to solve \eqref{e18} for $(q,h)$, yielding one pair of solutions related by $(q,h)\mapsto - (q,h)$,
\bea\label{qAdS2}
q(\tau_\ast) &=& \pm \left. \frac{\sqrt{\sqrt{(\partial_1 V)^2+(\partial_2 V)^2}+\partial_2 V} \left(\tau_2 \left(\partial_2 V-\sqrt{(\partial_1 V)^2+(\partial_2 V)^2}\right)+\tau_1 \partial_1 V \right)}{\partial_1 V}\right|_{\tau=\tau_\ast}\,,\\\label{hAdS2}
h(\tau_\ast) &=& \pm \left.\sqrt{\sqrt{(\partial_1 V)^2+(\partial_3 V)^2}+\partial_2 V}\right|_{\tau=\tau_\ast}\,.
\eea

Due to $SL(2,\mathbb{Z})$ invariance it suffices to scan with $\tau_\ast$ through the fundamental domain in order to find all $AdS_2$ fixed points and their corresponding values of $q$ and $h$. All images of the fundamental domain can then be obtained by acting with $SL(2,\mathbb{Z})$ on $\tau_\ast$ as usual, and on $(q,h)$
as in \eqref{Fxytransformation}-\eqref{Frttransformation}. From UV counting we expect the physics to depend on $q/h$ alone, hence we parametrize the fixed points by $q/h$ and $h$ itself. It is clear from \eqref{qAdS2} and \eqref{hAdS2} that for every point $\tau_\ast$ in the fundamental domain there exist exactly two $AdS_2$ fixed points with the same $q/h$, and with $h$ differing in the overall sign. However, it is neither clear from \eqref{qAdS2}-\eqref{hAdS2} how many solutions $\tau_\ast$ exist for each pair $(q/h,h)$, nor whether their position $\tau_\ast(q/h,q)=\tau_\ast(q/h)$ only depends on the filling fraction $q/h$. In fact, as Fig.~\ref{tau1qohh} shows, this is not the case if we scan with $\tau_\ast$ through the fundamental domain. In particular, $\tau_1(q/h,h)$ is independent of $h$ only for sufficiently large $|h| \geq 1$. If furthermore $|q/h|\leq \frac{1}{2}$, the axion seems to fulfill $q = \tau_{1\ast} h$.\footnote{We note that scanning over the T-duality related copies of the fundamental domain, the region $|h|\leq 1$, and in particular the region in which $\tau_\ast$ seems to vary wildly, is filled in more and more to form a slope with $\tau_{1\ast} = q/h$, and otherwise nearly $h$-independent. One can check that the fluctuations in that region stem from numerical uncertainties during the evaluation of the derivatives of the potential, which go to zero near the lower end of the fundamental domain quickly. Hence the values of $(q,h)$ in that region vary very quickly, and the sampled data points lie sparcely. Sampling more T-dual copies of the fundamental domain helps to fill in more data points in this region.} For $|q/h|\geq \frac{1}{2}$ and $|h|>1$  the distribution of $\tau_{1\ast}$ levels off, but still depends on $q/h$ in a weaker way, roughly $\tau_{1\ast} \sim - \frac{1}{3} \frac{q}{h}$.\footnote{This cannot be seen from Fig.~\ref{tau1qohh}, but from looking at the 3D plot from the side in Mathematica.}
On the other hand, $\tau_2(q/h,h)$ does depend more strongly on $h$ for larger $|h|$, and is nearly $|h|$-independent for smaller values (cf.~Fig.~\ref{tau2qohh}).

In order to see the role these additional gapless fixed points could play in the FQH physics of the model it is important to know the scaling exponents around these fixed points, which correspond to infrared dimensions of the operators dual to these fluctuations. Radially perturbing $\delta\phi$, $\delta \tau_1$, $\delta g_{tt}$ and $\delta g_{xx}$ with respective powers $\delta_\phi$, $\delta_a$, $\delta_t$, $\delta_x$, we find the following modes:\footnote{Due to the lengthy nature of the expressions, we refrain from giving the eigenvectors in this paper. They can be easily obtained by Mathematica.}

        \begin{tabular}{|c|c|}\hline
Power & Perturbed Fields\\\hline
One mode with $\delta_t = 2$ & $\delta g_{tt}$\\\hline
One mode with $\delta_t = 0$ & $\delta g_{tt}$\\\hline
One mode with $\delta_x = -1$ & $\delta \phi$, $\delta \tau_1$, $\delta g_{tt}$, $\delta g_{xx}$\\\hline
Four modes with $\delta_\phi = \delta_a = \delta_i$ given by {\protect\eqref{deteq},\eqref{soleq}} & $\delta\phi$, $\delta \tau_1$, $\delta g_{tt}$.\\\hline
        \end{tabular}

The matter modes $\delta_i$, $i=1,2,3,4$ are the most interesting ones, with exponents given by the fourth order polynomial equation
\begin{eqnarray}\nonumber
&& \delta_i^4 - 2 \delta_i^3 + \left( \frac{\gamma ^2 e^{2 \gamma  \phi_\ast } V^{(2,0)}\left(\tau _{1,\ast},\phi_\ast \right)+V^{(0,2)}\left(\tau _{1,\ast},\phi_\ast \right)}{V\left(\tau _{1,\ast},\phi_\ast \right)}+\gamma ^2 \left(\frac{1}{-\frac{e^{-2 \gamma  \phi_\ast } \left(Q-H \tau _{1,\ast}\right){}^2}{2 H^2}-\frac{1}{2}}-1\right)+1 \right)\delta_i^2\\\nonumber
&& + \left(\gamma ^2 \left(\frac{2}{\frac{e^{-2 \gamma_\ast \phi } \left(Q-H \tau _{1,\ast}\right){}^2}{H^2}+1}+1\right)-\frac{\gamma ^2 e^{2 \gamma  \phi_\ast } V^{(2,0)}\left(\tau _{1,\ast},\phi_\ast \right)+V^{(0,2)}\left(\tau _{1,\ast},\phi_\ast \right)}{V\left(\tau _{1,\ast},\phi_\ast \right)}\right)\delta_i\\\nonumber
&&-\frac{\gamma ^2 e^{2 \gamma  \phi_\ast }}{V\left(\tau _{1,\ast},\phi_\ast \right){}^2 \left(H^2 e^{2 \gamma  \phi_\ast }+\left(Q-H \tau _{1,\ast}\right){}^2\right)}\times \\\nonumber
&& \left( 
\frac{\left(V^{(1,1)}\left(\tau _{1,\ast},\phi_\ast \right) \left(H^2 e^{2 \gamma  \phi_\ast }+\left(Q-H \tau _{1,\ast}\right){}^2\right)+2 \gamma  H \left(H \tau _{1,\ast}-Q\right) V\left(\tau _{1,\ast},\phi_,\ast \right)\right){}^2}{H^2 e^{2 \gamma  \phi_\ast }+\left(Q-H \tau _{1,\ast}\right){}^2}
 \right.\\\nonumber
&&
\left. \left(\gamma ^2 V\left(\tau _{1,\ast},\phi_\ast \right)-V^{(0,2)}\left(\tau _{1,\ast},\phi_\ast \right)\right) \left(V^{(2,0)}\left(\tau _{1,\ast},\phi_\ast \right) \left(H^2 e^{2 \gamma  \phi_\ast }+\left(Q-H \tau _{1,\ast}\right){}^2\right)-2 H^2 V\left(\tau _{1,\ast},\phi_\ast \right)\right) \right)\\
&& = 0 \,.\label{deteq}
\end{eqnarray}
The four solutions to this equations are then given by the following expression, with four different choices of the two signs:
\begin{eqnarray}\label{soleq}
\delta_i &=& \frac{1}{2} \pm \sqrt{A \pm \sqrt{B}}\\\nonumber
A &=& \frac{V\left(\tau _{1,\ast},\phi_\ast \right) \left(\left(6 \gamma ^2+1\right) H^2 e^{2 \gamma  \phi_\ast }+\left(2 \gamma ^2+1\right) \left(Q-H \tau _{1,\ast}\right){}^2\right)}{4 V\left(\tau _{1,\ast},\phi_\ast \right) \left(H^2 e^{2 \gamma  \phi_\ast }+\left(Q-H \tau _{1,\ast}\right){}^2\right)}\\\nonumber&&
- 2 \frac{ \left(\gamma ^2 e^{2 \gamma  \phi_\ast } V^{(2,0)}\left(\tau _{1,\ast},\phi_\ast \right)+V^{(0,2)}\left(\tau _{1,\ast},\phi_\ast \right)\right) \left(H^2 e^{2 \gamma  \phi_\ast }+\left(Q-H \tau _{1,\ast}\right){}^2\right)}{4 V\left(\tau _{1,\ast},\phi_\ast \right) \left(H^2 e^{2 \gamma  \phi_\ast }+\left(Q-H \tau _{1,\ast}\right){}^2\right)}\\\nonumber
B &=& \frac{1}{4 V\left(\tau _{1,\ast},\phi _\ast\right){}^4 \left(H^2 e^{2 \gamma  \phi _\ast}+\left(Q-H \tau _{1,\ast}\right){}^2\right){}^4}\times\\\nonumber&&
\left(
\left(4 \gamma ^2 e^{2 \gamma  \phi _\ast} V^{(1,1)}\left(\tau _{1,\ast},\phi _\ast\right){}^2+\left(V^{(0,2)}\left(\tau _{1,\ast},\phi _\ast\right) 
-\gamma ^2 e^{2 \gamma  \phi _\ast} V^{(2,0)}\left(\tau _{1,\ast},\phi _\ast\right)\right){}^2\right)\right. \times\\\nonumber&&
\left(H^2 e^{2 \gamma  \phi _\ast}+\left(Q-H \tau _{1,\ast}\right){}^2\right){}^2
 +2 \gamma ^2 V\left(\tau _{1,\ast},\phi _\ast\right) \left(V^{(0,2)}\left(\tau _{1,\ast},\phi _\ast\right) \left(H^4 e^{4 \gamma  \phi _\ast}-\left(Q-H \tau _{1,\ast}\right){}^4\right)\right. \\\nonumber&&
\left.\left.+\gamma  e^{2 \gamma  \phi _\ast} \left(\gamma  V^{(2,0)}\left(\tau _{1,\ast},\phi _\ast\right) \left(\left(Q-H \tau _{1,\ast}\right){}^4-H^4 e^{4 \gamma  \phi _\ast}\right)\right.\right.\right.\\\nonumber&&\left.\left.\left.
+8 H \left(H \tau _{1,\ast}-Q\right) V^{(1,1)}\left(\tau _{1,\ast},\phi _\ast\right) \left(H^2 e^{2 \gamma  \phi _\ast}+\left(Q-H \tau _{1,\ast}\right){}^2\right)\right)\right) \right.\\\nonumber&&\left.
+\gamma ^4 V\left(\tau _{1,\ast},\phi _\ast\right){}^2 \left(H^4 e^{4 \gamma  \phi _\ast}+14 H^2 e^{2 \gamma  \phi _\ast} \left(Q-H \tau _{1,\ast}\right){}^2+\left(Q-H \tau _{1,\ast}\right){}^4\right)
\right)
\end{eqnarray}
We conclude that, depending on their position $(\phi_\ast, \tau_{1,\ast})$, these fixed points will generically have relevant deformations if $\delta_i>0$ and hence can not be stable IR end points of RG flows. There are however special cases where they are unstable under condensation of these neutral scalars, which happens if the $\delta_i$ all become complex. In this case these $AdS_2$ fixed points will be replaced by other, possibly more well-behaved IR fixed points, probably of general hyperscaling violating Lifshitz form. 
Furthermore, the distribution of fixed points $\tau_\ast$ is not obviously independent of the magnetic field, and whether the physics does not depend on $h$ is a dynamical question that can only be answered by shooting from different IR fixed points with the same $q/h$ and $h$ and observing whether or not the field theory quantities in the UV do or do not depend on $h$. We defer the analysis of both issues to future work, as they are connected to the possible role these $AdS_2$ fixed points may play as the quantum critical points in the Quantum Hall plateaux transitions.

\begin{figure}
	\centering
		\includegraphics[height=0.4\textheight]{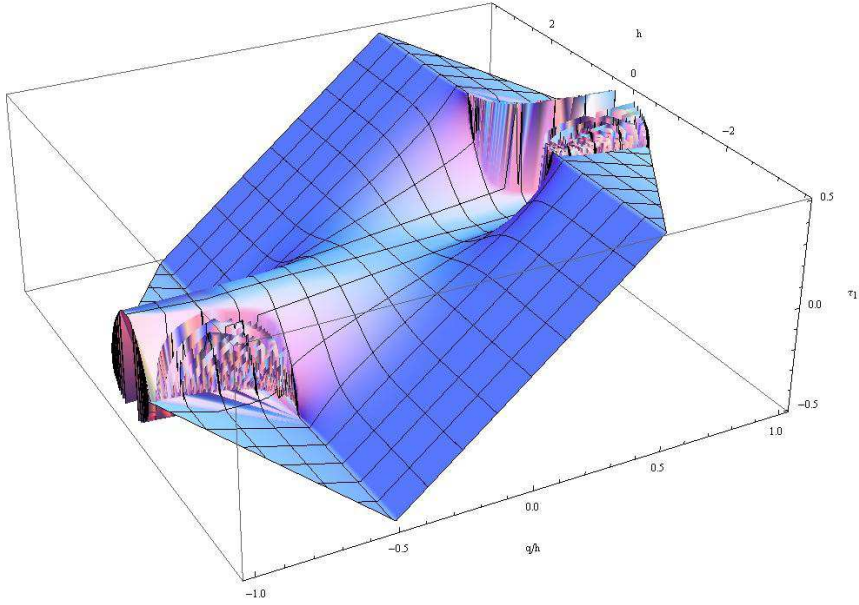}
	\caption{
	The position $\tau_{1\ast}(q/h,h)$ for $AdS_2$ fixed points, obtained from scanning with $\tau_\ast$ through the fundamental domain, evaluating \protect\eqref{qAdS2}-\protect\eqref{hAdS2}
	and inverting the data. We see that in the window $|h|\geq 1$, $|q/h|\leq \frac{1}{2}$ $\tau_{1\ast}$ does not depend on the magnetic field and that $\tau_{1\ast} = q/h$, but for smaller magnetic fields the position of the $AdS_2$ fixed point definitely depends on both $q/h$ and $h$.
	}
	\label{tau1qohh}
\end{figure}

\begin{figure}
	\centering
		\includegraphics[height=0.4\textheight]{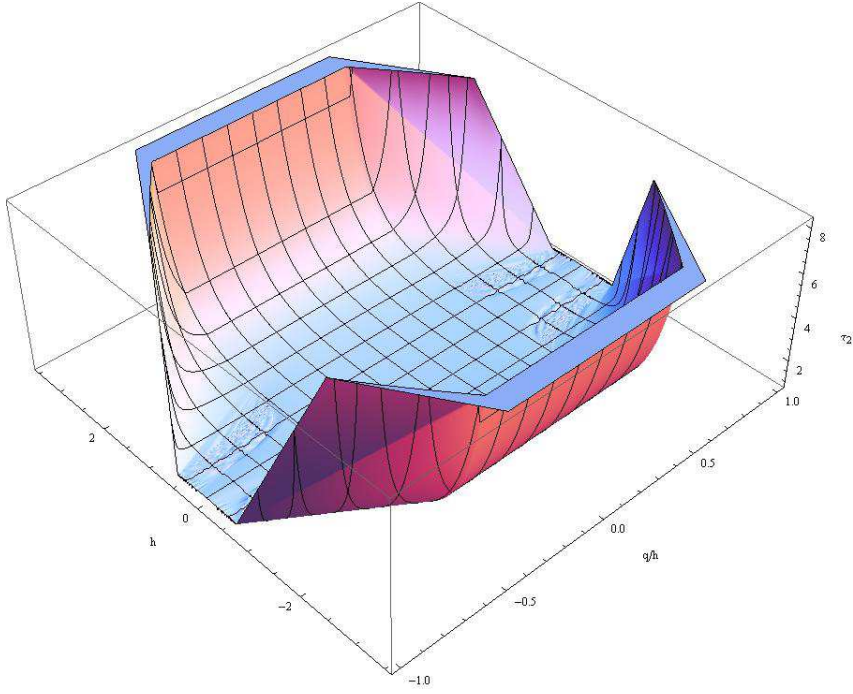}
	\caption{
	The position $\tau_{2\ast}(q/h,h)$ for $AdS_2$ fixed points, obtained from scanning with $\tau_\ast$ through the fundamental domain, evaluating \protect\eqref{qAdS2}-\protect\eqref{hAdS2}  and inverting the data. We see that $\tau_2$ is nearly independent of $q/h$ and $h$ for $|h|\leq 1$, but the $AdS_2$ fixed point shifts quickly upwards in the fundamental domain for larger values of $h$.
	}
	\label{tau2qohh}
\end{figure}



\subsection{Scaling solutions}

We rewrite the equations as
\footnote{In the absence of a magnetic field and axion $(h=0,\tau_1=0)$, (\ref{generalTrr+T33}) can be integrated to
\be
{\cal Q}=qA_t -C\sqrt{D\over B}\left({D'\over D}-{C'\over c}\right)={CZ\over \sqrt{BD}}A_tA_t'-
C\sqrt{D\over B}\left({D'\over D}-{C'\over c}\right)\sp {d{\cal Q}\over dr}=0
\ee
The constant ${\cal Q}$ is zero at extremality, therefore it must be a function of temperature vanishing when $T=0$.
In the presence of $h$ but for constant $\a$, \cite{KrausdHoker} find a similar constant.
We suspect that a generalization of this result also holds if the axion $\a$ is not constant.}
\be
BV={1\over 2}{D''\over D}+{1\over 2}{C''\over C}+{D'\over 4D}\left(3{C'\over C}-{D'\over D}\right)-{1\over 4}{B'\over B}\left( {C'\over C}+{D'\over D}\right)
\label{e21}\ee
\be
{W\over 2}\a'^2=-{C''\over C}-{1\over 2}\phi'^2+{1\over 2}\left({B'\over B}+{D'\over D}\right){C'\over C}
+{1\over 2}{C'^2\over C^2}
\label{e22}\ee
\be
{B\over C^2}\left({(q-h\a)^2\over Z}+{h^2Z}\right)={D''\over D}-{C''\over C}-{1\over 2}\left({D'\over D}-{C'\over C}\right)\left({D'\over D}+{B'\over B}\right)
\label{e23}\ee
the dilaton equation
\be
\sqrt{B\over DC^2}\partial_r\left(\sqrt{DC^2\over B}\phi'\right)+B{\pa V\over \pa \phi}+{B\over 2C^2}{Z'\over Z}\left[{(q-h\a)^2\over Z}-h^2Z\right]-{W'\a'^2\over 2}=0
\label{e24}\ee
and the axion equation
\be
W\sqrt{B\over DC^2}\partial_r\left(\sqrt{DC^2\over B}\alpha'\right)+B{\pa V\over \pa \alpha}+{W'\alpha'\phi'}+{B\over C^2}{h(q-h\a)\over Z}=0
\label{e12}\ee

Considering a scaling metric
\be
B=b_0 r^{b}\sp D=r^d\sp C=r^c\sp \phi=\kappa \log r\sp \a=a_0 r^{\lambda}
\label{e25}\ee
and $\pa_{\phi}V\simeq \gamma sV$, $\pa_a V\simeq 0$, ${Z'\over Z}=\gamma$, ${W'\over W}=-\gamma$, $Z=e^{\g \phi}$, $V=V_0 e^{\g s\phi}$.

Substituting in the equations we obtain
\be
r^2BV=k_1\sp r^2{a'^2\over Z^2}=\gamma^2 k_2\sp {r^2B\over C^2}{(q-h\a)^2\over Z}={k_3+k_4\over 2}
\sp {r^2B\over C^2}{h^2Z}={k_3-k_4\over 2}
\label{e26}\ee
\be
k_5 a+{r^2B\over C^2}h(q-ha)Z=0
\label{e27}\ee
with
\be
k_1={(c+d) (-2-b+2 c+d)\over 4}\sp k_2=c(2+b-c+d)-\kappa^2\sp k_3={(c-d) (2+b-2 c-d)\over 2}
\label{e28}\ee
\be
k_4={(\kappa(-2-b+2c+d)\over \gamma}-2sk_1-2k_2\sp k_5={\lambda(-2-b+2c+d+2\lambda-2\gamma \kappa)\over 2\g^2}
\label{e29}\ee
We must have
\be
k_1\geq 0\sp k_2\geq 0\sp k_3\geq 0\sp k_3\geq k_4
\label{e30}\ee
The second equation in (\ref{e26}) implies that
\be
a=\pm {\g\over \l}\sqrt{k_2}Z
\ee
so that the ratio $a/Z$ is asymptotically constant if $k_2\not=0$. Therefore they both grow or they both vanish in the IR.
The case where they both vanish can be mapped by an $SL(2,{\mathbb Z})$ transformation to one where they both grow; we will therefore assume without loss of generality that
$a$ and $Z$ grow indefinitely in the IR.
There are two possibilities now:

\paragraph{The case $h=0$} In this case the equations imply
\be
k_3=k_4\sp k_5=0\sp {r^2B\over ZC^2}={2k_3\over q^2}\sp r^2BV=k_1
\ee
which in turn give
\be
b_0={2k_3\over q^2}\sp 2k_3V_0=q^2k_1\sp 2b=\g\k+2c\sp \g(\k+2s)=-2(2+c)
\ee
This case should include the hyperscaling violating metrics we
have been using in this work, but also the neutral scaling violating metrics of \cite{Charmousis:2010zz}, where the
gauge field is subdominant in the equations \cite{lines}.

\paragraph{The case with magnetic field $h\not=0$} In this case the equations imply
\be
{r^2BZ\over C^2}={k_3-k_4\over 2h^2}\sp \l^2(k_3+k_4)=\g^2k_2(k_3-k_4)\sp \pm 2\g k_5\sqrt{k_2}=\l(k_3-k_4)
\ee
\be
V_0b_0=k_1\sp 2+b+\g\k s=0\sp 2h^2k_1=V_0(k_3-k_4)\sp b-2c+2+\g\k=0
\ee
These are five equations for five unknowns, and we therefore expect a generic solution.
However, these might not be the only solutions. There may be special cases, and in order to find them we need to
\begin{enumerate}
\item One must choose a subset of the terms $BV$, $W\a'^2$, ${B(q-h\a)^2\over ZC^2}$, ${Bh^2Z\over C^2}$ and assume they vanish to leading order in the IR  solution. This already imposes some constraints on the powers. For example if $BV$ is subleading, then from (\ref{e26}) we obtain the equation  $k_1=0$.
and so on. We then must solve the remaining equations (\ref{e26}), (\ref{e27}) as done above.
\item After a solution is found, its consistency must be checked: Does the solution satisfy all equations above (even redundant ones)?
After calculating the leading power of the terms we set to zero by perturbation theory of the leading equations, is this power indeed subleading?
\end{enumerate}
This procedure has been used in \cite{lines}, and it will be interesting to complete this classification together with an investigation of the quantum Hall plateaux transitions in a future work.


\end{document}